\documentclass[floatfix,preprint,rmp,apsrmp,showpacs,,amsmath]{revtex4}
\usepackage{graphicx}
\usepackage{epsfig}
\def \b{{\cal B}}
\def \bea{\begin{eqnarray}}
\def \beq{\begin{equation}}

\def \eea{\end{eqnarray}}
\def \eeq{\end{equation}}

\def \ket#1{| #1 \rangle}
\def \s{\sqrt{2}}
\def \st{\sqrt{3}}
\def \sx{\sqrt{6}}

\begin{document}
\renewcommand{\thetable}{\Roman{table}}

\preprint{CLNS 07/1988}
\preprint{EFI-06-15}
\preprint{FERMILAB-PUB-07-006-T}
\preprint{hep-ph/0701208}

\title{Quarkonia and their transitions}
\author{Estia Eichten}
\affiliation{Fermilab, \\ P. O. Box 500, Batavia, IL 60510}
\author{Stephen Godfrey}
\affiliation{Ottawa Carleton Institute for Physics,
Department of Physics, \\
Carleton University,
1125 Colonel By Drive, Ottawa, ON K1S 5B6, Canada}
\author{Hanna Mahlke}
\affiliation{Laboratory for Elementary-Particle Physics, \\
Cornell University, Ithaca, NY 14853}
\author{Jonathan L. Rosner}
\affiliation{Enrico Fermi Institute and Department of Physics, \\
University of Chicago, 5640 South Ellis Avenue, Chicago, IL 60637}

\date{September 10, 2007}

\begin{abstract}
Valuable data on quarkonia (the bound states of a heavy quark $Q=c,b$ and the
corresponding antiquark) have recently been provided by a variety of sources,
mainly $e^+ e^-$ collisions, but also hadronic interactions.  This permits
a thorough updating of the experimental and theoretical status of
electromagnetic and strong transitions in quarkonia.  We discuss $Q \bar Q$
transitions to other $Q \bar Q$ states, with some reference to processes
involving $Q \bar Q$ annihilation.
\end{abstract}

\maketitle

\bigskip

\tableofcontents

\section{Introduction}
\bigskip

Quarkonium spectroscopy has celebrated a great resurgence in the past few
years thanks to a wealth of new information, primarily from electron-positron
colliders, but also from hadronic interactions.  Transitions between quarkonium
states shed light on aspects of quantum chromodynamics (QCD), the theory
of the strong interactions, in both the perturbative and the non-perturbative
regimes.  In the present article we review the new information on these states
and their transitions and indicate theoretical implications, updating
earlier discussions such as those in
\citet{Kwong:1987mj,Kwong:1988ae,Kwong:1987ak,Godfrey:2001eb,Godfrey:2001vc,Godfrey:2002rp,Brambilla:2004wf,Eichten:2004uh,Barnes:2003vb}
(which may be consulted for explicit formulae).

We shall deal with states composed of a heavy quark $Q = c$ or $b$ and the
corresponding antiquark $\bar Q$.  We shall discuss $Q \bar Q$ transitions
primarily to other $Q \bar Q$ states, with some reference to processes
involving $Q \bar Q$ annihilation, and will largely bypass decays to open
flavor (treated, for example, 
in~\citet{Brambilla:2004wf,Eichten:2004uh,Barnes:2003vb,Eichten:2005ga,Barnes:2005pb}).

A brief overview of the data on the charmonium and $\Upsilon$ systems is
provided in Section~\ref{sec:overview}.
We then review theoretical underpinnings in Section~\ref{sec:theo_underpinnings}
discussing quarks and potential models, lattice gauge theory approaches,
perturbative QCD and decays involving gluons, and hadronic transitions of the
form $Q \bar Q \to (Q \bar Q)' +$ (light hadrons).
Section~\ref{sec:charmonium} is devoted to
charmonium and Section~\ref{sec:bottomonium}
to the $b \bar b$ levels and includes a brief mention of
interpolation to the $b \bar c$ system.
Section~\ref{sec:summary} summarizes.

\section{Overview of quarkonium levels}
\label{sec:overview}

\begin{figure}
\begin{center}
\includegraphics[width=5in]{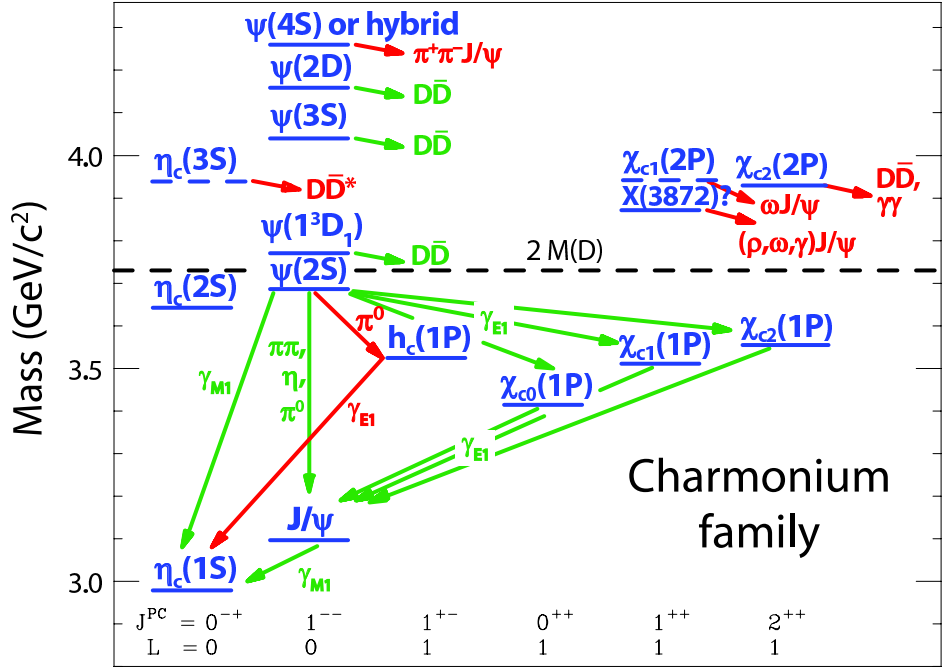}
\caption{Known charmonium states and candidates, with selected decay
modes and transitions.
Red (dark) arrows denote recent observations.
\label{fig:charmon}}
\end{center}
\end{figure}

\begin{figure}
\begin{center}
\includegraphics[height=0.38\textheight]{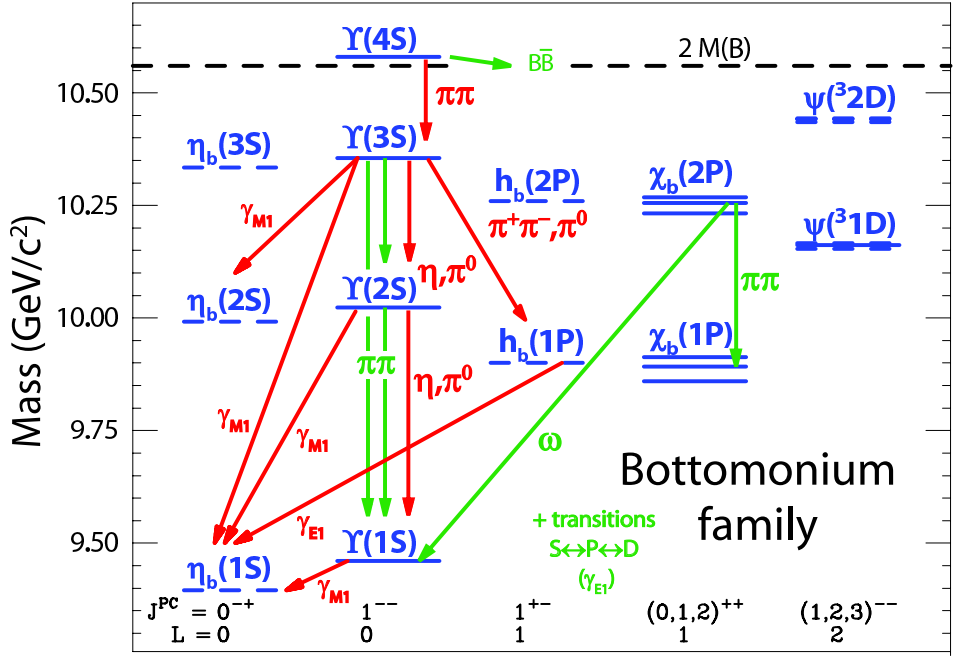}
\end{center}
\caption{Transitions among $b \bar b$ levels.  There are also numerous
electric dipole transitions $S \leftrightarrow P \leftrightarrow D$ (not
shown).  Red (dark) arrows denote objects of recent searches.
\label{fig:ups}}
\end{figure}

Since the discovery of the $J/\psi$ more than thirty years ago, information
on quarkonium levels has grown to the point that more is known about the $c
\bar c$ and $b \bar b$ systems than about their namesake positronium, the
bound state of an electron and a positron.  The present status of charmonium
$(c \bar c)$ levels is shown in Fig.\ \ref{fig:charmon}, while that of
bottomonium ($b \bar b$) levels is shown in Fig.\ \ref{fig:ups}.
The best-established states are summarized in Tables~\ref{tab:charmonium}
and~\ref{tab:bottomonium}.

The levels are labeled by $S$, $P$, $D$, corresponding to relative orbital
angular momentum $L = 0,~1,~2$ between quark and antiquark.  (No candidates
for $L \ge 3$ states have been seen yet.)  The spin of the quark and antiquark
can couple to either $S=0$ (spin-singlet) or $S=1$ (spin-triplet) states.
The parity of a quark-antiquark state with orbital angular momentum $L$ is
$P = (-1)^{L+1}$; the charge-conjugation eigenvalue is $C = (-1)^{L+S}$.
Values of $J^{PC}$ are shown at the bottom of each figure.  States are
often denoted by $^{2S+1}[L]_J$, with $[L] = S,~P,~D, \ldots$.  Thus, $L=0$
states can be $^1S_0$ or $^3S_1$; $L=1$ states can be $^1P_1$ or
$^3P_{0,1,2}$; $L=2$ states can be $^1D_2$ or $^3D_{1,2,3}$, and so on.
The radial quantum number is denoted by $n$.

\begin{table*}
\begin{minipage}{\textwidth}
\caption{Observed charmonium states. All numbers are quoted
from \citet{Yao:2006px}. More recent information is
included in the text, where available.
\label{tab:charmonium}}
\begin{center}
\begin{tabular}{c|c|c|c|c|c|c} \hline \hline
\multicolumn{4}{c|}{Quantum numbers}
& Name         & Mass & Width \\
$n$  & $L$ & $J^{PC}$ & $n^{2S+1}L_J$
&              &  MeV &  MeV\footnote{
Unless noted otherwise.} \\
\hline
1    &  0  & $0^{-+}$ & $1^1S_0$ & $\eta_c(1S)$ & $2980.4 \pm 1.2 $
& $25.5 \pm 3.4$ \\
1    &  0  & $1^{--}$ & $1^3S_1$ & $J/\psi$     & $3096.916 \pm 0.011$
& $93.4 \pm 2.1$ keV \\
1    &  1  & $0^{++}$ & $1^3P_0$ & $\chi_{c0}(1P)$  & $ 3414.76 \pm 0.35 $
& $10.4 \pm 0.7$ \\
1    &  1  & $1^{++}$ & $1^3P_1$ & $\chi_{c1}(1P)$  & $ 3510.66\pm 0.07 $
& $0.89 \pm 0.05$ \\
1    &  1  & $2^{++}$ & $1^3P_2$ & $\chi_{c2}(1P)$  & $ 3556.20\pm 0.09 $
& $2.06\pm 0.12$ \\
1    &  1  & $1^{+-}$ & $1^1P_1$ & $h_c(1P)$  & $ 3525.93 \pm 0.27$
& $<1$ \\
1    &  2  & $1^{--}$ & $1^3D_1$ & $\psi(3770)$ & $ 3771.1 \pm 2.4 $
& $23.0 \pm 2.7$ \\
2    &  0  & $0^{-+}$ & $2^1S_0$ & $\eta_c(2S)$ & $3638\pm 4 $
& $14 \pm 7 $ \\
2    &  0  & $1^{--}$ & $2^3S_1$ & $\psi(2S)$   & $3686.093 \pm 0.034$
& $337\pm 13$ keV \\
2    &  1  & $2^{++}$ & $2^3P_2$ & $\chi_{c2}(2P)$  & $ 3929 \pm 5$
& $ 29 \pm 10 $ \\
\hline \hline
\end{tabular}
\end{center}
\end{minipage}
\end{table*}

\begin{table*}
\begin{minipage}{\textwidth}
\caption{Observed bottomonium states. All numbers are quoted
from \citet{Yao:2006px}. More recent information is
included in the text, where available.
\label{tab:bottomonium}}
\begin{center}
\begin{tabular}{c|c|c|c|c|c|c} \hline \hline
\multicolumn{4}{c|}{Quantum numbers}
& Name         & Mass & Width \\
$n$  & $L$ & $J^{PC}$ & $n^{2S+1}L_J$
&              &  MeV &  \\
\hline
1    &  0  & $1^{--}$ & $1^3S_1$ & $\Upsilon(1S)$     & $9460.30 \pm 0.26$
& $54.02 \pm 1.25$ keV \\
1    &  1  & $0^{++}$ & $1^3P_0$ & $\chi_{b0}(1P)$  & $ 9859.44 \pm 0.52 $
& unknown \\
1    &  1  & $1^{++}$ & $1^3P_1$ & $\chi_{b1}(1P)$  & $ 9892.78 \pm 0.40$
& unknown \\
1    &  1  & $2^{++}$ & $1^3P_2$ & $\chi_{b2}(1P)$  & $ 9912.21 \pm 0.40$
& unknown \\
1    &  2  & $2^{--}$ & $1^3D_J$\ \footnote{Probably all or mostly $J=2$.}
& $\Upsilon(1D)$ &  $ 10161.1 \pm 1.7$
& unknown \\
2    &  0  & $1^{--}$ & $2^3S_1$ & $\Upsilon(2S)$     & $10023.26 \pm 0.31$
& $31.98 \pm 2.63$ keV \\
2    &  1  & $0^{++}$ & $2^3P_0$ & $\chi_{b0}(2P)$  & $ 10232.5 \pm 0.6$
& unknown \\
2    &  1  & $1^{++}$ & $2^3P_1$ & $\chi_{b1}(2P)$  & $ 10255.46 \pm 0.55$
& unknown \\
2    &  1  & $2^{++}$ & $2^3P_2$ & $\chi_{b2}(2P)$  & $ 10268.65 \pm 0.55$
& unknown \\
3    &  0  & $1^{--}$ & $3^3S_1$ & $\Upsilon(3S)$     & $10355.2 \pm 0.5$
& $20.32 \pm 1.85$ keV \\
4    &  0  & $1^{--}$ & $4^3S_1$ & $\Upsilon(4S)$     & $10579.4 \pm 1.2$
& $20.5 \pm 2.5$ MeV \\
\hline \hline
\end{tabular}
\end{center}
\end{minipage}
\end{table*}

\section{Theoretical underpinnings}
\label{sec:theo_underpinnings}

\subsection{Quarks and potential models}

An approximate picture of quarkonium states may be obtained by describing
them as bound by an interquark force whose short-distance behavior is
approximately Coulombic (with an appropriate logarithmic modification of
coupling strength to account for asymptotic freedom) and whose long-distance
behavior is linear to account for quark confinement.  An example of this
approach is found in \citet{Eichten:1974af,Eichten:1975ag,Eichten:1978tg,Eichten:1979ms}; early reviews may be found in 
\citet{Novikov:1977dq,Appelquist:1978aq,Quigg:1979vr,Grosse:1979xm}.
\citet{Radford:2007vd} presents more recent results.

\subsubsection{Validity of nonrelativistic description}

In order to estimate whether a nonrelativistic (NR) quarkonium description makes
sense, ``cartoon'' versions of $c \bar c$ and $b \bar b$ spectra may be
constructed by noting that the level spacings are remarkably similar in the
two cases.  They would be exactly equal if the interquark potential were of the
form $V(r) = C \log(r/r_0)$ (see \citet{Quigg:1977dd}), 
which may be regarded as a
phenomenological interpolation between the short-distance $\sim -1/r$ and
long-distance $\sim r$ behaviors expected from QCD.  In such a potential
the expectation value of the kinetic energy $\langle T \rangle = (r/2)
(dV/dr)$ is just $C/2 \simeq 0.37$ GeV with $C=0.733$ as found
in \citet{Quigg:1979vr}.  Since $\langle T \rangle = 2
\cdot (1/2) m_Q \langle v^2 \rangle$, one has $\langle v^2 \rangle \simeq
0.24$ for a charmed quark of mass $m_c \simeq 1.5$ GeV/$c^2$ (roughly
half the $J/\psi$ mass), and $\langle v^2 \rangle \simeq 0.08$ for a $b$
quark of mass $m_b \simeq 4.9$ GeV/$c^2$ (roughly half the $\Upsilon(1S)$
mass).  Thus a nonrelativistic description for charmonium is quite crude,
whereas it is substantially better for $b \bar b$ states.

\subsubsection{Role of leptonic partial widths:  $|\Psi(0)|^2$}

The partial widths for $^3S_1$ states to decay to a lepton pair through a
virtual photon are a probe of the squares $|\Psi_n(0)|^2$ of the relative
$n^3S_1$ wave functions at the origin through the relation 
\cite{VanRoyen:1967nq}
\beq
\Gamma(n^3S_1 \to e^+ e^-) =   \frac{16 \pi \alpha^2 e_Q^2 |\Psi_n(0)|^2} {M_n^2}
\times
\left( 1 - \frac{16 \alpha_S}{3 \pi} + \ldots \right)~~~,
\eeq
where $e_Q = 2/3$ or $-1/3$ is the quark charge, $M_n$ is the mass of the
$n^3S_1$ state, and the last term is a QCD correction \cite{Kwong:1987ak}.
Thus leptonic partial widths are a probe of the compactness of
the quarkonium system, and provide important information complementary to
level spacings.  Indeed, for the phenomenologically adequate potential
$V(r) = C \log(r/r_0)$, a change in the quark mass $m_Q$ can be compensated by
a change in $r_0$ without affecting quarkonium mass predictions
($r_0$ can be viewed as setting the overall energy scale), whereas a larger
quark mass will lead to a spatially more compact bound state and hence
to an increased value of $|\Psi(0)|^2$ for each state.  A more general form is
the power-law potential, $V(r) \sim {\rm sgn}(\nu) r^\nu$, which approaches the
logarithmic potential in the limit of $\nu \to 0$.  One can show that in the
power-law potential lengths scale as $m_Q^{-1/(2+\nu)}$ and hence $|\Psi(0)|^2$
scales as $m_Q^{3/(2+\nu)}$, or $\sim m_Q^3, m_Q^{3/2}, m_Q$ for $\nu=-1,0,1$
\cite{Quigg:1979vr}.  (In charmonium and bottomonium the ground states have
sizes of about 0.4--0.5 fm and 0.2 fm, respectively \cite{Quigg:1981bj}.)
Thus the effective quark mass in a potential
description is constrained by measured leptonic widths.  One can expect that
in descriptions such as lattice gauge theories, to be
discussed in Section~\ref{sec:TheoryLQCD}, similar constraints will hold.

The scaling of leptonic widths from the charmonium to the bottomonium family
can be roughly estimated using the above discussion, assuming an effective
power $\nu \simeq 0$.  In that case the leptonic width for each $n$ scales
as $\Gamma_{ee}(nS) \propto e_Q^2 |\Psi(0)|^2 /m_Q^2 \propto e_Q^2/ m_Q^{1/2}$.
As the QCD correction in Eq.\ (1) is appreciable [as are relativistic
corrections, particularly for charmonium], this is only an approximate rule.

The important role of leptonic widths is particularly evident in constructions
of the interquark potential based on inverse-scattering methods
\cite{Quigg:1981bj,Thacker:1977aq,Thacker:1977ar,Schonfeld:1979cd,Kwong:1985ti}.
The reduced radial wave functions $u_{nS}(r) = r \Psi_{nS}(r)$ on the interval
$0 \le r < \infty$ for an $S$-wave Schr\"odinger equation with central potential
$V(r)$ may be regarded as the odd-parity levels (since they must vanish at
$r=0$) in a symmetric potential $V(-r) = V(r)$ on the interval $-\infty < r <
\infty$.
The even-parity levels [with $u(0)\neq 0$] do not correspond to bound
states but, rather, equivalent information is provided by the leptonic
widths of the $nS$ levels, which gives the quantities
$|\Psi(0)|=|u'_{nS}(0)|$.
Thus, if QCD and relativistic corrections can be
brought under control, leptonic widths of the $S$-wave levels are every bit as
crucial as their masses.

A recent prediction of the leptonic width ratio $\Gamma_{ee}[\Upsilon(2S)]/
\Gamma_{ee}[\Upsilon(1S)] = 0.43 \pm 0.05$ in lattice QCD \cite{Gray:2005ur}
raises the question of what constitutes useful measurement and prediction
precisions, both for ratios and for absolute leptonic widths.  (For comparison,
the CLEO Collaboration has measured this ratio to be $0.457 \pm 0.006$
\cite{Rosner:2005eu}.)  Potential models have little trouble in predicting
ratios $\Gamma_{ee}(n'S)/\Gamma_{ee}(nS)$ to an accuracy of a few percent, and
one would thus hope for lattice approaches eventually to be capable of similar
accuracy.  Much more uncertainty is encountered by potential modes in
predicting {\it absolute} leptonic widths as a result of QCD and relativistic
corrections (see, for example, the inverse-scattering approach of 
\citet{Quigg:1981bj}).  Measurements with better than a few percent accuracy,
such as those in \citet{Rosner:2005eu} and others to be discussed
presently, thus outstrip present theoretical capabilities.

\subsubsection{Spin-dependent interactions}

Hyperfine and fine-structure splittings in quarkonium are sensitive to
the Lorentz structure of the interquark
interaction~\cite{Kwong:1987mj,Brambilla:2004wf,Novikov:1977dq,Appelquist:1978aq}.
One may regard the effective potential $V(r)$ as the sum of Lorentz vector
$V_V$ and Lorentz scalar $V_S$ contributions.  The spin-spin interaction is
due entirely to the Lorentz vector:
\beq
V_{SS}(r) = \frac{\sigma_Q \cdot \sigma_{\bar Q}}{6 m_Q^2} \nabla^2 V_V(r)~~~,
\eeq
where $\sigma_Q$ and $\sigma_{\bar Q}$ are Pauli matrices acting on the spins
of the quark and antiquark, respectively.  For a Coulomb-like potential $\sim
-1/r$ the Laplacian is proportional to $\delta^3(r)$, so that $V_{SS}(r)$
contributes to hyperfine splittings only for $S$ waves, whose wave functions
are non-zero at the origin.  In QCD the coupling constant undergoes slow
(logarithmic) variation with distance, leading to small non-zero contributions
to hyperfine splittings for $L > 0$ states.  Relativistic corrections also
result in small non-zero contributions to these splittings.

Both spin-orbit and tensor forces affect states with $L > 0$.
The spin-orbit potential is
\beq
V_{LS}(r) = \frac{L \cdot S}{2 m_Q^2 r}
\left(3 \frac{dV_V}{dr} - \frac{dV_S}{dr} \right)~~~,
\eeq
where $L$ is the relative orbital angular momentum of $Q$ and $\bar Q$, while
$S$ is the total quark spin.
The tensor potential is~\cite{Messiah:Vol2,Radford:2007vd}
\beq
V_T(r) = \frac{S_{T}}{12 m_Q^2} \left( \frac{1}{r} \frac{dV_V}{dr}
- \frac{d^2V_V}{dr^2} \right)~~~,
\eeq
with $S_{T} \equiv 2[3(S \cdot \hat{r})(S \cdot \hat{r}) - S^2]$
(where $S = S_Q + S_{\bar Q}$ is the total spin operator and $\hat r$ is
a unit vector) has
non-zero expectation values only for $L > 0$
[e.g., $-4,2,-2/5$ for $^3P_{0,1,2}$ states].

\subsection{QCD on the lattice}
\label{sec:TheoryLQCD}

At momentum scales less than about 2 GeV/$c$ (distance scales greater than
about 0.1 fm) the QCD coupling constant $\alpha_S(Q^2)$ becomes large enough
that perturbation theory cannot be used.  The value $\alpha_S(m_\tau^2) = 0.345
\pm 0.010$ \cite{Bethke:2006ac,Kluth:2006vf,Davier:2007ym} is just about at the limit of
usefulness of perturbation theory, and $\alpha_S(Q^2)$ increases rapidly below
this scale.  One must resort to non-perturbative methods to describe
long-distance hadronic interactions.

If space-time is discretized, one can overcome the dependence in QCD on
perturbation theory.  Quark confinement is established using this lattice
gauge theory approach.    For low-lying heavy quarkonium states,
below the threshold for Zweig-allowed decay to open heavy flavor
mesons, an accurate description of the
spectrum can be obtained,
once one takes account of the degrees of freedom
associated with the production of pairs of light ($u,d,s$) quarks
\cite{Davies:2003ik}. For example,  recent lattice calculations of the
spin-splitting between $J/\psi$ and $\eta_c$ yield
$111\pm 5$ MeV \cite{Follana:2006rc},
while the experimental value is  $117.1 \pm 1.2$ MeV.

Above threshold, the situation is more challenging:
(1) Heavy quarkonium states have more typical-size hadronic widths;
(2) Such states are usually not
the ground state for a given set of quantum numbers; and
(3) These resonances are embedded in a multibody continuum.
In the lattice approach, information is extracted from
Euclidean correlation functions. This makes dealing with
excited-state resonances in a multibody continuum particularly
difficult \cite{Bulava:2007ka}.

Lattice QCD also provides a theoretical underpinning for the phenomenological
potential model approach.  The well-measured static energy between a heavy
quark-antiquark pair justifies the form of the nonrelativistic potential
\cite{Bali:2000gf}.  Recently, high-accuracy lattice calculations of the
spin-dependent potentials have also been made \cite{Koma:2005nq,Koma:2006fw}.  This
approach allows the direct determination of the spin-orbit, spin-spin and
tensor potentials as well.  At present, these spin-dependent
potential calculations have not yet included the effects of light quark loops.

\subsection{Electromagnetic transitions}
\label{sec:emagntrans}

The theory of electromagnetic (EM) transitions between quarkonium states is
straightforward, with terminology and techniques familiar from the
study of EM transitions in atomic and nuclear systems.  Although
electromagnetic transition amplitudes can be computed from first
principles in lattice QCD, these calculations are in their infancy. At
the present time, only potential model approaches provide the detailed
predictions that can be compared to experimental results.  In this
approach, the spatial dependence of EM transition amplitudes reduces to
functions of quark position and momentum between the initial and final
state wave functions. Expanding the matrix elements in powers of
photon momentum generates the electric and magnetic multipole moments
and is also an expansion in powers of velocity. The leading order
transition amplitudes are electric dipole (E1) and magnetic dipole (M1).
In what follows we shall take $m_c = 1.5$ GeV/$c^2$ and
$m_b = 4.9$ GeV/$c^2$ \cite{Kwong:1987ak}, which are considered
``constituent-quark'' values, appropriate to the non-perturbative
regime found in charmonium and bottomonium.

\subsubsection{Magnetic dipole transitions}

Magnetic dipole transitions flip the quark spin, so their amplitudes are
proportional to the quark magnetic moment and therefore inversely proportional
to the constituent quark mass.  At leading order the magnetic dipole (M1)
amplitudes between $S$-wave states are independent of the potential model:  The
orthogonality of states guarantees, in the limit of zero recoil, that the
spatial overlap is one for states within the same multiplet and zero for
transitions between multiplets which have different radial quantum numbers.

Including relativistic corrections due to spin dependence in the
Hamiltonian spoils this simple scenario and induces a small overlap
between states with different radial quantum numbers.  Such $n\neq n'$
transitions are referred to as ``hindered''.  Including finite size
corrections the rates are given by \cite{Eichten:1974af,Eichten:1975ag,Eichten:1978tg,Eichten:1979ms}
\begin{widetext}
\beq \label{eqn:M1}
\left\{ \begin{array}{c} \Gamma(n^3S_1 \to {n'}^1S_0 + \gamma) \\
\Gamma(n^1S_0 \to {n'}^3S_1 + \gamma) \end{array} \right\}
= 4\alpha e_Q^2 k^3 (2 J_f +1) | \langle f | j_0(kr/2) | i \rangle |^2/3 m_Q^2
~~~,
\eeq
\end{widetext}
where $e_Q = 2/3$ or $-1/3$ is the quark charge, $k$ is the photon energy,
$j_0(x) = \sin x / x$, and $m_Q$ is the quark mass.  The only M1~transitions
between quarkonia states so far observed occur in charmonium, but the
corresponding transitions in $b \bar b$ systems are the objects of current
searches.  For small $k$, $j_0(kr/2) \to 1$, so that transitions with $n' = n$
have favored matrix elements, though the corresponding partial decay widths
are suppressed by smaller $k^3$ factors.

Numerous papers have studied these M1 transitions including full
relativistic corrections
\cite{Godfrey:2001eb,Zambetakis:1983te,Grotch:1984gf,Godfrey:1985xj,Zhang:1991et,Ebert:2002pp,Lahde:2002wj}.  They depend
explicitly on the Lorentz structure of the nonrelativistic potential.
Several sources of uncertainty make M1 transitions particularly difficult
to calculate.  In addition to
issues of relativistic corrections and what are known as ``exchange currents,''
there is the possibility of an
anomalous magnetic moment of the quark ($\kappa_Q$).  Furthermore,
the leading-order results depend explicitly on the constituent
quark masses, and corrections depend on the Lorentz structure of the potential.

\subsubsection{Electric dipole transitions}

The partial widths for electric dipole (E1) transitions between
states $^{2S+1}3{L_i}_{J_i}$ and $^{2S+1}{L_f}_{J_f}$ are
given by \cite{Eichten:1974af,Eichten:1975ag,Eichten:1978tg,Eichten:1979ms}
\begin{widetext}
\beq \label{eqn:E1}
\Gamma(n^{2S+1}{L_i}_{J_i} \to n'^{2S+1}{L_f}_{J_f} + \gamma)
= \frac{4 \alpha e_Q^2 k^3}{3} (2 J_f +1) {\cal S}_{if}| \langle f | r | i \rangle |^2
~~~.
\eeq
\end{widetext}
The statistical factor ${\cal S}_{if}$ is
\beq \label{enq:E1S}
{\cal S}_{if} = {\cal S}_{fi} = {\rm max}(L_i, L_f)
\left\{ \begin{array}{ccc} J_i & 1 & J_f \\ L_f & S & L_i \end{array}  \right\}.
\eeq
For transitions between spin-triplet $S$-wave and $P$-wave states, ${\cal S}_{if} =
\frac{1}{9}$.  Expressions for $P \leftrightarrow D$ transitions, which have
also been observed both in charmonium and in the $b \bar b$ system, are given,
for example, in \citet{Kwong:1988ae}.

The leading corrections for electric dipole corrections
have been considered by a number of authors
\cite{Zambetakis:1983te,Grotch:1984gf,Godfrey:1985xj,Ebert:2002pp,Lahde:2002wj,Feinberg:1975hk,Sucher:1978wq,Kang:1978yw,Moxhay:1983vu,McClary:1983xw,Grotch:1982bi}.
A general form was derived by Grotch, Owen and Sebastian \cite{Grotch:1984gf}.
There are three main types of corrections:
relativistic modification of the nonrelativistic wave functions,
relativistic modification of the electromagnetic transition operator,
and finite-size corrections.  In addition to these there are
additional corrections arising from the quark anomalous magnetic
moment. For the $^3P_J \leftrightarrow \mbox{}^3S_1$ transitions in which we
are primarily interested, the dominant relativistic corrections arise from
modifications of the wavefunctions and are included by the quarkonium analog of
Siegert's theorem \cite{McClary:1983xw,Siegert:1937yt}.  We
will find that differences in theoretical assumptions of the various
potential models make it difficult to draw sharp conclusions from the
level of agreement of a particular model with experimental data.
However, there is usually very little model variation in the NR
predictions if the models are fit to the same states
\cite{Kwong:1988ae}.  The only exceptions are transitions where the
dipole matrix element exhibits large dynamical cancellations, for
instance when higher radial excitations are involved which have nodes
in their wavefunctions.

\subsubsection{Higher multipole contributions in charmonium}

Magnetic quadrupole (M2) amplitudes are higher order in $v^2/c^2$.  They are
of interest because they provide an indirect measure of the charmed quark's
magnetic moment \cite{Karl:1975jp,Karl:1980wm} and are sensitive to $D$-wave admixtures in
$S$-wave states, providing another means of studying the $1^3D_1-2^3S_1$ mixing
in the $\psi(3770) - \psi(2S)$ states
\cite{Godfrey:1985ei,Sebastian:1992xq}.  They
affect angular distributions in decays such as
$\psi(2S) \to \chi_{cJ} + \gamma$
and $\chi_{cJ} \to J/\psi + \gamma$ and become experimentally accessible
through interference with the dominant E1~amplitudes.

The $\chi_{cJ} \to \gamma J/\psi$ or $\psi(2S) \to \gamma \chi_{cJ}$ decays may
be described by the respective helicity amplitudes $A_\lambda$ or $A'_\lambda$,
in which $\lambda$ labels the projection of the spin of the $\chi_{cJ}$ parallel
(for $A_\lambda$) or antiparallel (for $A'_\lambda$) to the photon, which is
assumed to have helicity $+1$.  The radiative widths are given in terms of
these amplitudes by
\bea
\Gamma(\psi(2S) \to \gamma \chi_{cJ}) & = & \frac{E_\gamma^3}{3}
\sum_{\lambda \ge 0} |A'_\lambda|^2~~,\\
\Gamma(\chi_{cJ} \to J/\psi) & = & \frac{E_\gamma^3}{2J+1}
\sum_{\lambda \ge 0} |A_\lambda|^2~~.
\eea
In terms of a parameter $\epsilon \equiv \xi E_\gamma/(4 m_c)$, where $\xi =
-1$ for $\psi(2S) \to \gamma \chi_{cJ}$ and $\xi = +1$ for $\chi_{cJ} \to \gamma
J/\psi$, the predicted helicity amplitudes $A_\lambda$ or $A'_\lambda$ are in
the relative proportions \cite{Karl:1975jp,Karl:1980wm}:
\bea
\chi_{c2}:~~ A_2 & = & \sx [1 + \epsilon(1 + \kappa_c)] \\
A_1 & = & \st [1 - \epsilon(1 + \kappa_c)] \\
A_0 & = &  \quad [ 1 - 3 \epsilon(1 + \kappa_c)] \\
\chi_{c1}:~~ A_1 & = & \st [1 + \epsilon(1 + \kappa_c)] \\
A_0 & = & \st [1 - \epsilon(1 + \kappa_c)] \\
\chi_{c0}:~~ A_0 & = & \s [1 - 2\epsilon(1 + \kappa_c)] ~~~.
\eea
Here an overall E1 amplitude has been factored out, and $\kappa_c$ is the
charmed quark's anomalous magnetic moment.

\subsection{Perturbative QCD and decays involving gluons}

Many quarkonium decays proceed through annihilation of $Q \bar Q$ into
final states consisting of gluons and possibly photons and light-quark
pairs.  Expressions for partial widths of color-singlet $Q \bar Q$ systems
are given in \citet{Kwong:1987ak}, and have been updated in 
\citet{Petrelli:1997ge}.  In that work, annihilation rates are also given
for the color-octet component of the $Q \bar Q$ system, which appears
necessary for successful description of $Q \bar Q$ production in hadronic
interactions.  We shall confine our discussion to the effects of the
color-singlet $Q \bar Q$ component in decays.  Discrepancies between theory
and experiment can be ascribed in part to neglected relativistic effects
(particularly in charmonium) and in part to the neglected color-octet
component.

\subsection{Hadronic transitions [$Q \bar Q \to (Q \bar Q)' +$ (light
hadrons)]}

A number of transitions from one $Q \bar Q$ state to another occur with the
emission of light hadrons.  So far, the observed transitions in charmonium
include
$\psi(2S) \to J/\psi \pi^+ \pi^-$,
$\psi(2S) \to J/\psi \pi^0 \pi^0$,
$\psi(2S) \to J/\psi\eta$,
$\psi(2S) \to J/\psi \pi^0$,
and $\psi(2S) \to h_c \pi^0$.
In addition, above charm threshold a state $X(3872)$ decays to $J/\psi
\pi^+ \pi^-$, and a state $Y(3940)$ decays to $J/\psi \omega$.  The observed
transitions in the $b \bar b$ system include
$\Upsilon(2S) \to \Upsilon(1S) \pi \pi$,
$\Upsilon(3S) \to \Upsilon(1S,2S) \pi \pi$,
$\chi(2P)_{b1,2} \to \Upsilon(1S) \omega$, and
$\chi(2P)_{bJ} \to \chi_{bJ} \pi \pi$.  Many of
these transitions have been observed only in the past few years
(see later sections for experimental data).

The theoretical description of hadronic transitions uses a multipole expansion
for gluon emission developed in 
\citet{Gottfried:1977gp,Bhanot:1979af,Peskin:1979va,Bhanot:1979vb,Voloshin:1978hc,Yan:1980uh},
Formally, it resembles the usual multipole expansion for photonic transitions
discussed in Section~\ref{sec:emagntrans}.  The interaction for color electric
and magnetic emission from a heavy quark is given by
\begin{equation}
H_I = \int d^3x Q^{\dagger}(x){\bf t^a}[{\bf x \cdot
E_a(x)} + {\bf \sigma \cdot B_a(x)}]Q(x) + ...~~,
\end{equation}
where ${\bf t^a}~(a=1,\ldots,8)$ is a generator of color SU(3), and
the $(\bar{Q})Q$ and ${\bf E}, {\bf B}$ are dressed (anti)quarks
and color electric and magnetic fields~\cite{Yan:1980uh}.
As usual, the multipole expansion arises from expanding
the color-electric and color-magnetic fields about their values
at the center of mass of the initial quarkonium state.
However, unlike EM transitions, a single interaction of $H_I$
changes a color singlet $Q\bar Q$ initial state $(i)$ into some
color octet $Q\bar Q$ state. Therefore, a second interaction $H_I$
is required to return to a color singlet $Q\bar Q$ final state $(f)$.
In the overall process at least two gluons are emitted.
Assuming factorization for the quarkonium systems~\cite{Kuang:1981se},
the full transition amplitude can be expressed as a product
of two subamplitudes: One that acts on the quarkonium system to
produce the multipole transition and a second that creates the
final light hadrons ($H$) from the action of the gluonic
operators on the vacuum state.

In non-relativistic QCD (NRQCD)~\cite{Caswell:1985ui,Bodwin:1994jh,Luke:1996hj}, the strength of the
various interactions can be ordered in powers of the heavy quark velocity $v$.
The leading behavior comes from two color-electric (E1)
gluon emissions.
This amplitude can be written in the factorized form~\cite{Kuang:1981se}:
\begin{equation}
\sum_{{\cal O}} \frac{\langle i|{\bf r^j t^a}|{\cal O} \rangle
\langle {\cal O}|{\bf r^k t^b}|f\rangle}{E_i - E_{{\cal O}}}
\langle 0 |{\bf E^j_a E^k_b}|H \rangle
\end{equation}
The sum runs over allowed $Q\bar Q$ octet intermediate states ${\cal O}$.
Phenomenological models (e.g.~the Buchm\"uller-Tye vibrating string
model~\cite{Buchmuller:1979gy}) are used to estimate this quarkonium
overlap amplitude.  The quantum numbers of the initial and final
quarkonium states determine which terms in the multipole expansion
may contribute.  For the light hadron amplitude the the states
allowed are determined by the overall symmetries.
In transitions between various $^3S_1$ quarkonium states the
leading term in the multipole expansion has two color-electric
(E1) interactions. The lowest-mass light hadron state allowed
is a two-pion state with either an $S$- or $D$-wave
relative angular momentum. The form of the light hadron
amplitude is determined by chiral symmetry
considerations~\cite{Brown:1975dz}:
\begin{widetext}
\begin{equation}
\langle 0|{\bf E^j_a E^k_b}| \pi(k_1)\pi(k_2) \rangle =
\delta_{ab} [c_1 \delta^{jk}k_1\cdot k_2 +
c_2 (k_1^j k_2^k + k_2^j k_1^k -
\frac{2}{3} \delta^{jk} k_1\cdot k_2)].
\end{equation}
\end{widetext}
The two unknowns ($c_1, c_2$) are the coefficients of the
$S$-wave and $D$-wave two-pion systems. Their values are determined
from experiment. Additional terms can arise in higher orders
in~$v$~\cite{Voloshin:2006ce}.

Hadronic transitions which can flip the heavy quark spins
first occur in  amplitudes with one color-electric (E1)
and one color-magnetic (M1) interaction. These transitions are
suppressed by an additional power of $v$ relative to the
purely electric transitions.
Transitions involving two color-magnetic interactions (M1)
are suppressed by an additional power of $v$.
Many detailed predictions for hadronic transition rates
can be found in \citet{Kuang:1981se,Voloshin:2006ce,Voloshin:1985em,Kuang:1988bz,Kuang:1989ub,Kuang:2002hz,Voloshin:2003kn,Kuang:2006me}.

\section{Charmonium}
\label{sec:charmonium}

In what follows we shall quote masses and partial widths from 
\citet{Yao:2006px} unless otherwise noted.  The masses are
used to calculate photon transition energies.
We shall use an electromagnetic coupling constant
$\alpha = 1/137$ in all cases.
For gluon emission in $Q \bar Q$ annihilation we
shall use a momentum-dependent strong coupling constant $\alpha_S(Q^2)$
evaluated at $Q^2 = m_Q^2$.  The QCD corrections to the decay widths we quote
are performed for this scale choice \cite{Kwong:1987ak}.
Typical values are $\alpha_S(m_c^2) \simeq 0.3$, $\alpha_S(m_b^2) \simeq 0.2$
\cite{Kwong:1987ak}. A different scale choice would lead to different
${\cal O}(\alpha_S)$ corrections \cite{Brodsky:1982gc}.

\subsection{The $J/\psi$}

The $J/\psi$ was the first charmonium state discovered, in 1974
\cite{Aubert:1974js,Augustin:1974xw}.
It is the lowest $^3S_1$ $c \bar c$~state and thus can couple directly
to virtual photons produced in $e^+ e^-$ collisions.
The most precise mass determination to date comes from the KEDR
Collaboration~\cite{Aulchenko:2003qq},
$m(J/\psi) = 3096.917 \pm 0.010 \pm 0.007\,\mathrm{MeV}$, a relative
uncertainty of $4 \times 10^{-6}$.

The $J/\psi$ intrinsic width originally was
determined indirectly. The history of these measurements shows values
below $70\,\mathrm{MeV}$~\cite{Bai:1995ik}. A direct determination by
measuring the excitation curve in
$p \bar p \to e^+e^-$~\cite{Armstrong:1992wu}
was the first to result in a substantially higher value,
albeit still with considerable statistical uncertainty:
$\Gamma(J/\psi) = 99 \pm 12 \pm 6 \,\mbox{keV}$.
Recent indirect measurements, resulting in uncertainties
of $3-4\,\mbox{keV}$, were carried
out~\cite{Adams:2005mp,Aubert:2003sv}
using the radiative return process $e^+e^- \to \gamma e^+ e^-
\to \gamma J/\psi \to \gamma (\mu^+\mu^-)$.
The experimental observable is the radiative cross-section,
a convolution of the photon emission probability and the
$J/\psi$ Breit-Wigner resonance shape. It is
calculable and proportional to the coupling
of the $J/\psi$ to the annihilating $e^+e^-$ pair and
the $J/\psi$ decay branching fraction,
$\Gamma_{ee} \times {\cal B}(J/\psi \to \mu^+\mu^-)$.
Interference with the QED process
$e^+e^- \to \gamma \mu^+\mu^-$ introduces an asymmetry around
the $J/\psi$ peak in $m(\mu^+\mu^-)$ and must be taken into account.
${\cal B}(J/\psi \to \mu^+\mu^-)$ is known well, hence
the product gives access to $\Gamma_{ee}$ and, together with
${\cal B}(J/\psi \to \mu^+\mu^-)$, to $\Gamma_{\mathrm{tot}}$.
The current world average is
$\Gamma(J/\psi) = 93.4 \pm 2.1\,\mathrm{keV}$~\cite{Yao:2006px}.

The largest data sample now consists of 58 million $J/\psi$ collected by the
BES-II Collaboration.
Decays from the $\psi(2S)$ state, in particular $\psi(2S) \to \pi^+\pi^- J/\psi
\to \pi^+\pi^- \mbox{hadrons}$, offer a very clean avenue to study $J/\psi$
final states, yielding one $\pi^+\pi^- J/\psi$ event per three $\psi(2S)$
produced.  Experimentally, this can be handled by requiring a $\pi^+\pi^-$ pair
recoiling against a system of $m(J/\psi)$, without further identification of
the $J/\psi$ decay products.  This path also eliminates contamination of the
sample by continuum production of a final state under study, $e^+ e^- \to
\gamma^* \to \mbox{hadrons}$.  Other $J/\psi$ production mechanisms include
$p \bar p$ collisions and radiative return from $e^+ e^-$ collisions with
center-of-mass energy $> m(J/\psi)$.
Many decays of $J/\psi$ to specific states of light hadrons provide valuable
information on light-hadron spectroscopy.  Here we shall be concerned primarily
with its decay to the $\eta_c(1^1S_0)$, the lightest charmonium state of all;
its annihilation into lepton pairs; and its annihilation into three gluons, two
gluons and a photon, and three photons.

\subsubsection{$J/\psi \to \gamma \eta_c$}

The rate predicted for the process $J/\psi \to \gamma \eta_c$
on the basis of Eq.\ (\ref{eqn:M1}) is
$\Gamma(J/\psi \to \gamma \eta_c) = 2.85$ keV.  Here we have taken the photon
energy to be 114.3 MeV based on $m(J/\psi) = 3096.916$ MeV and $m(\eta_c) =
2980.4$ MeV, and have assumed that the matrix element of $j_0(kr/2)$ between
initial and final states is 1. With $\Gamma_{\rm tot}(J/\psi) = (93.4 \pm 2.1)$
keV, this implies a branching ratio ${\cal B}(J/\psi \to \gamma \eta_c) =
(3.05\pm0.07)\%$. The branching ratio observed in \citet{Gaiser:1985ix} is
considerably less, ${\cal B}_{\rm exp} (J/\psi \to \gamma \eta_c) = (1.27 \pm
0.36)\%$, calling for re-examination both of theory and experiment.

One might be tempted to ascribe the discrepancy to relativistic corrections or
the lack of wave function overlap generated by a relatively strong hyperfine
splitting.  A calculation based on lattice QCD does not yet provide a
definitive answer \cite{Dudek:2006ej}, though it tends to favor a larger
decay rate.
Theoretical progress may also be made using a NRQCD approach \cite{Brambilla:2005zw}.
Part of the ambiguity is associated with the effective value of
the charmed quark mass, which we take to be 1.5 GeV/$c^2$.

\subsubsection{New measurements of leptonic branching ratios}

New leptonic $J/\psi$ branching ratios were measured by the CLEO Collaboration
\cite{Li:2005uga} by comparing the transitions $\psi(2S) \to
\pi^+ \pi^- J/\psi(1S) \to \pi^+ \pi^- X$ with $\psi(2S) \to \pi^+ \pi^-
J/\psi(1S) \to \pi^+ \pi^- \ell^+ \ell^-$.  The results,
${\cal B}(J/\psi \to e^+ e^-) = (5.945 \pm 0.067 \pm 0.042)\%$,
${\cal B}(J/\psi \to \mu^+ \mu^-) = (5.960 \pm 0.065 \pm 0.050)\%$, and
${\cal B}(J/\psi \to \ell^+ \ell^-) = (5.953 \pm 0.056 \pm 0.042)\%$, are
all consistent with, but more precise than, previous measurements.

\subsubsection{Hadronic, $g g \gamma$, and $\gamma \gamma \gamma$ decays:
Extraction of $\alpha_S$}

The partial decay rate of $J/\psi$ to hadrons through the three-gluon final
state in principle provides information on $\alpha_S(m_c^2)$ through the ratio
\beq \label{eqn:cgggmumu}
\frac{\Gamma(J/\psi \to ggg)}{\Gamma(J/\psi \to \ell^+ \ell^-)} = \frac{5}{18}
\left[ \frac{m(J/\psi)}{2 m_c} \right]^2 \frac{(\pi^2-9) [\alpha_S(m_c^2)]^3}
{\pi \alpha^2} \left[ 1 + 1.6 \frac{\alpha_S}{\pi} \right]~~.
\eeq
Both processes are governed by $|\Psi(0)|^2$, the squared magnitude of the
$S$-wave charmonium wave function at the origin.
In \citet{Kwong:1987ak} a value of $\alpha_S(m_c^2) = 0.175 \pm 0.008$ was
extracted from this ratio, which at the time was measured to be $9.0 \pm
1.3$.  This is far below what one expects from the running of $\alpha_S$
down to low momentum scales ($\alpha_S(m_c^2) \simeq 0.3$
\cite{Kwong:1987ak,Bethke:2006ac,Kluth:2006vf,Davier:2007ym}), highlighting the
importance of relativistic corrections to Eq.\ (\ref{eqn:cgggmumu}).  We shall
update the value of the ratio as extracted from data, but the qualitative
conclusion will remain the same.

The branching ratio ${\cal B}(J/\psi \to ggg)$ is inferred by counting all
other decays, to $\gamma \eta_c$, $\ell^+ \ell^-$, $\gamma^* \to {\rm
hadrons}$, and $\gamma g g$.  As mentioned earlier, we have ${\cal B}(J/\psi
\to \gamma \eta_c) = (1.27 \pm 0.36)\%$ \cite{Gaiser:1985ix} and ${\cal B}
(J/\psi \to \ell^+ \ell^-) = (5.953 \pm 0.056 \pm 0.042)\%$ \cite{Li:2005uga}
for $\ell = e,\mu$.  We use the value 
$R_{e^+ e^-} = 2.28 \pm 0.04$ at $E_{\rm cm}/c^2 = m(J/\psi)$  
\cite{Seth:2004qc}  and the leptonic branching ratio to estimate
\beq
{\cal B}(J/\psi \to \gamma^* \to {\rm hadrons})
= R_{e^+ e^-}{\cal B}(J/\psi \to \ell^+ \ell^-)
= (13.6 \pm 0.3)\%~~~.
\eeq
Thus the branching ratio of $J/\psi$ to states other than $ggg + gg \gamma$ is
$[(1.27 \pm 0.36) + (2 + 2.28 \pm 0.04)(5.953\pm0.070) = (26.75\pm0.53)]\%$.
Finally, we use 
$\Gamma(J/\psi \to \gamma g g)/ \Gamma(J/\psi \to ggg)
= (10 \pm 4)\%$ \cite{Lepage:1983tk}
to infer $\Gamma(J/\psi \to ggg)
= (66.6 \pm 2.5)\% \Gamma_{\rm tot}(J/\psi)$.  Then
\beq
\frac{\Gamma(J/\psi \to ggg)}{\Gamma(J/\psi \to \ell^+ \ell^-)}
= \frac{66.6 \pm 2.5}{5.953 \pm 0.070} = 11.2 \pm 0.4
\eeq
implying $\alpha_S(m_c^2) = 0.188^{+0.002}_{-0.003}$.  Although somewhat higher
than the earlier estimate, this is still far below what we will estimate from
other decays, and indicates that the small hadronic width of the $J/\psi$
remains a problem within a nonrelativistic approach.  As mentioned earlier,
this could have been anticipated. In particular the contribution of color-octet
$Q\bar Q$ components is expected to be large \cite{Petrelli:1997ge,Maltoni:2000km}.
In any event, the hadronic width of the $J/\psi$
provides a useful testing ground for any approach which seeks to treat relativistic effects
in charmonium quantitatively.  The ratio
\beq
\frac{\Gamma(J/\psi \to \gamma g g)}{\Gamma(J/\psi \to ggg)} = \frac{16}{5}
\frac{\alpha}{\alpha_S(m_c^2)} \left[1 - 2.9 \frac{\alpha_S}{\pi} \right]
= (10 \pm 4)\%
\eeq
itself provides information on $\alpha_S(m_c^2)$ within a much larger range,
yielding $\alpha_S(m_c^2) = 0.19^{+0.10}_{-0.05}$ as found in 
\citet{Kwong:1987ak}.

The decay $J/\psi \to \gamma \gamma \gamma$ is also governed by $|\Psi(0)|^2$.
The ratio of its rate to that for $J/\psi \to ggg$ is \cite{Kwong:1987ak}
\beq
\frac{\Gamma(J/\psi \to \gamma \gamma \gamma)}{\Gamma(J/\psi \to ggg)} =
\frac{54}{5} e_Q^6 \left(\frac{\alpha}{\alpha_S}\right)^3
\frac{1 - 12.7 \alpha_S/\pi}{1 - 3.7 \alpha_S/\pi} =
\frac{128}{135} \left(\frac{\alpha}{\alpha_S}\right)^3
\frac{1 - 12.7 \alpha_S/\pi}{1 - 3.7 \alpha_S/\pi}~~.
\eeq
The last ratio is a QCD correction; $e_Q = 2/3$ for the charmed quark's charge.
(For the $\Upsilon(1S)$ ratio, take $e_Q = -1/3$ and replace 3.7 by 4.9 in the
denominator of the QCD correction term.)  With $\alpha_S(m_c^2) = 0.3$, the
uncorrected ratio is $1.4 \times 10^{-5}$.  The large negative QCD correction
indicates that this is only a rough estimate but probably an upper bound.

\subsection{The $\eta_c$}
\label{subsec:etac}

Some progress has been made in pinning down properties of the $\eta_c(1S)$,
but better measurements of its mass, total width, and two-photon partial
width would still be welcome.

The mass has been determined through fits to the invariant mass spectrum
of $\eta_c(1S)$ decay products in reactions such as
$\gamma\gamma \to \eta_c(1S)$~\cite{Asner:2003wv,Aubert:2003pt},
$B \to \eta_c(1S) K$~\cite{Fang:2002gi},
and $J/\psi,\ \psi(2S) \to \gamma \eta_c(1S)$~\cite{Bai:2003et,Bai:2000sr}
using all-charged or dominantly charged final states, and in
$p \bar p \to \eta_c(1S) \to \gamma\gamma$~\cite{Ambrogiani:2003md}.
All these recent measurements have uncertainties in the few-MeV range,
but do not agree with each other particularly well.  The averaged value is
$m(\eta_c(1S)) = (2980.4 \pm 1.2)\,\mbox{MeV}$~\cite{Yao:2006px},
which includes an error inflation of $S=1.5$ to account for the spread
of results.  The observed splitting of $116.5 \pm 1.2$ MeV between $J/\psi$ and
$\eta_c(1S)$ is consistent with an unquenched lattice QCD prediction
of $\simeq 110$~MeV \cite{Davies:2006tx}.

The square of the wave function at the origin cancels out in the ratio
of partial widths \cite{Kwong:1987ak},
\beq \label{eqn:ecpsi}
\frac{\Gamma(\eta_c \to \gamma \gamma)}{\Gamma(J/\psi \to \mu^+ \mu^-)}
= \frac{4}{3} \left[1 + 1.96 \frac{\alpha_S}{\pi} \right]~~~.
\eeq
Using the ``evaluated'' partial widths in \citet{Yao:2006px},
$\Gamma(\eta_c \to \gamma \gamma) = (7.2 \pm 0.7 \pm 2.0)$ keV and
$\Gamma(J/\psi \to \mu^+ \mu^-) = (5.55 \pm 0.14 \pm 0.02)$ keV, one finds that
$(3/4)\Gamma(\eta_c \to \gamma \gamma)/\Gamma(J/\psi \to \mu^+ \mu^-) =
0.97 \pm 0.29$, which is consistent with Eq.\ (\ref{eqn:ecpsi}) but still not
precisely enough determined to test the QCD correction.  A more precise test
would have taken into account $m(J/\psi) \ne 2 m_c$ and the running of
$\alpha_S$.

The total width of $\eta_c$ is dominated by the $gg$ final state.  Its value
has not remained particularly stable over the years, with
\citet{Yao:2006px} quoting $\Gamma_{\rm tot}(\eta_c) = (25.5 \pm
3.4)$ MeV.  This value is $(3.54 \pm 1.14) \times 10^3$ that of $\Gamma(\eta_c
\to \gamma \gamma)$. The $gg/\gamma \gamma$ ratio is predicted \cite{Kwong:1987ak} to be
\beq
\frac{\Gamma(\eta_c \to gg)}{\Gamma(\eta_c \to \gamma \gamma)} = \frac{9
[\alpha_S(m_c^2)]^2}{8 \alpha^2} \left[ 1 + 8.2\frac{\alpha_S}{\pi} \right]~~~,
\eeq
leading to $\alpha_S(m_c^2)=0.30^{+0.03}_{-0.05}$.  This value should be
regarded with caution in view of the large QCD correction factor
$1 + 8.2 \alpha_S/\pi \sim 1.8$.

New measurements have been reported of the product of the two-photon widths
and branching ratios to selected four-meson final states for the
$\eta_c$~\cite{Uehara:2007vb}. Combining with branching ratios from the
Particle Data Group, one obtains $\Gamma(\eta_c \to \gamma \gamma) = 2.46 \pm
0.60$ keV~\cite{Metreveli:2007sj}, a value considerably lower than that
just quoted, and disagreeing with the prediction in Eq.\ (\ref{eqn:ecpsi}).

\subsection{$P$-wave $\chi_{cJ}$ states}

The $1P$ states of charmonium, $\chi_{cJ}$, were first seen
in radiative decays from the $\psi(2S)$.
The $\chi_{cJ}$ states lie 128 / 171 / 261~MeV
($J=2/1/0$) below the $\psi(2S)$.
Their masses can most accurately be determined in $p \bar p$
collisions~\cite{Armstrong:1991he,Andreotti:2005ts,Andreotti:2003sk}
with $\chi_{cJ} \to \gamma J/\psi \to \gamma (e^+e^-)$ or
$\chi_{c0} \to \pi^0\pi^0$ by measuring the excitation curve,
where the well-known and small beam energy spread
results in very low systematic uncertainty
(${\cal O}(100\,\mbox{keV}$~\cite{Andreotti:2003sk}).
In principle, a precise measurement of the photon energy
in $\psi(2S) \to \gamma \chi_{cJ}$ allows a mass measurement
as well, given that the $\psi(2S)$ mass is very
well known.
BES used the decay $\psi(2S) \to \gamma \chi_{cJ}$ followed by
photon conversions $\gamma \to e^+e^-$
to improve upon the photon energy resolution~\cite{Ablikim:2005yd}.

The $J=0$ state is wide, about 10~MeV, while
the $J=1$ and $J=2$ states are narrower
($0.89 \pm 0.05\,\mathrm{MeV}$ and $2.06 \pm 0.12\,\mathrm{MeV}$,
respectively~\cite{Yao:2006px}), which is below detector resolution
for most exclusive $\chi_{cJ}$ decays.
The most accurate width determinations to date come from $p \bar p$
experiments, again from fits to the excitation
curve~\cite{Andreotti:2003sk,Andreotti:2005ts}.

\subsubsection{Production and decay via E1 transitions}

E1 transitions have played an important role in quarkonium physics with the
initial theoretical papers describing charmonium suggesting that
the triplet $1P$ states could be observed through the E1 transitions
from the $\psi(2S)$ resonance \cite{Eichten:1974af,Eichten:1975ag,Eichten:1978tg,Eichten:1979ms,Appelquist:1974yr}.  It is a
great success of this picture that the initial calculations by the
Cornell group \cite{Eichten:1974af,Eichten:1975ag,Eichten:1978tg,Eichten:1979ms} agree within 25\% of the present
experimental values.

New studies have been performed by the CLEO Collaboration of the rates for
$\psi(2S) \to \gamma \chi_{c0,1,2}$ \cite{Athar:2004dn} and $\psi(2S) \to
\gamma \chi_{c0,1,2} \to \gamma \gamma J/\psi$ \cite{Adam:2005uh}.  We shall
use these data to extract the magnitudes of electric dipole matrix elements
and compare them with various predictions.

The inclusive branching ratios and inferred rates for $\psi(2S) \to
\gamma \chi_{cJ}$ are summarized in Table~\ref{tab:2Schic}.  Photon energies
are based on masses quoted in \citet{Yao:2006px}.  Branching ratios are
from \citet{Athar:2004dn}.  Partial widths are obtained from these using
$\Gamma_{\rm tot}[\psi(2S)] = 337 \pm 13$ keV \cite{Yao:2006px}.  The E1 matrix
elements $|\langle 1P |r|2S  \rangle|$ extracted using the nonrelativistic
expression (\ref{eqn:E1}) are shown in the last column.

\begin{figure}
\begin{center}
\includegraphics[height=0.67\textheight]{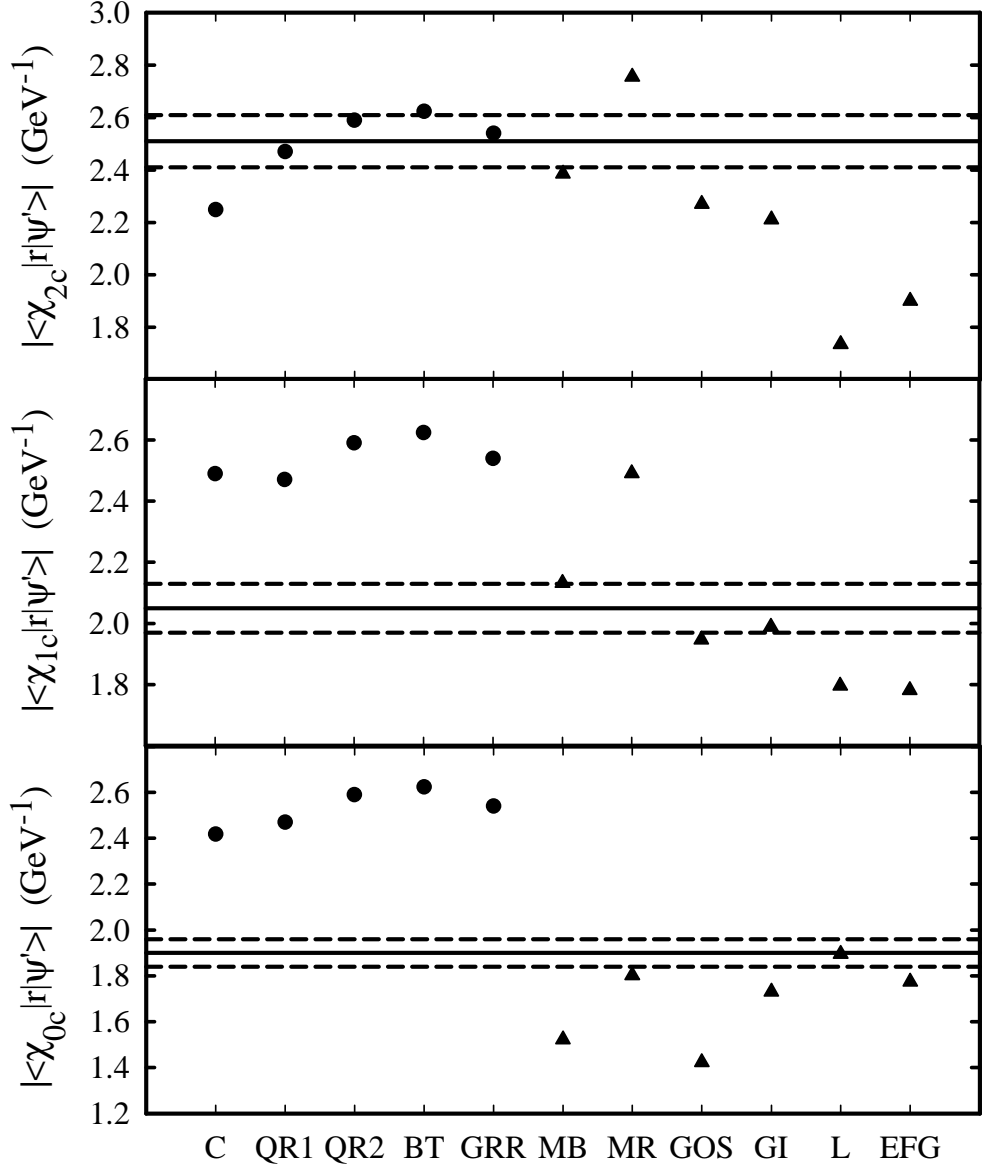}
\caption{E1 dipole transition matrix elements for the charmonium decays
$2^3S_1\to 1^3P_J$.
The horizontal bands indicate the experimental results.
The circles designate nonrelativistic predictions and
the triangles relativistic predictions.  Within these subsets the
results are given in chronological order of the publication date.  The
labels refer to C-Cornell Model \cite{Eichten:1974af,Eichten:1975ag,Eichten:1978tg,Eichten:1979ms},
QR-Quigg Rosner, $c\bar{c}$ $\rho=2$ and $b\bar{b}$ potentials
\cite{Quigg:1981bj},
BT-Buchm\"uller Tye \cite{Buchmuller:1980su},
GRR-Gupta Radford Repko \cite{Gupta:1986xt},
MB-McClary Byers \cite{McClary:1983xw},
MR-Moxhay Rosner \cite{Moxhay:1983vu},
GOS-Grotch Owen Sebastian \cite{Grotch:1984gf},
GI-Godfrey Isgur, calculated using the wavefunctions of
\citet{Godfrey:1985xj},
L-Lahde, DYN column \cite{Lahde:2002wj},
EFG-Ebert Faustov Galkin \cite{Ebert:2002pp}.
\label{fig:cc2sto1p}}
\end{center}
\end{figure}

In the nonrelativistic limit the dipole matrix elements in $^3S_1 \to~^3P_J$
transitions,  $|\langle r \rangle_{\rm NR}|$, for different $J$ values are
independent of $J$.  Predictions of specific nonrelativistic potential models
sit in a small range from 2.4 to 2.7 GeV$^{-1}$ (see Fig.~\ref{fig:cc2sto1p}),
with a slightly larger range obtained using potentials constructed from
charmonium and $b \bar b$ data using inverse-scattering methods
\cite{Quigg:1981bj}.
However the magnitudes of the matrix elements are observed with the ordering
$|\langle \chi_{c2}|r| \psi(2S) \rangle| >
|\langle \chi_{c1}|r| \psi(2S) \rangle| >
|\langle \chi_{c0}|r| \psi(2S) \rangle|$.
This is in accord with predictions that
take into account relativistic corrections
\cite{Grotch:1984gf,Godfrey:1985xj,Ebert:2002pp,Moxhay:1983vu,McClary:1983xw}.
Figure~\ref{fig:cc2sto1p} shows that at least some models are in good agreement
with the observed rates so that we can conclude that relativistic corrections
can explain the observed rates.  However, it is probably premature to say that
the transitions are totally understood given the large scatter of the
predictions around the observed values.

\begin{table*}
\caption{Properties of $\psi(2S) \to \gamma \chi_{cJ}$ decays,
using results from \citet{Yao:2006px} and branching fractions
${\cal B}$ from \citet{Athar:2004dn}, as well as Eq.\ (\ref{eqn:E1}).
\label{tab:2Schic}}
\begin{center}
\begin{tabular}{c c c c c} \hline \hline
$J$ & $k_\gamma$ & ${\cal B}$ &
$\Gamma[\psi(2S) \to \gamma \chi_{cJ}]$ & $|\langle 1P |r| 2S \rangle|$ \\
& (MeV) & (\%) & (keV) & (GeV$^{-1}$) \\ \hline
2 & 127.60$\pm$0.09 & 9.33$\pm$0.14$\pm$0.61 & 31.4$\pm$2.4  &
2.51$\pm$0.10  \\
1 & 171.26$\pm$0.07 & 9.07$\pm$0.11$\pm$0.54 & 30.6$\pm$2.2  &
2.05$\pm$0.08  \\
0 & 261.35$\pm$0.33 & 9.22$\pm$0.11$\pm$0.46 & 31.1$\pm$2.0  &
1.90$\pm$0.06  \\
\hline \hline
\end{tabular}
\end{center}
\end{table*}

Information on the electromagnetic cascades $\psi(2S) \to \gamma \chi_cJ \to
\gamma \gamma J/\psi$ is summarized in Table~\ref{tab:2Schic1S}.  The products
${\cal B}_1 {\cal B}_2 \equiv {\cal B}[\psi(2S) \to \gamma \chi_cJ] {\cal B}
[\chi_{cJ} \to \gamma J/\psi]$ are taken from \citet{Adam:2005uh}.  These
and prior measurements may be combined with values of ${\cal B}_1$ from
\citet{Athar:2004dn} and previous references to obtain the values of
${\cal B}_2$ in the Table~\cite{Yao:2006px}.  Other data come from
the high-statistics studies of Fermilab Experiment E835
\cite{Andreotti:2005ts}, who also measure total $\chi_{cJ}$ widths and
present partial widths for $\chi_{cJ} \to \gamma J/\psi$.

\begin{table}
\caption{Properties of the exclusive transitions $\psi(2S) \to \gamma \chi_cJ
\to \gamma \gamma J/\psi$.
\label{tab:2Schic1S}}
\begin{center}
\begin{tabular}{c c c c} \hline \hline
$J$ & ${\cal B}_1 {\cal B}_2$ (\%) & 
      ${\cal B}_2$ (\%) & $\Gamma_{\rm tot}$ (MeV) \\ 
    & \cite{Adam:2005uh} & 
\multicolumn{2}{c}{\cite{Yao:2006px}} \\ \hline
2  & 1.85$\pm$0.04$\pm$0.07 & 20.1$\pm$1.0 & 2.06$\pm$0.12 \\
1  & 3.44$\pm$0.06$\pm$0.13 & 35.6$\pm$1.9 & 0.89$\pm$0.05 \\
0  & 0.18$\pm$0.01$\pm$0.02 & 1.30$\pm$0.11 & 10.4$\pm$0.7 \\
\hline \hline
\end{tabular}
\end{center}
\end{table}

The partial widths for $\chi_{cJ} \to \gamma J/\psi$ extracted from PDG
averages for ${\cal B}_2$ and the values of $\Gamma_{\rm tot}(\chi_{c2,1,0})$
mentioned above are summarized in Table~\ref{tab:chic1S}.  The dipole matrix
elements have been extracted using Eq.\ (\ref{eqn:E1}) using photon energies
obtained from the $\chi_{cJ}$ and $J/\psi$ masses in \citet{Yao:2006px}.

\begin{table}
\caption{Properties of the transitions $\chi_{cJ} \to \gamma J/\psi$
(\citet{Yao:2006px}; Eq.~(\ref{eqn:E1})).
\label{tab:chic1S}}
\begin{center}
\begin{tabular}{c c c c} \hline \hline
$J$ & $k_\gamma$ & $\Gamma(\chi_{cJ} \to \gamma J/\psi)$
&   $|\langle 1S |r| 1P \rangle|$ \\
& (MeV) & (keV) &  (GeV)$^{-1}$ \\ \hline
2 & 429.63$\pm$0.08 & 416$\pm$32 &  1.91$\pm$0.07 \\
1 & 389.36$\pm$0.07 & 317$\pm$25 &  1.93$\pm$0.08 \\
0 & 303.05$\pm$0.32 & 135$\pm$15 &  1.84$\pm$0.10 \\
\hline \hline
\end{tabular}
\end{center}
\end{table}

\begin{figure}
\begin{center}
\includegraphics[height=0.78\textheight]{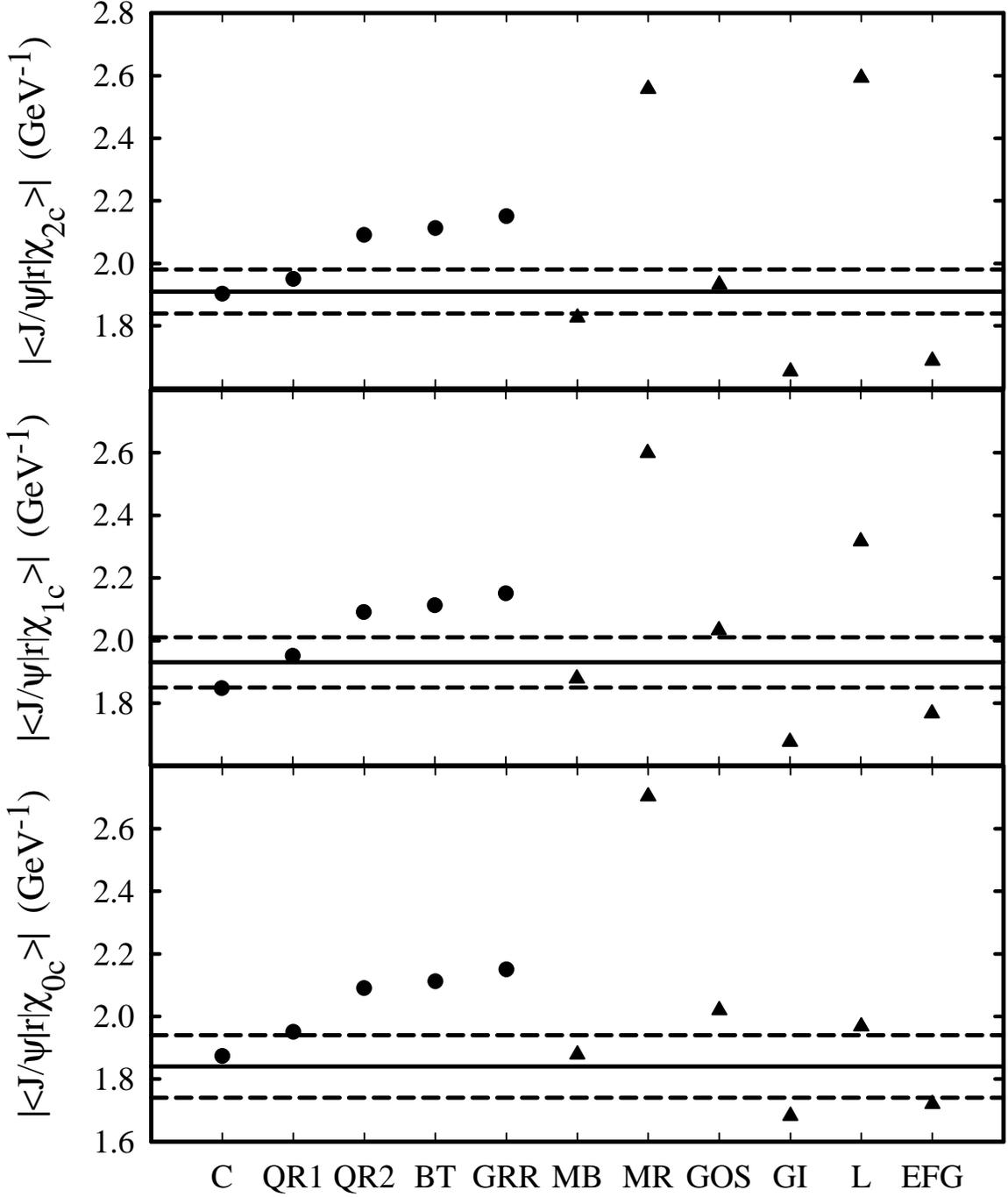}
\caption{E1 dipole transition matrix elements for the charmonium decays
$1^3P_J\to 1^3S_1$.  Labels are as in Fig.\ \ref{fig:cc2sto1p}.
\label{fig:cc1pto1s}}
\end{center}
\end{figure}

Predictions from both nonrelativistic and relativistic
calculations are shown in Fig.~\ref{fig:cc1pto1s}.  Overall the nonrelativistic
calculations, with typical values of 1.9 to 2.2~GeV$^{-1}$,
are in reasonable agreement with the observed values
reflecting their relative $J$-independence.  The predictions including
relativistic corrections are generally poorer which is surprising
because both the $1P$ and $1S$ wavefunctions have no nodes so that the
integrals should be relatively insensitive to details of the calculation.

\subsubsection{Search for M2 transitions}

Attempts have been made to observe magnetic quadrupole (M2) transitions in
charmonium through their interference with the dominant E1 amplitudes.  These
are not yet conclusive \cite{Oreglia:1981fx,Ambrogiani:2001jw}.  The best
prospects are expected for the most energetic photons, i.e., those in
$\chi_{cJ} \to \gamma J/\psi$.  Using the notation of \cite{Ambrogiani:2001jw},
the expected normalized M2/E1 amplitude ratios $a_2$ for these decays are
\bea
a_2(\chi_{c1}) & = & E_{\gamma_1}(1 + \kappa_c)/(4 m_c)~~,\\
a_2(\chi_{c2}) & = & (3/\sqrt{5})E_{\gamma_2}(1 + \kappa_c)/(4 m_c)~~,
\eea
and are shown in Table~\ref{tab:a2}.  These values are based on averages
\cite{Yao:2006px} of those in \citet{Oreglia:1981fx} and
\citet{Ambrogiani:2001jw}.
We note that a comparison between the {\sl ratios} of the two
decays would yield a more stringent test due to the cancellation
of the charm quark mass (theory) and possible systematic
uncertainties (experiment).

\begin{table}
\caption{Predicted and observed M2/(E1$^2$ + M2$^2$)$^{1/2}$ ratios for
the transitions $\chi_{cJ} \to \gamma J/\psi$.
\label{tab:a2}}
\begin{center}
\begin{tabular}{c c c} \hline \hline
\renewcommand{\arraystretch}{1.4}
State & Prediction                 & Experiment \\
      & \cite{Ambrogiani:2001jw}   & \cite{Yao:2006px}\\
\hline
$\chi_{c1}$ & $-0.065(1+\kappa_c)$ & $-0.002^{+0.008}_{-0.017}$ \\
$\chi_{c2}$ & $-0.096(1+\kappa_c)$ & $-0.13 \pm 0.05$ \\ \hline \hline
\end{tabular}
\end{center}
\end{table}

\subsubsection{Hadronic and $\gamma \gamma$ decays}

In principle the measured $\chi_{cJ}$ widths~\cite{Yao:2006px}
can be used to determine $\alpha_S(m_c^2)$ if the value of
the derivative of the $L=1$ radial wave function for zero separation,
$|R'_{nP}(0)|$, is known.  Potential models or lattice gauge theories can be
used to estimate such quantities.  However, they cancel out in ratios of
partial widths to various final states.  We shall concentrate on the
ratios $\Gamma_{\gamma \gamma}(\chi_{cJ})/\Gamma_{gg}(\chi_{cJ})$ for
$J=2,0$ ($\chi_{c1}$ cannot decay into two photons).
These are predicted to be \cite{Kwong:1987ak,Ebert:2003mu}
\beq \label{eqn:gmgmgg}
\frac{\Gamma_{\gamma \gamma}(\chi_{cJ})}{\Gamma_{gg}(\chi_{cJ})} =
\frac{8 \alpha^2}{9[\alpha_S(m_c^2)]^2}C_J~;~~
C_2 = \frac{1 - (16 \alpha_S)/(3 \pi)}{1 - (2.2 \alpha_S)/\pi}~,~~
C_0 = \frac{1 + (0.2 \alpha_S)/\pi}{1 + (9.5 \alpha_S)/\pi}~~.
\eeq
Here we have exhibited the corrections separately to the $\gamma \gamma$
partial widths (numerators) and $gg$ partial widths (denominators).

CLEO has reported a measurement of
$\Gamma(\chi_{c2} \to \gamma \gamma) =559\pm57\pm48\pm36$ eV based on
14.4 fb$^{-1}$ of $e^+ e^-$ data
at $\sqrt{s}= 9.46$--11.30 GeV \cite{Dobbs:2005yk}.
The result is compatible with other
measurements when they are corrected for CLEO's
${\cal B}(\chi_{c2} \to \gamma J/\psi)$ and
${\cal B}(J/\psi \to \ell^+ \ell^-)$.
The errors given are statistical, systematic, and
$\Delta {\cal B}(\chi_{c2} \to \gamma J/\psi)$.
One can average the CLEO measurement with a
Belle result \cite{Abe:2002va} that is likewise corrected for updated
input branching fractions
to obtain
$\Gamma(\chi_{c2} \to \gamma \gamma) = 565 \pm 62$ eV.
Using $\Gamma_{\mathrm{tot}}(\chi_{c2}) = 2.06 \pm 0.12$ MeV~\cite{Yao:2006px}
and ${\cal B}(\chi_{c2} \to \gamma J/\psi)=(20.2 \pm 1.0)\%$~\cite{Yao:2006px}
one finds
$\Gamma(\chi_{c2} \to gg) \approx \Gamma(\chi_{c2} \to$ light hadrons)
= $1.64\pm0.10$ MeV.
This can be compared to $\Gamma(\chi_{c2} \to \gamma \gamma)$,
taking account of the QCD radiative corrections noted above,
to obtain $\alpha_S(m_c^2) = 0.296^{+0.016}_{-0.019}$.

The decay $\chi_{c0} \to \gamma \gamma$ also has been measured.
Results from the Fermilab E835 Collaboration \cite{Ambrogiani:2000vc,Andreotti:2004ru}
are combined with other data to yield ${\cal B}(\chi_{c0} \to
\gamma \gamma) = (2.76 \pm 0.33) \times 10^{-4}$ \cite{Yao:2006px}, or, with
$\Gamma_{\rm tot}(\chi_{c0}) = 10.4 \pm 0.7$ MeV \cite{Yao:2006px},
$\Gamma(\chi_{c0} \to \gamma \gamma) = (2.87 \pm 0.39)$ keV.  Taking account of
the $(1.30 \pm 0.11)\%$ branching ratio of
$\chi_{c0}$ to $\gamma J/\psi$~\cite{Yao:2006px}
one estimates $\Gamma(\chi_{c0} \to gg) = 10.3 \pm 0.7$ MeV and hence
${\cal B}(\chi_{c0} \to \gamma \gamma)/{\cal B}(\chi_{c0} \to gg) = (2.80 \pm
0.42) \times 10^{-4}$.  Using Eq.\ (\ref{eqn:gmgmgg}) one then finds
$\alpha_S(m_c^2) = 0.32 \pm 0.02$, compatible both with the value
found from the corresponding $\chi_{c2}$ ratio and with a slightly higher
value obtained by extrapolation from higher momentum scales
\cite{Bethke:2006ac,Kluth:2006vf,Davier:2007ym}.

The success of the above picture must be regarded with some caution, as the
experimental values of the ratios
\beq
R_{\gamma \gamma} \equiv \frac{\Gamma(\chi_{c2} \to \gamma \gamma)}
{\Gamma(\chi_{c0} \to \gamma \gamma)}~,~~
R_{gg} \equiv \frac{\Gamma(\chi_{c2} \to gg)}{\Gamma(\chi_{c0} \to gg)}~,
\eeq
namely $R_{\gamma \gamma} = 0.197 \pm 0.034$, $R_{gg} = 0.159 \pm 0.015$,
are far from their predicted values
\beq
R_{\gamma \gamma} = \frac{4}{15} \cdot
\frac{1 - 1.70 \alpha_s}{1 + 0.06 \alpha_s}~;
~~R_{gg} = \frac{4}{15} \cdot \frac{1 - 0.70 \alpha_s}{1 + 3.02 \alpha_s}
\eeq
for the nominal value $\alpha_S(m_c^2)=0.3$, which are $R_{\gamma \gamma} =
0.128$ and $R_{gg} = 0.110$.  This may be due to the large values of some of
the first-order QCD corrections (particularly for $\chi_{c2} \to \gamma \gamma$
and $\chi_{c0} \to gg$), rendering a perturbation expansion unreliable; it
could signify effects of neglected color-octet components of the $\chi_{cJ}$
wavefunctions \cite{Petrelli:1997ge,Maltoni:2000km}; or it
could signify that the values of $|R'_{nP}(0)|$ differ for the $^3P_2$ and
$^3P_0$ states.  It would be interesting to see if lattice gauge theories
could shed light on this last possibility.

The measurements of the product of two-photon widths and branching ratios
to $2(\pi^+\pi^-)$, $K^+K^- \pi^+\pi^-$, and $2(K^+K^-)$ in~\cite{Uehara:2007vb}
for $\chi_{c0,2}$ lead (combining with the relevant branching
branching ratios from the Particle Data Group \cite{Yao:2006px}) to
$\Gamma(\chi_{c0} \to \gamma \gamma) = (1.99 \pm 0.24)$ keV and
$\Gamma(\chi_{c2} \to \gamma \gamma) = (0.44 \pm 0.06)$ keV.
The
results entail a value $R_{\gamma \gamma} = 0.22 \pm 0.04$, even farther
from the prediction based on first-order QCD corrections.

\subsection{The $\psi(2S)$}

The $\psi(2S)$ resonance was discovered at SLAC in $e^+ e^-$ collisions
within days after the announcement of the $J/\psi$ \cite{Abrams:1974yy}.

The most precise $\psi(2S)$ mass measurement to date comes, as for
the $J/\psi$, from KEDR~\cite{Aulchenko:2003qq}, at a relative
uncertainty of $7 \times 10^{-6}$. The current world average is
$m(\psi(2S)) = 3686.093 \pm 0.034 \,\mathrm{MeV}$.

The total $\psi(2S)$ width has been determined in direct $p \bar p$ production
(E760 from the shape of the resonance curve~\cite{Armstrong:1992wu})
as well as in $e^+e^-$ collisions (BES~\cite{Bai:2002zn} from a fit to
the cross-sections $\psi(2S) \to \mbox{hadrons}$, $\pi^+\pi^-J/\psi$,
and $\mu^+\mu^-$ to obtain the corresponding partial widths;
the total width is computed as the sum of hadronic and leptonic widths).
The PDG average of these two ``direct'' measurements is
$277 \pm 22 \,\mathrm{keV}$.  (Not included in the average is a recent
value of $290 \pm 25 \pm 4$ keV based on a measurement of the shape of the
resonance curve by Fermilab Experiment E835 \cite{Andreotti:2007ur}.)
Another estimation comes from the PDG's global fit~\cite{Yao:2006px},
which among many other measurements takes a measurement of $\Gamma_{ee}$
into account.  As for the $J/\psi$, the radiative return process can be
used~\cite{Adam:2005mr}; the decay chain presented there is
$e^+e^- \to \gamma \psi(2S) \to \gamma (X + J/\psi)$,
which holds for any decay $\psi(2S) \to X J/\psi$.
The observed cross-section is proportional to
$\Gamma_{ee}(\psi(2S)) \times {\cal B}(J/\psi \to X J/\psi)$, where
$X = \pi^+\pi^-,\ \pi^0\pi^0,\ \eta$ were used.
The result of the global fit is $337 \pm 13\,\mathrm{keV}$.

The two largest modern on-resonance samples are 29~M $\psi(2S)$ decays
from the CLEO detector and a 14~M sample collected with the BES~II detector.
We have already discussed the transitions $\psi(2S) \to \gamma \chi_{cJ}$ in
the previous subsection.  Here we treat a variety of other electromagnetic and
hadronic transitions of the $\psi(2S)$.
We also briefly comment on $\psi(2S)$ decay via $c \bar c$~annihilation.

\subsubsection{Decay to $\gamma \eta_c(1S)$}

The decay $\psi(2S) \to \gamma \eta_c(1S)$ is a forbidden magnetic dipole (M1)
transition, which would vanish in the limit of zero photon energy because of
the orthogonality of 1S and 2S wave functions.  The photon energy is 638 MeV,
leading to a non-zero matrix element $\langle 1S | j_0(kr/2) | 2S \rangle$.
The decay was first observed by the Crystal Ball Collaboration
\cite{Gaiser:1985ix} in the inclusive photon spectrum of $\psi(2S)$ decays
with branching ratio $(2.8 \pm 0.6) \times 10^{-3}$.  The CLEO Collaboration
measures ${\cal B}[\psi(2S) \to \gamma \eta_c(1S)] = (3.2 \pm 0.4 \pm 0.6)
\times 10^{-3}$, also using the inclusive $\psi(2S)$ photon spectrum.
We note that the yield fit depends considerably on the $\eta_c$ width.  The
Crystal Ball Collaboration arrived at a width that is substantially below more
recent experimental data, $11.5 \pm 4.5$~MeV as opposed to about 25~MeV.
CLEO's result is for a nominal width of $24.8 \pm 4.9$~MeV;
rescaled to the width found by Crystal Ball the CLEO result becomes
${\cal B}[\psi(2S) \to \gamma \eta_c(1S)] = (2.5 \pm 0.6) \times 10^{-3}$.
We average the two primary results and arrive at $(3.0 \pm 0.5) \times
10^{-3}$.  When combined with $\Gamma_{\rm tot}[\psi(2S)] =
(337 \pm 13)$ keV, this implies $\Gamma[\psi(2S) \to \gamma \eta_c(1S)] =
(1.00 \pm 0.16)$ keV, and hence [via Eq.\ (\ref{eqn:M1})] $|\langle 1S |
j_0(kr/2) | 2S \rangle | = 0.045 \pm 0.004$.
While this result is in agreement with some quark model predictions -- 
for example \citet{Eichten:1974af,Eichten:1975ag,Eichten:1978tg,Eichten:1979ms} 
and \citet{Ebert:2002pp} give 0.053 and 0.042, respectively --, 
there is a wide scatter of predictions
\cite{Zambetakis:1983te,Grotch:1984gf,Godfrey:1985xj,Zhang:1991et,Lahde:2002wj,Kang:1978yw}.
It would therefore be useful to have a
prediction from lattice QCD for this matrix element, as well as for
corresponding forbidden matrix elements in the $b \bar b$ system.

\subsubsection{Decay to $\gamma \eta_c(2S)$}

The decay $\psi(2S) \to \gamma \eta_c(2S)$ is an allowed M1 transition and thus
should be characterized by a matrix element $\langle 2S | j_0(kr/2) | 2S
\rangle$ of order unity in the limit of small~$k$.  One may estimate the
branching ratio ${\cal B}[\psi(2S) \to \gamma \eta_c(2S)]$ by scaling from
$J/\psi \to \gamma \eta_c(1S)$.

With ${\cal B}(J/\psi \to \gamma \eta_c) = (1.27 \pm 0.36)\%$
\cite{Gaiser:1985ix} and $\Gamma_{\rm tot}(J/\psi) = (93.4 \pm 2.1)$ keV
\cite{Yao:2006px}, one has $\Gamma(J/\psi \to \gamma \eta_c) = (1.19 \pm
0.34)$ keV.
Assuming that the matrix elements for $\psi(2S) \to \gamma \eta_c(2S)$ and
$J/\psi(1S) \to \gamma \eta_c(1S)$ are equal, the $2S \to 2S$ rate should be
$[E_\gamma(2S \to 2S)/E_\gamma(1S \to 1S)]^3$ times that for $1S \to 1S$.
With photon energies of 47.8 MeV for $2S \to 2S$ and 114.3 MeV for $1S \to 1S$,
this factor is 0.073, giving a predicted partial width $\Gamma[\psi(2S) \to
\gamma \eta_c(2S)] = (87 \pm 25)$ eV (compare, for example with $170-210$~eV in
\citet{Barnes:2005pb}).  Using $\Gamma_{\rm tot}(\psi(2S) = (337
\pm 13)$ keV \cite{Yao:2006px}, one then finds ${\cal B}[\psi(2S) \to \gamma
\eta_c(2S)] = (2.6 \pm 0.7) \times 10^{-4}$, below the sensitivity
of current experiments.

\subsubsection{Hadronic transitions from $\psi(2S)$ to $J/\psi$}

The transitions $\psi(2S) \to \pi^+ \pi^- J/\psi$ and $\psi(2S) \to
\pi^0 \pi^0 J/\psi$ are thought to proceed via electric dipole emission of
a pair of gluons followed by hadronization of the gluon pair into $\pi \pi$
\cite{Gottfried:1977gp,Bhanot:1979af,Peskin:1979va,Bhanot:1979vb,Voloshin:1978hc}.  In addition, the hadronic transitions $\psi(2S) \to
\eta J/\psi$ and $\psi(2S) \to \pi^0 J/\psi$ have been observed.  Recent
CLEO measurements of the branching ratios for these transitions
\cite{Adam:2005uh} are summarized
in Table~\ref{tab:2Shad1S}.  (We have already quoted the branching ratios to
$J/\psi$ via the $\chi_{cJ}$ states in Table~\ref{tab:2Schic1S}.)

\begin{table}
\caption{Branching ratios for hadronic transitions $\psi(2S) \to J/\psi X$
\cite{Adam:2005uh}.
\label{tab:2Shad1S}}
\begin{center}
\begin{tabular}{l c} \hline \hline
Channel & ${\cal B}$ (\%) \\ \hline
$\pi^+ \pi^- J/\psi$ & 33.54$\pm$0.14$\pm$1.10 \\
$\pi^0 \pi^0 J/\psi$ & 16.52$\pm$0.14$\pm$0.58 \\
$\eta J/\psi$        &  3.25$\pm$0.06$\pm$0.11 \\
$\pi^0 J/\psi$       &  0.13$\pm$0.01$\pm$0.01 \\
$X J/\psi$           & 59.50$\pm$0.15$\pm$1.90 \\ \hline \hline
\end{tabular}
\end{center}
\end{table}

Isospin predicts the $\pi^0 \pi^0$ rate to be one-half that of $\pi^+ \pi^-$.
CLEO determines ${\cal B}(\pi^0 \pi^0 J/\psi)/{\cal B}(\pi^+ \pi^- J/\psi)
= (49.24 \pm 0.47 \pm 0.86)\%$~\cite{Adam:2005uh},
taking cancellations of common uncertainties into account.
Two other direct measurements of this ratio are:
$(57.0 \pm 0.9 \pm 2.6)\%$ (BES, \citet{Ablikim:2004mv}),
$(57.1 \pm 1.8 \pm 4.4)\%$ (E835, \citet{Andreotti:2005pf});
the PDG fit result is $(51.7 \pm 1.8)\%$~\cite{Yao:2006px}.
The $\pi^0/\eta$ ratio has been measured as 
$(4.1\pm0.4\pm0.1)\%$ (CLEO, \citet{Adam:2005uh}) 
and $(4.8\pm0.5)\%$ (BES, \citet{Bai:2004cg}).
These results are somewhat above theoretical expectations,
for example 1.6\% quoted in \citet{Bai:2004cg} based on \citet{Miller:1990iz}, 
or 3.4\% from \citet{Ioffe:1980mx,Ioffe:1981qa,Kuang:1988bz,Maltman:1990mp}. 
The inclusive branching ratio for $\psi(2S) \to J/\psi X$, ${\cal B} = (59.50
\pm 0.15 \pm 1.90)\%$, is to be compared with the sum of known modes $(58.9
\pm 0.2 \pm 2.0)\%$.  Thus there is no evidence for any ``missing'' modes.
The results imply ${\cal B}[\psi(2S) \to {\rm~light~hadrons}] = (16.9 \pm 2.6)
\%$, whose significance will be discussed presently.

\subsubsection{Light-hadron decays}

Decays to light hadrons proceed via annihilation of the $c \bar c$ pair
into either three gluons or a virtual photon. This includes production
of baryons.
Such studies can receive substantial background due to continuum production
of the same final state, $e^+ e^- \to \gamma^* \to \mathrm{hadrons}$.
When interpreting the observed rate on the $\psi(2S)$, interference
effects between on-resonance and continuum production can complicate
the picture.

CLEO-c has collected a sample of $20.7\,\mathrm{pb}^{-1}$ at $\sqrt s
= 3.67\,\mathrm{GeV}$, while BES's below-$\psi(2S)$ continuum data,
$6.6\,\mathrm{pb}^{-1}$,
were taken at $\sqrt s = 3.65\,\mathrm{GeV}$.
At the two center-of-mass energies, the $\psi(2S)$ tail is of order
1/1000 [1/5000] compared to the peak cross-section for the two
experiments (this number depends on the collider's beam energy spread).

One expects $Q \equiv {\cal B}[\psi(2S) \to f]/{\cal B}(J/\psi \to f)$
to be comparable to ${\cal B}[\psi(2S) \to \ell^+ \ell^-]/
{\cal B}(J/\psi \to \ell^+ \ell^-) =(12.4 \pm 0.3)\%$ (the ``12\% rule''),
since light-quark decays are presumably governed by $|\Psi(0)|^2$ as are
leptonic decays.  In fact, $Q$ is much smaller than 12\% for most VP and VT
modes, where P=pseudoscalar, V=vector, T=tensor, and severely so in some cases
\cite{Adam:2004pr,Bai:2003vf}.  For example, $Q(\rho \pi)$=(1.9$\pm$0.6)$\times
10^{-3}$, with a similar suppression for $K^{*\pm}K^\mp$.
Many models have been brought forward to explain this behavior.  Another
interesting observation is that the Dalitz plot for the decay to $\pi^+ \pi^-
\pi^0$ looks quite different for $J/\psi$, $\psi(2S)$, and the continuum below
the $\psi(2S)$ \cite{Ablikim:2005jy}: In the case of the $J/\psi$, the $\rho$
bands dominate, while at the two higher energies the $m(\pi\pi)$ distributions
tend towards higher values.  Studies of $\psi(2S) \to VP$ states by CLEO
\cite{Adam:2004pr} and BES \cite{Ablikim:2004sf,Ablikim:2004kv,Ablikim:2004ky}
show that the 12\% rule is much-better obeyed for $VP$ decays forbidden by
G-parity and hence proceeding via electromagnetism (e.g., $\psi(2S) \to \omega
\pi^0, \rho \eta, \rho \eta')$.
The AP (A for axial-vector) final state
$b_1 \pi$ obeys the scaling prediction for both the charged and the
neutral isospin configuration~\cite{Yao:2006px}.

Investigation of decays of the kind $\psi(2S) \to PP$ for $P=\pi^+$, $K^+$,
and $K^0$ allow one to extract the relative phase and strength ratio between
the $\psi(2S) \to ggg$ and $\psi(2S) \to \gamma^*$ amplitudes. This has
been done by the CLEO and BES Collaborations~(\citet{Dobbs:2006fj} and
references therein).

CLEO has studied many exclusive multi-body final states of $\psi(2S)$
\cite{Briere:2005rc}, several of which have not been reported before.  Mode by
mode, deviations from the 12\% rule rarely amount to more than a factor of two.
Moreover, the ratio of
${\cal B}[\psi(2S) \to {\rm light~hadrons}] = (16.9 \pm 2.6)\%$
to
${\cal B}[J/\psi \to {\rm light~hadrons}] = (86.8 \pm 0.4)\%$~\cite{Yao:2006px} is
$(19.4 \pm 3.1)\%$, which {\it exceeds} the
aforementioned corresponding ratio for lepton pairs, $(12.4 \pm 0.3)\%$,
by $2.3 \sigma$.
The suppression of hadronic $\psi(2S)$ final states
thus appears to be confined to certain species such as $\rho \pi, K^* \bar K$.

The CLEO Collaboration has measured decays of $\psi(2S)$ to baryon-antibaryon
pairs \cite{Pedlar:2005px}, as has the BES Collaboration~\cite{Ablikim:2006aw}.
The branching ratios indicate that flavor SU(3) seems approximately valid for
octet-baryon pair production.  In all measured channels, the values of $Q$ are
either compatible with or greater than the expected 12\% value.

No clear pattern emerges, with some channels obeying the
12\%~rule while others fail drastically, and so the conclusion at
this point is that the simplified picture as painted by the 12\%~rule
is not adequate, and more refined models are necessary.

\subsection{The $h_c(1^1P_1)$}

\begin{figure}
\begin{center}
\includegraphics[width=0.65\textwidth]{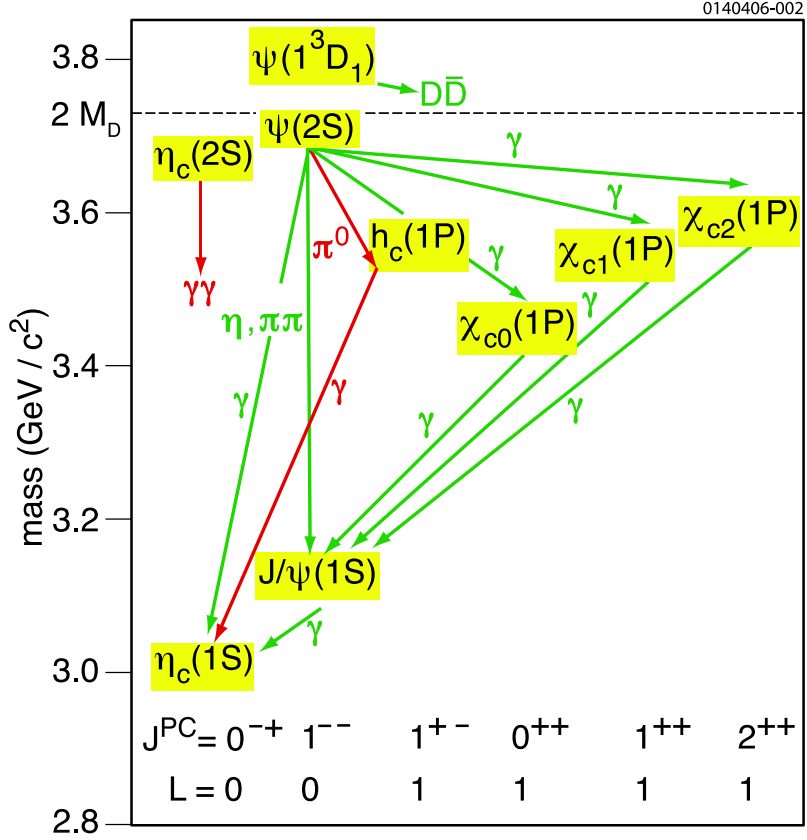}
\end{center}
\caption{Transitions among low-lying charmonium states. From 
\citet{Cassel:2006ccour}.
\label{fig:ccour}}
\end{figure}

The $h_c(1^1P_1)$ state of charmonium has been observed by CLEO
\cite{Rosner:2005ry,Rubin:2005px} via $\psi(2S) \to \pi^0 h_c$ with $h_c \to
\gamma \eta_c$.  These transitions are denoted by red (dark) arrows in Fig.\
\ref{fig:ccour} \cite{Cassel:2006ccour}. It has also been seen by Fermilab Experiment E835
\cite{Andreotti:2005vu} in the reaction $\bar p p \to h_c \to \gamma \eta_c
\to \gamma \gamma \gamma$, with 13 candidate events.  A search for the
decay $B^\pm \to h_c K^\pm$ by the Belle Collaboration, however, has
resulted only in an upper limit on the branching ratio \cite{Fang:2006bz}
${\cal B}(B^\pm \to h_c K^\pm) < 3.8 \times 10^{-5}$ for $m(h_c) = 3527$
MeV and ${\cal B}(h_c \to \gamma \eta_c) = 0.5$.  Attempts at
previous observations are documented in \citet{Rubin:2005px}.

\subsubsection{Significance of $h_c$ mass measurement}

Hyperfine splittings test the spin-dependence and spatial behavior of the $Q
\bar Q$ force.  Whereas these splittings are $m(J/\psi) - m(\eta_c) = 116.5 \pm
1.2$ MeV for 1S and
$m(\psi(2S)) - m(\eta_c(2S))=48 \pm 4$ MeV for 2S levels, $P$-wave
splittings should be less than a few MeV since the potential is proportional to
$\delta^3(\vec{r})$ for a Coulomb-like $c \bar c$ interaction.  Lattice QCD
\cite{Manke:2000dg,Okamoto:2001jb} and relativistic potential \cite{Ebert:2002pp} calculations confirm
this expectation.  One expects $m(h_c) \equiv m(1^1P_1) \simeq
\langle m(^3P_J) \rangle = 3525.36 \pm 0.06$ MeV.

\subsubsection{Detection in $\psi(2S) \to
\pi^0 h_c \to \pi^0 \gamma \eta_c$}

\begin{figure}
\mbox{
\includegraphics[width=0.57\textwidth, bb = 0 20 474 346]{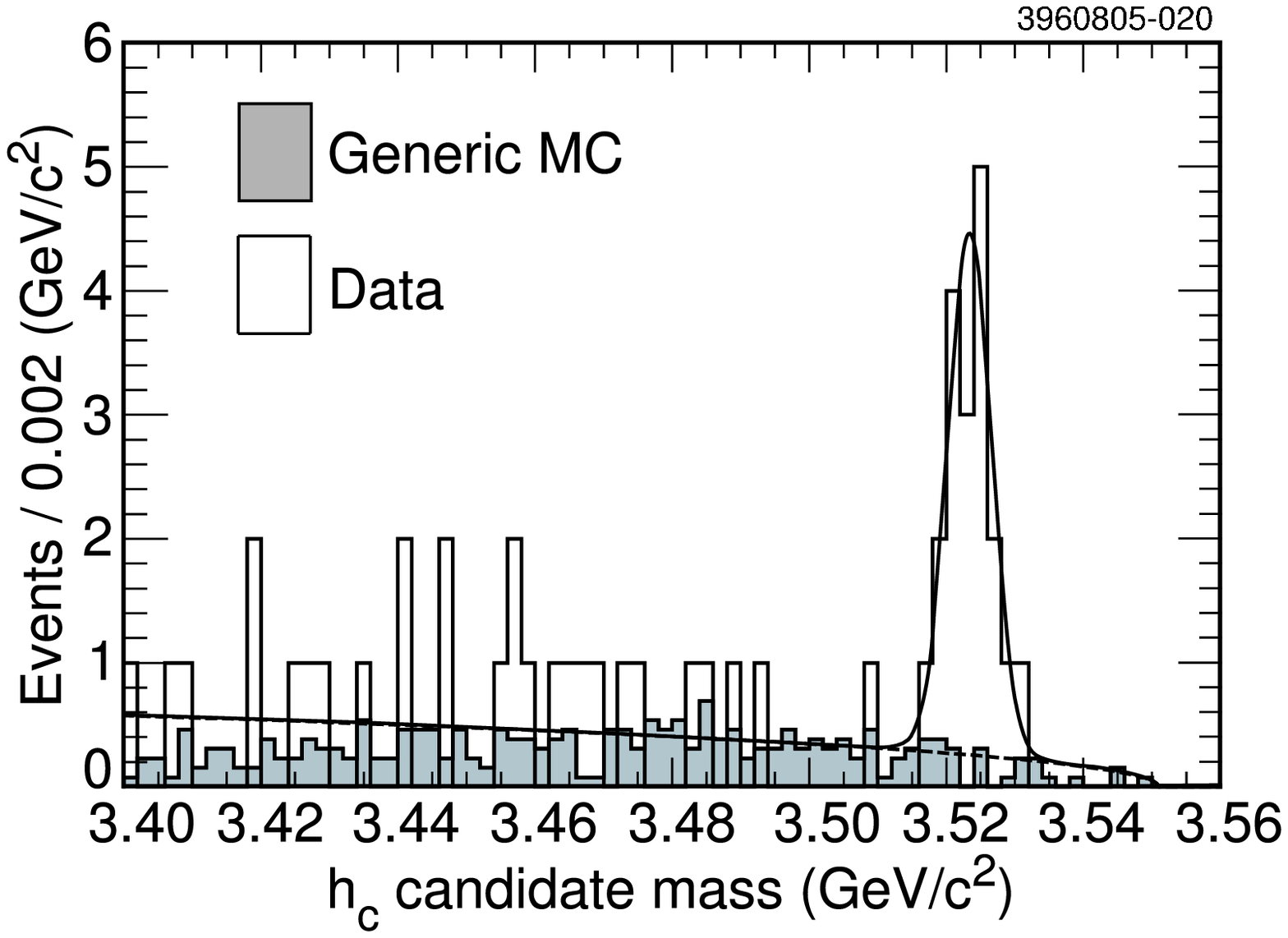}
\includegraphics[width=0.4\textwidth]{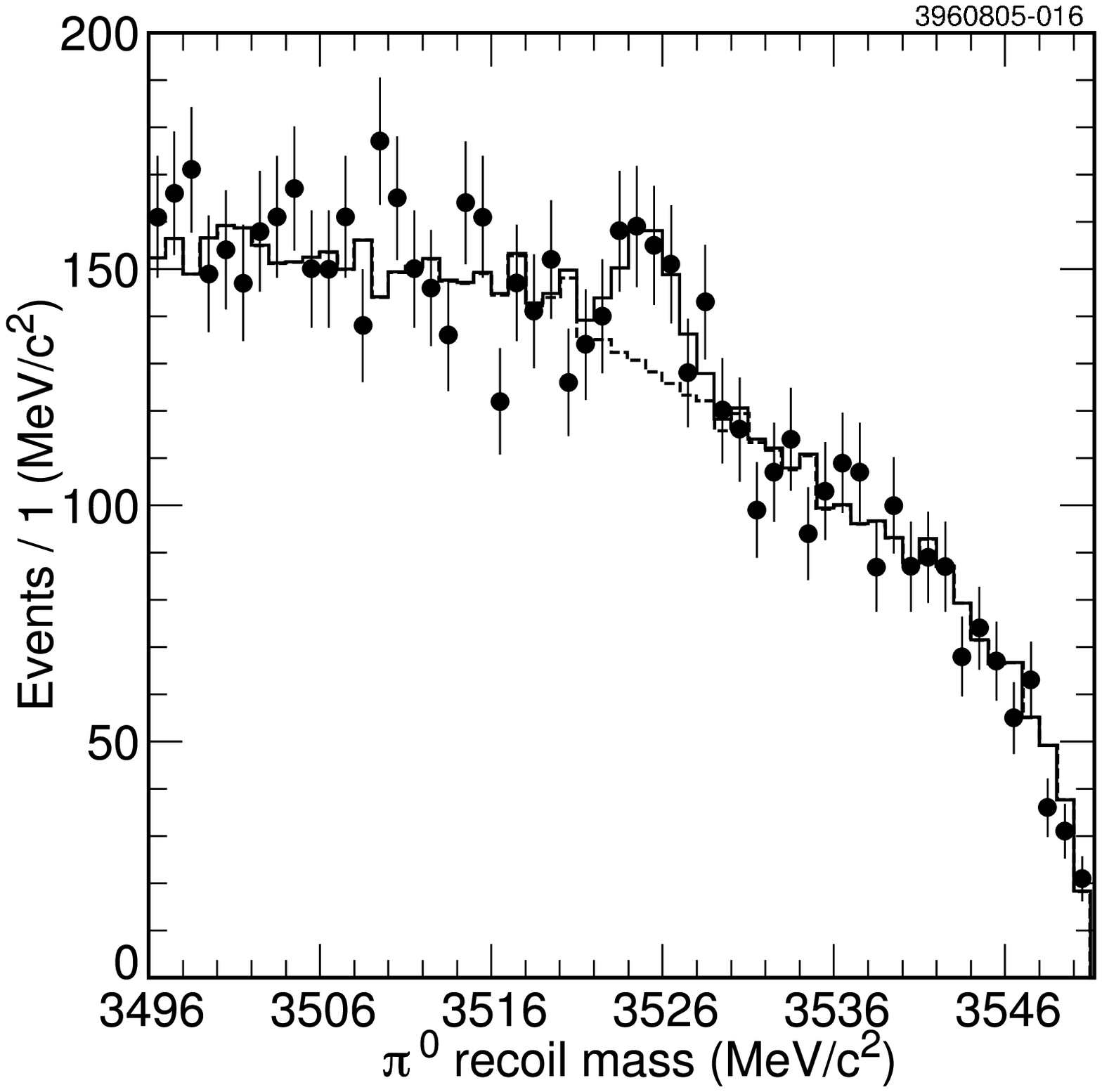}
}
\caption{Left: Exclusive $h_c$ signal from CLEO (3 million $\psi(2S)$ decays).
Data events correspond to open histogram; Monte Carlo background estimate is
denoted by shaded histogram.  The signal shape is a double Gaussian, obtained
from signal Monte Carlo.  The background shape is an ARGUS function.
Right: Inclusive $h_c$ signal from CLEO (3 million $\psi(2S)$ decays).
The curve denotes the background function based on generic Monte Carlo plus
signal.  The dashed line shows the contribution of background alone.
Both figures are from \citet{Rubin:2005px}.
\label{fig:hc}}
\end{figure}

In the CLEO data, both inclusive and exclusive analyses saw a signal near
$\langle m(^3P_J) \rangle$.  The exclusive analysis reconstructed $\eta_c$ in 7
decay modes, while no $\eta_c$ reconstruction was performed in the inclusive
analysis.  The exclusive signal is shown on the left in Fig.\ \ref{fig:hc}.  A
total of 19 candidates were identified, with a signal of $17.5 \pm 4.5$ events
above background.  The result of one of two inclusive analyses is shown on the
right in Fig.\ \ref{fig:hc}.  Combining
exclusive and inclusive results yields $m(h_c) = (3524.4 \pm 0.6 \pm 0.4)$ MeV,
${\cal B}_1 {\cal B}_2 = (4.0 \pm 0.8 \pm 0.7) \times 10^{-4}$.  The $h_c$ mass
is $(1.0 \pm 0.6 \pm 0.4)$ MeV below $\langle m(^3P_J) \rangle$, at the edge of
the (nonrelativistic) bound \cite{Stubbe:1991qw} $m(h_c) \ge
\langle m(^3P_J) \rangle$ and indicating little $P$-wave hyperfine splitting in
charmonium.  The value of ${\cal B}_1 {\cal B}_2$ agrees with theoretical
estimates \cite{Godfrey:2002rp} of $(10^{-3} \cdot 0.4)$.

\subsubsection{Detection in the exclusive process $p \bar p \to h_c \to
\gamma \eta_c \to 3 \gamma$}

The Fermilab E835 Collaboration \cite{Andreotti:2005vu} studied a number
of charmonium resonances accessible in the direct $\bar p p$ channel using
the carefully controlled $\bar p$ energy of the Fermilab Accumulator ring
and a gas-jet fixed target.  The signal of 13 events sits above an estimated
background of 3 events and corresponds to a mass $m(h_c) = 3525.8 \pm 0.2 \pm
0.2$ MeV.  The signal strength is evaluated to be $\Gamma_{\bar p p} {\cal B}
_{\eta_c \gamma} = (10.0 \pm 3.5,~12.0 \pm 4.5)$ eV for $\Gamma_{\rm tot}(h_c)
= (0.5,~1.0)$ MeV.  With ${\cal B}_{\eta_c \gamma} = 0.4$ this would correspond
to $\Gamma_{h_c \to \bar p p} = (25,~30)$ eV.  (Kuang, Tuan, and Yan predicted
$\Gamma_{h_c \to \bar p p} = 186$ eV \cite{Kuang:1988bz}.) For comparison the partial
widths of $\eta_c,J/\psi,\chi_{c0,1,2},$ and $\psi(2S)$ to $\bar p p$ are
roughly
$(33 \pm 11)~{\rm keV}$,
$(203 \pm 9)~{\rm eV}$,
$(2.25 \pm 0.25)~{\rm keV}$\footnote{Using
${\cal B}(\chi_{c0} \to p \bar p)= (2.16 \pm 0.19)\times 10^{-4}$.},
$(60 \pm 6)~{\rm eV}$,
$(136 \pm 13)$ eV, and
$(89 \pm 8)$ eV, where we have used
branching ratios and total widths from \citet{Yao:2006px}.

\subsection{The $\eta_c(2S)$}

The claim by the Crystal Ball Collaboration \cite{Edwards:1981mq} for the first
radial excitation of the~$\eta_c$, the $\eta_c(2S)$, at a mass of
$3594 \pm 5$ MeV, remained unconfirmed for 20 years.  Then, the Belle
Collaboration observed a candidate for $\eta_c(2S)$ in
$B \to K (K_S K \pi)$
\cite{Choi:2002na} and $e^+ e^- \to J/\psi + X$ \cite{Abe:2002rb} at a
significantly higher mass.  An upper limit on the
decay $\psi(2S) \to \gamma \eta_c(2S)$ by the CLEO Collaboration
\cite{Athar:2004dn} failed to confirm the Crystal Ball state at 3594 MeV.
The Belle result stimulated a study of what other charmonium states could
be produced in $B$ decays \cite{Eichten:2002qv}.

By studying its production in photon-photon collisions,
CLEO \cite{Asner:2003wv}
confirmed the presence of the new $\eta_c(2S)$ candidate, as did the BaBar
Collaboration \cite{Aubert:2003pt}.  The mass of the $\eta_c(2S)$
is found to be only $48 \pm 4$ MeV/$c^2$ below the corresponding spin-triplet
$\psi(2S)$ state, a hyperfine splitting which is considerably less than the
$116.5 \pm 1.2$ MeV/$c^2$ difference seen in the 1S charmonium states ({\it
i.e.}, between the $J/\psi$ and the $\eta_c(1S)$).  While potential models
predict the $\psi(2S)-\eta_c(2S)$ splitting to be less than the
$J/\psi-\eta_c$ splitting due to the smaller wavefunction at the
origin for the $2S$ state compared to the $1S$ state, most models
(e.g., \citet{Ebert:2002pp,Eichten:1994gt,Fulcher:1991dm,Gupta:1995ps}),
but not all \cite{Godfrey:1985xj,Zeng:1994vj}, predict a much larger splitting than
what is observed.  It is likely that the proximity of the charmed meson pair
threshold, which can lower the $\psi(2S)$ mass by tens of MeV/$c^2$
\cite{Eichten:2004uh,Eichten:2005ga,Martin:1982nw}, plays an
important role in the $\psi(2S)-\eta_c(2S)$ splitting.

The width of the $\eta_c(2S)$ can be determined in similar ways as that of
the $\eta_c(1S)$ in principle. In practice, the M1~photon $\psi(2S) \to
\gamma_{\mathrm{M1}} \eta_c(2S)$ is difficult to measure well
in an inclusive measurement $\eta_c(2S) \to X$ due to background,
and all channels $\eta_c(2S) \to Y$ that can be faked by
$\psi(2S)$ decay, which is several orders of magnitude more copious,
will not be helpful in an exclusive measurement either. The only
available width measurements come from two-photon
reactions: $\Gamma = (6.3 \pm 12.4 \pm 4.0)\,\mbox{Mev}$~\cite{Asner:2003wv},
$\Gamma = (17.0 \pm 8.3 \pm 2.5)\,\mbox{MeV}$~\cite{Aubert:2003pt}, leading
to an average of $\Gamma = (14 \pm 7)\,\mbox{MeV}$~\cite{Yao:2006px}.
The measurements are dominated by statistical uncertainties, which is dictated
by the need to identify an $\eta_c(2S)$ final state.  To date, the only known
decay modes of the $\eta_c(2S)$ are $K_S K^\pm \pi^\mp$ and $\gamma\gamma$.
One may be led to try the same modes to which the $\eta_c(1S)$ decays; the
listed two measurements use $\eta_c(2S) \to K_S K^\pm \pi^\mp$.

The CLEO Collaboration found that the product $\Gamma(\eta_c(2S) \to \gamma
\gamma) {\cal B}(\eta_c(2S) \to K_S K \pi)$ is only
$0.18 \pm 0.05 \pm 0.02$ times the corresponding product for $\eta_c(1S)$.
This could pose a problem for descriptions of charmonium if the branching
ratios to $K_S K \pi$ are equal.  More likely, the heavier $\eta_c(2S)$ has
more decay modes available to it, so its branching ratio to $K_S K \pi$ is
likely to be less than that of the $\eta_c(1S)$.

\subsection{The $\psi(3770)$}

The $\psi(3770)$ is primarily a $1^3D_1$ state with small admixtures
of $n^3S_1$ states [notably $\psi(2S)$]
\cite{Eichten:2004uh,Eichten:2005ga,Rosner:2004wy}.
It is most easily produced in $e^+e^-$ collisions, where it
appears at 3770~MeV as a broad resonance
($23.0 \pm 2.7\,\mathrm{MeV}$~\cite{Yao:2006px}).
Both Belle~\cite{Abe:2003zv} and BaBar~\cite{Aubert:2005vi}
observed $\psi(3770)$ in $B$~decay.  The broadness of the state is
due to the fact that decay to open charm $D \bar D$ is kinematically
available and also allowed by quantum numbers. Final states involving
$D^*$ and $D_s$ are not accessible at this energy. The mass and
width have are most accurately determined in a scan.
\citet{Ablikim:2006zq} achieves uncertainties of below 1~MeV
for the mass and below 10\% relative on the width.
The leptonic width
can be determined via a hadron production rate measurement as the cross-section
is proportional to the coupling, $\Gamma_{ee}$.
The BES Collaboration has been studying its decays to
charmed and non-charmed final states (see, e.g., \citet{Ablikim:2004ck}),
and for the past few years it has been the subject of dedicated studies by the
CLEO Collaboration \cite{Briere:2001CLEOc}.

\subsubsection{$\psi(3770)$ as a ``charm factory''}

The fact that $\psi(3770)$ lies so close to charm threshold [only about 40 MeV
above $2 m(D^0)$] makes it a
a source of charmed particle pairs in a well-defined quantum state
(without additional pions) in $e^+ e^-$ collisions.  An interesting question is
whether the total cross section $\sigma(e^+ e^- \to \psi(3770))$ is nearly
saturated by $D \bar D$.  If not, there could be noticeable non-$D \bar D$
decays of the $\psi(3770)$ \cite{Rosner:2004wy}.  A CLEO measurement
\cite{Besson:2005hm}, $\sigma(\psi(3770)) = (6.38 \pm 0.08 ^{+0.41}_{-0.30})$
nb, appears very close to the CLEO value $\sigma(D \bar D) =
(6.39\pm0.10^{+0.17}_{-0.08})$ nb \cite{He:2005bs}, leaving little room for
non-$D \bar D$ decays.  Some question has nonetheless been raised by
BES analyses \cite{Ablikim:2006aj,Ablikim:2006zq,Ablikim:2006md} in which a
substantial non-$D \bar D$ component could still be present.

As a result of the difference between $D^0$ and $D^-$ masses, the $\psi(3770)$
decays to $D^0 \bar D^0$ more frequently than to $D^+ D^-$.  For example,
\citet{He:2005bs} finds
$\sigma(e^+ e^- \to \psi(3770) \to D^+ D^-)/ \sigma
(e^+ e^- \to \psi(3770) \to D^0 \bar D^0) = 0.776 \pm 0.024^{+0.014}_{-0.006}$.
This ratio reflects not only the effect of differing phase space, but also
different final-state electromagnetic interactions \cite{Voloshin:2004nu}, and
is expected to vary somewhat as center-of-mass energy is varied over the
resonance peak.

\subsubsection{Leptonic width and mixing}

The CLEO measurement of $\sigma(\psi(3770))$
mentioned above \cite{Besson:2005hm}
also leads to a more precise value for the $\psi(3770)$ leptonic width,
$\Gamma_{ee}(\psi(3770)) = (0.204\pm0.003^{+0.041}_{-0.027})$ keV.  This enters
into the quoted average \cite{Yao:2006px} of $(0.242^{+0.027}_{-0.024})$ keV.
Subsequent results are
$(0.251\pm0.026\pm0.011)$ keV \cite{Ablikim:2006zq} and
$(0.279\pm0.011\pm0.013)$ keV \cite{Ablikim:2006md} from BES-II.
These improvements allow a
more precise estimate for the angle $\phi$ describing the mixing between
$1D$ and $2S$ states in $\psi(2S)$ and $\psi(3770)$:
\beq
\psi(2S)   = - \sin \phi \ket{1^3D_1} + \cos \phi \ket{2^3S_1}~~,~~~
\psi(3770) = \cos \phi \ket{1^3D_1} + \sin \phi \ket{2^3S_1}~~.
\eeq
This mixing affects the ratio $R_{\psi(3770)/\psi(2S)}$
of leptonic widths of $\psi(2S)$ and $\psi(3770)$
and their predicted rates for E1 transitions to the $\chi_{cJ}$ states
\cite{Rosner:2001nm,Kuang:2002hz}.
A previous analysis based on $\Gamma_{ee}(\psi(3770)) = 0.26 \pm 0.04$ keV
\cite{Rosner:2004wy} gave $\phi = (12\pm2)^\circ$, while the present leptonic
width will give smaller errors on $\phi$.  The large present and anticipated
CLEO-c $\psi(3770)$ data sample will further constrain this value.
A solution with negative $\phi$ consistent with $R_{\psi(3770)/\psi(2S)}$
gives an unphysically large rate for $\psi(2S) \to \gamma \chi_{c0}$.

As noted earlier, the nonrelativistic predictions for
the $\psi(2S)$ rates are generally too high, indicating the limitations of a
nonrelativistic approach.  We shall see that the predicted rate for $\psi(3770)
\to \gamma \chi_{c0}$, which has recently been observed by the CLEO
Collaboration \cite{Briere:2006ff}, is also a factor of 2 too high in a
nonrelativistic approach but is satisfactory when relativistic and
coupled-channel effects are taken into account.

\subsubsection{$\psi(3770)$ transitions to $\pi \pi J/\psi$}

The rates for transitions of $\psi(3770)$ to $\pi \pi J/\psi$
have been predicted on the assumption that it is mainly a $D$-wave state
with a small $S$-wave
admixture as in the above example \cite{Kuang:2006me}.  (The sign convention
for the mixing angle in \citet{Kuang:2006me} is opposite to ours.)  A wide
range of partial widths, $\Gamma(\psi(3770) \to \pi^+ \pi^- J/\psi) = 26$ to 147
keV, corresponding to branching ratios ranging from about 0.1\% to 0.7\%,
is predicted.

The BES Collaboration \cite{Bai:2003hv} finds ${\cal B}(\psi(3770) \to \pi^+ \pi^-
J/\psi) = (0.34 \pm 0.14 \pm 0.09)\%$.  The CLEO Collaboration has measured
a number of branching ratios for $\psi(3770) \to X J/\psi$ \cite{Adam:2005mr}:
${\cal B}(\psi(3770) \to \pi^+ \pi^- J/\psi) =(0.189\pm0.020\pm0.020)\%$,
${\cal B}(\psi(3770) \to \pi^0 \pi^0 J/\psi) =(0.080\pm0.025\pm0.016)\%$,
${\cal B}(\psi(3770) \to \eta J/\psi) = (0.087\pm0.033\pm0.022)\%$, and
${\cal B}(\psi(3770) \to \pi^0 J/\psi) < 0.028\%$.
Together these account for less than 1/2\% of the total $\psi(3770)$ decays.
In these analyses, the contribution from the tail of the $\psi(2S)$ decaying
to the same final states has been subtracted incoherently.

\subsubsection{$\psi(3770)$ transitions to $\gamma \chi_{cJ}$}

CLEO has recently reported results on $\psi(3770) \to \gamma \chi_{cJ}$ partial
widths, based on the exclusive process $\psi(3770) \to \gamma \chi_{c1,2} \to
\gamma \gamma J/\psi \to \gamma \gamma \ell^+ \ell^-$ \cite{Coan:2005ps} and
reconstruction of exclusive $\chi_{cJ}$ decays \cite{Briere:2006ff}.  The
results are shown in Table \ref{tab:psipprad}, implying
$\sum_J{\cal B}(\psi(3770) \to \gamma \chi_{cJ}) = {\cal O}$(1\%).
Recent calculations \cite{Eichten:2004uh,Barnes:2005pb} including relativistic
corrections are in good agreement with these measurements while nonrelativistic
treatments overestimate $\Gamma(\psi(3770) \to \gamma \chi_{c0})$.
The contribution from the tail of the $\psi(2S)$ decaying
to the same final states has been subtracted incoherently.

\begin{table}
\caption{Radiative decays $\psi(3770) \to \gamma \chi_{cJ}$:
energies, predicted and measured partial widths.
Theoretical predictions of \citet{Eichten:2004uh} are (a) without and
(b) with coupled-channel effects; nonrelativistic (c) and relativistic (d)
predictions of \citet{Barnes:2005pb}; (e) shows predictions of
\citet{Rosner:2001nm}.
\label{tab:psipprad}}
\begin{center}
\begin{tabular}{|c|c|ccccc|c|} \hline \hline
Mode & $E_\gamma$ (MeV)    & \multicolumn{5}{c|}{Predicted (keV)} & CLEO (keV) \\ \cline{3-7}
& \cite{Yao:2006px}   & (a) & (b) & (c) & (d) & (e) & \cite{Briere:2006ff} \\ \hline
$\gamma \chi_{c2}$ & 208.8 & 3.2 & 3.9 & 4.9 & 3.3 & 24$\pm$4 & $<21$ \\
$\gamma \chi_{c1}$ & 251.4 & 183 & 59 & 125 & 77 & $73\pm9$ & $70\pm17$ \\
$\gamma \chi_{c0}$ & 339.5 & 254 & 225 & 403 & 213 & $523 \pm 12$ & $172\pm30$ \\
\hline \hline
\end{tabular}
\end{center}
\end{table}

\subsubsection{$\psi(3770)$ transitions to light-hadron final states}

Several searches for $\psi(3770) \to ({\rm light~ hadrons})$, including VP
\cite{Adams:2005ks,Zhu:2006uc}, $K_L K_S$
\cite{Ablikim:2004bc,CroninHennessy:2006su},
and multi-body \cite{Huang:2005fx,Ablikim:2007ss} final states
have been performed.  No evidence was seen for any light-hadron $\psi(3770)$
mode above
expectations from continuum production except for a marginally significant
branching ratio ${\cal B}(\psi(3770) \to \phi \eta) = (3.1 \pm 0.7) \times
10^{-4}$ \cite{Adams:2005ks},
indicating no obvious signature of non-$D \bar D$ $\psi(3770)$ decays.

\subsection{Missing charmonium 1D states}
In addition to the $\psi(3770)$, three more charmonium $1D$ states
are expected:
the spin triplet $^3D_2$ and $^3D_3$ states and a spin singlet $^1D_2$ state.
All these remaining states are expected to be narrow.

The masses of the remaining states are expected to be slightly above the
$\psi(3770)$. Using the usual spin-dependent potentials we expect the
$^3D_2, ^1D_2, ^3D_3$ to lie about (+20, +20,  +30) MeV, respectively,
above the $\psi(3770)$ mass \cite{Brambilla:2004wf}.  The effects of
coupling to decay channels may also produce important mass splittings.
In one model these additional shifts are (+37, +44, +59) MeV respectively
\cite{Eichten:2004uh,Eichten:2005ga}.

The $J=2$ states ($^3D_2$ and $^1D_2$) are forbidden by parity to decay
into two pseudoscalar $D$~mesons.  Hence these states are quite narrow.
The principal decay modes for the $^3D_2$ state are expected to be:
radiative transitions ($\gamma ^3P_1$ and $\gamma ^3P_2$),
hadronic transitions ($\pi \pi J /\psi$), and to light hadrons ($ggg$).
The total width is expected to be about
400~keV \cite{Eichten:2002qv,Barnes:2003vb,Eichten:2004uh}
The principal decay modes for the spin-singlet $^1D_2$ state are similar:
radiative transitions ($\gamma ^1P_1$),
hadronic transitions ($\pi \pi \eta_c(1S)$), and light hadrons ($gg$).
The total width is expected to be about
460~keV \cite{Eichten:2002qv,Barnes:2003vb,Eichten:2004uh}.

Finally, the $^3D_3$ state has a Zweig-allowed strong decay to
$D\bar D$ but only in an $F$-wave \cite{Barnes:2003vb,Eichten:2004uh}.
Hence the expected rate of this dominant decay is small.
For example, at a mass of 3868~MeV this decay width is only
0.8~MeV \cite{Eichten:2004uh,Eichten:2005ga,Barnes:2005pb}.
Thus other decay modes such as $\pi \pi J/\psi$ and $\gamma ^3P_2$
may be observable.

Production rates for these remaining $1D$ states in hadronic collisions or
$B$~meson decays are expected to be not significantly larger than those for
$\psi(3770)$. Qualitatively, this is based on the assumption
that the production of $c \bar c$ states with large relative orbital
momentum is suppressed, and the states in question do not mix with $S$- or
$P$-wave charmonium states.

\subsection{$\psi(4040)$, $\psi(4160)$, and $\psi(4415)$}

The $\psi(4040)$ and $\psi(4160)$ resonances appear as elevations
in the measurement of $R = \sigma(\mbox{hadrons})/\sigma(\mu^+\mu^-)$.
They are commonly identified with the $3S$ and $2D$ states of charmonium
(Fig.~\ref{fig:charmon}).  Their parameters have undergone
some refinement as a result of a recent analysis in \citet{Seth:2005ny}.
Results using initial state radiation events
from Belle~\cite{Abe:2006fj} indicate that the $D^* \bar D$ and
$D^* \bar D^*$ final states are populated throughout this energy region,
making interference effects between the resonances inevitable.
BES has re-evaluated earlier published data from a scan in
the region $2-5\,\mathrm{GeV}$ in center-of-mass energy~\cite{Ablikim:2007gd}
to arrive at estimates of masses, total widths, partial electronic widths, and
relative phases of $\psi(3770)$, $\psi(4040)$, $\psi(4160)$, and $\psi(4415)$.
The analysis is the first to take interference between the states into account.
Doing so affects especially the parameters extracted for the three upper states
significantly.  In summary, the treatment of charmonium states above threshold
from inclusive decays is not unambiguous, and parameters must be seen within
the context of the method that was used to obtain them.
Belle has reported a result~\cite{Pakhlova:2007fq}
for the first exclusive decay of $\psi(4415)$,
$\psi(4415) \to D D_2^*(2460) \to DD\pi$.
Belle determines mass and total width of the $\psi(4415)$ from the
$m(D D_2^*(2460))$ distribution, achieving a result in
agreement with~\cite{Ablikim:2007gd}.

Data taken at the $\psi(4040)$ and the $\psi(4160)$ can be useful to
search for the $2P$ states through radiative decays $\psi(4160) \to \gamma
\chi_{c0,1,2}(2P)$.
Identifying the transition photon in the inclusive photon spectrum requires
excellent background suppression and is therefore a challenge.
The E1~branching fractions listed in~\cite{Barnes:2006xq} are, calculated for
$\chi_{cJ}(2P)$ masses chosen to be\footnote{
The motivation for this choice will become apparent in Section~\ref{sec:xyz}.
}
3929/3940/3940~MeV for $J=2/1/0$:\newline
\mbox{\qquad}
$\psi(4040) \to \gamma \chi_{c2,1,0}(2P)$:
\qquad
$0.7 / 0.3 / 0.1 \times 10^{-3}$, \newline
\mbox{\qquad}
$\psi(4160) \to \gamma \chi_{c2,1,0}(2P)$:
\qquad
$0.1 / 1.3 / 1.7 \times 10^{-3}$.

The $J=0$ and $J=1$ states
can be distinguished since the decays $\chi_{c0} \to D \bar D$ and $\chi_{c1}
\to D \bar D^*$ are possible but not the reverse. $\chi_{c2}(2P)$ can
decay to either, where the relative rate depends on the amount of
phase space, which in turn depends on the mass.
Exclusive decays to charmonium have not been observed, though
CLEO has set upper limits on a number of final states involving
charmonium~\cite{Coan:2006rv}.

\subsection{New charmonium-like states}
\label{sec:xyz}

Many new charmonium states above $D \bar D$ threshold have recently
been observed.  While some of these states appear to be consistent
with conventional $c\bar{c}$ states, others do not.  Here we give a
brief survey of the new states and their possible interpretations.
Reviews may be found in \citet{Rosner:2005gf,Godfrey:2006pd,Swanson:2006ap}.
In all cases, the picture is not entirely clear.
This situation could be remedied by
a coherent search of the decay pattern to $D \bar D^{(*)}$,
search for production in two-photon fusion and ISR, the
study of radiative decays of $\psi(4160)$, and of course
tighter uncertainties by way of improved statistical precision upon
the current measurements.

\subsubsection{$X(3872)$}

The $X(3872)$, discovered by Belle in $B$ decays \cite{Choi:2003ue} and
confirmed by BaBar \cite{Aubert:2004ns} and in hadronic production by CDF
\cite{Acosta:2003zx} and D0 \cite{Abazov:2004kp}, is a narrow state of mass
3872~MeV that was first seen decaying to $J/\psi \pi^+ \pi^-$.
No signal at this mass was seen in $B \to X^- K$, $X^- \to \pi^-\pi^0 J/\psi$
\cite{Aubert:2004zr}, which would have implied a charged partner of $X(3872)$.
It was not observed in two-photon production or initial state radiation
\cite{Dobbs:2004di}. Subsequent studies focused on determining the mass, width,
and decay properties in order to establish its quantum numbers and possible
position in the charmonium system of states.  To date, decays to $\pi^+\pi^-
J/\psi$, $\gamma J/\psi$, $\pi^+\pi^-\pi^0 J/\psi$ and possibly
$D^0 \bar D^0 \pi^0$ have been reported.  Results on
decay modes of $X(3872)$ are summarized in Table~\ref{tab:x3872_decay}.

\begin{table}
\caption{Summary of the $X(3872)$ decay modes and searches.
The two entries for $D^0 \bar D^0 \pi^0$ are both from Belle
and are based on samples of
$88\,\mathrm{fb}^{-1}$ and $414\,\mathrm{fb}^{-1}$,
respectively.
\label{tab:x3872_decay}}
\begin{minipage}{\textwidth}
\begin{center}
\begin{tabular}{c c c c} \hline \hline
final state          & $X(3872)$ branching fraction
& reference             \\
\hline
$\pi^+\pi^- J/\psi$  & $(11.6 \pm 1.9) \times 10^{-6}
/{\cal B}_{B^+ \to X(3872)K^+}$ ($>10\sigma$)
& \cite{Aubert:2005vi}      \\
$\pi^-\pi^0 J/\psi$  & not seen    & \cite{Aubert:2004zr} \\
$\gamma\chi_{c1}$    & $< 0.9 \times {\cal B}_{\pi^+\pi^-J/\psi}$
& \cite{Choi:2003ue}    \\
$\gamma J/\psi$      & $(3.3 \pm 1.0 \pm 0.3) \times 10^{-6}
/{\cal B}_{B \to X(3872)K^+}$ ($>4\sigma$)
& \cite{Aubert:2006aj}  \\
& $ (0.14 \pm 0.05)
\times {\cal B}_{X(3872) \to \pi^+\pi^- J/\psi}$
($4.0\sigma$)
& \cite{Abe:2005ix}                   \\
$\eta J/\psi$        & $< 7.7 \times 10^{-6}/{\cal B}_{B \to X(3872)K^+}$
& \cite{Aubert:2004fc}   \\
$\pi^+ \pi^- \pi^0 J/\psi$
& $ (1.0 \pm 0.4 \pm 0.3)
\times {\cal B}_{X(3872) \to \pi^+\pi^- J/\psi}$
($4.3\sigma$)
& \cite{Abe:2005ix}      \\
$D^0 \bar D^0$       & $< 6 \times 10^{-5}/{\cal B}_{B^+ \to X(3872)K^+}$
& \cite{Abe:2003zv}   \\
$D^+ D^-$            & $< 4 \times 10^{-5}/{\cal B}_{B^+ \to X(3872)K^+}$
& \cite{Abe:2003zv}   \\
$D^0 \bar D^0 \pi^0$ & $< 6 \times 10^{-5}/{\cal B}_{B^+ \to X(3872)K^+}$
& \cite{Abe:2003zv} \\
& $(12.2 \pm 3.1 ^{+2.3}_{-3.0})
\times 10^{-5}/{\cal B}_{B^+ \to X(3872)K^+}$\footnote{
Belle report the quoted number as branching fraction
at the peak. They find a peak position that is slightly
above that seen by other experiments for other
$X(3872)$ decays.}
($6.4\sigma$)
& \cite{Gokhroo:2006bt}   \\
\hline \hline
\end{tabular}
\end{center}
\end{minipage}
\end{table}

\begin{figure}
\begin{center}
\includegraphics[width=0.95\textwidth]{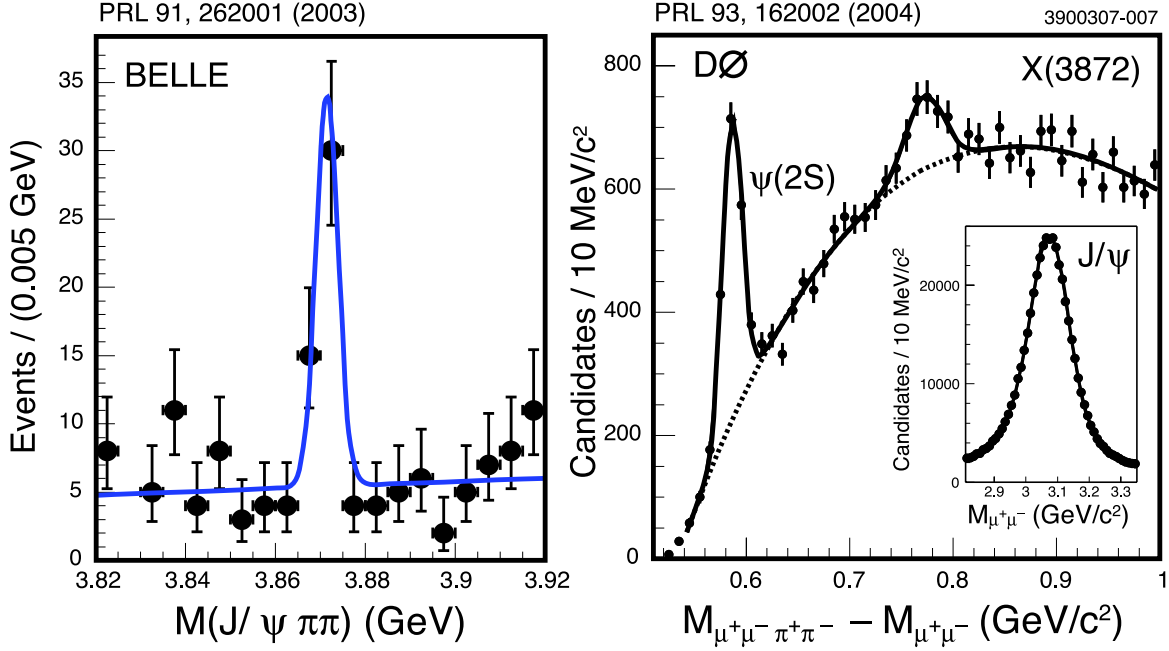}
\caption{Observation of $X(3872) \to \pi^+\pi^- J/\psi$
in $B$~decay (example from Belle~\cite{Choi:2003ue}) and in
$p \bar p$~collisions (example from D0~\cite{Abazov:2004kp}).
\label{fig:x3872_obs}}
\end{center}
\end{figure}

\begin{figure}
\begin{center}
\includegraphics[height=0.55\textheight]{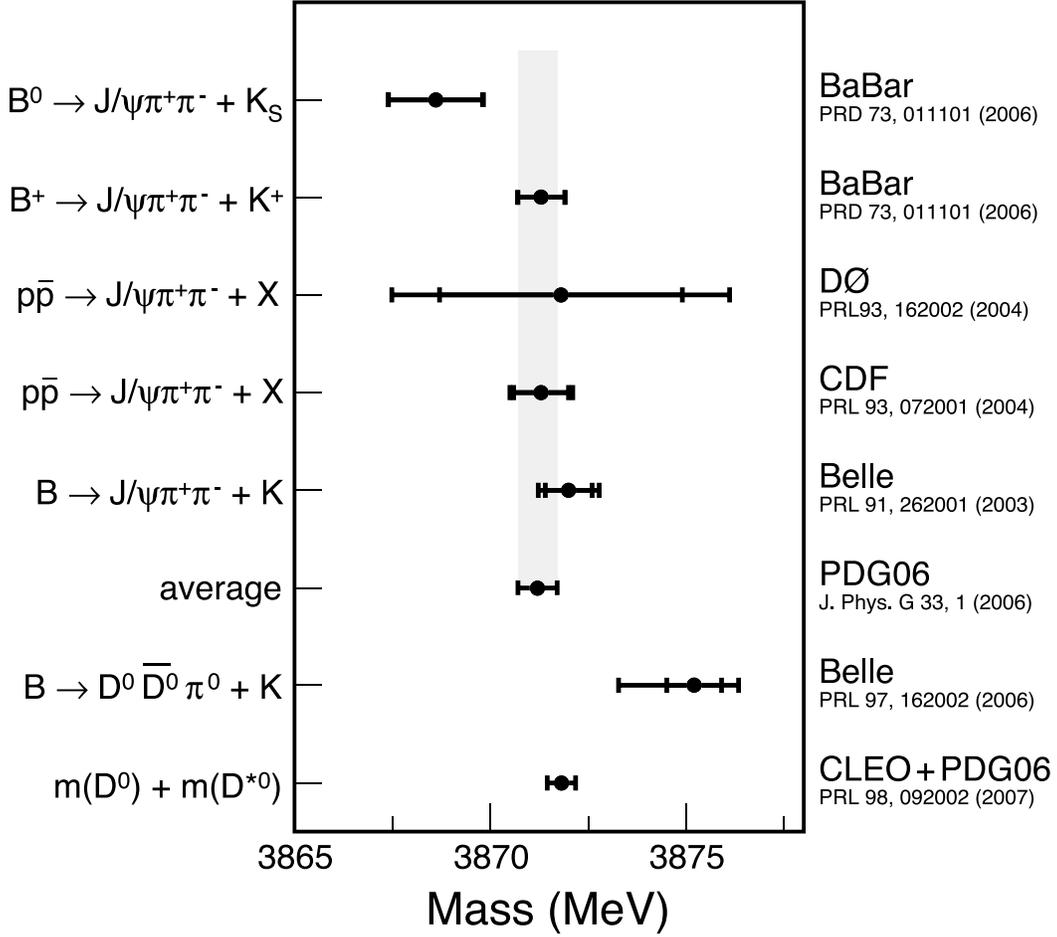}
\caption{Comparison of mass determinations: From
$X(3872) \to \pi^+\pi^- J/\psi$, their weighted average
as computed by PDG, an observed threshold enhancement
in $B \to D^0 \bar D^0 \pi^0 + K$ (at $2\sigma$ deviation from
the average), and the sum of the $D^0$ and $D^{*}$ mass~\cite{Cawlfield:2007dw}.
\label{fig:x3872_mass}}
\end{center}
\end{figure}

The averaged mass of this state is $M = 3871.2 \pm 0.5$~MeV~\cite{Yao:2006px};
the width is determined to be $\Gamma < 2.3$~MeV (90\%~C.L.)
\cite{Choi:2003ue}, below detector resolution.  Signal distributions
from two experiments are shown in Figure~\ref{fig:x3872_obs},
and mass measurements (including $m(D^0) + m(D^{*0})$ \cite{Cawlfield:2007dw})
are compared in Figure~\ref{fig:x3872_mass}.

The combined branching fraction product from Belle and BaBar is
\linebreak
${\cal B}[B^+ \to K^+ X(3872)]
\times {\cal B} [X(3872) \to \pi^+\pi^- J/\psi]
= (11.4 \pm 2.0) \times 10^{-6}$~\cite{Yao:2006px}.
After setting a limit of
${\cal B }[B^+ \to K^+ X(3872)] < 3.2 \times 10^{-4}$ (90\% C.L.),
BaBar \cite{Aubert:2005vi}
derives ${\cal B}[X(3872) \to \pi^+ \pi^- J/\psi] > 4.2\%$ (90\% C.L.).
For comparison, examples of other states above open flavor threshold are
${\cal B}[\psi(3770) \to \pi^+ \pi^- J/\psi] = (1.93 \pm 0.28)\times 10^{-3}$
\cite{Yao:2006px} (partial width 46~keV) and limits
${\cal B}[\psi(4040, \ 4160 ) \to \pi^+ \pi^- J/\psi]$
of order $10^{-3}$~\cite{Yao:2006px} (partial widths $\sim 100\,\mbox{keV}$).

Decay into a pair of $D$~mesons has not been observed, and upper limits
on the rate are in the range of a few times that for $\pi^+ \pi^- J/\psi$
\cite{Abe:2003zv}. A signal in $B \to (D^0 \bar D^0 \pi^0) K$ with $m(D^0
\bar D^0 \pi^0)$ in the right range is the first candidate for open-charm
decays of $X(3872)$.  The observed
rate is an order of magnitude above that for $\pi^+ \pi^- J/\psi$.

The dipion mass distribution favors high $m(\pi^+\pi^-)$ values.  This is not
untypical for charmonium states ({\sl cf.} $\psi(2S) \to \pi^+\pi^- J/\psi$),
but could be an indication that the pion pair might even be produced in a
$\rho$ configuration; if that were indeed the case the $X(3872)$ could not be a
charmonium state.

The decay $X(3872) \to \pi^+ \pi^- \pi^0 J/\psi$ was observed
at a rate comparable to that of $\pi^+ \pi^- J/\psi$~\cite{Abe:2005ix}
(preliminary). The $m(\pi^+ \pi^- \pi^0)$ distribution is concentrated
at the highest values, coinciding with the kinematic limit,
which spurred speculations that the decay might
proceed through (the low-side tail of) an $\omega$.
In any case, if confirmed, the co-existence of both the $X(3872) \to \pi^+
\pi^- J/\psi$ and $X(3872) \to \pi^+ \pi^- \pi^0 J/\psi$ transitions implies
that the $X(3872)$ is a mixture of both I=0 and I=1.

Since the $X(3872)$ lies well above $D \bar D$ threshold but is narrower than
experimental resolution, unnatural $J^P = 0^-,1^+, 2^-$ is favored.
An angular distribution analysis by the Belle Collaboration, utilizing in part
suggestions in \citet{Rosner:2004ac}, favors $J^{PC}=1^{++}$
\cite{Abe:2005iya}, although a higher-statistics analysis by CDF cannot
distinguish between $J^{PC} = 1^{++}$ or $2^{-+}$
\cite{Abulencia:2006ma} (see also
\cite{Swanson:2006ap,Marsiske:2006mh,Kravchenko:2006qx}).
$J^{PC} = 2^{-+}$ is disfavored by Belle's observation \cite{Gokhroo:2006bt} of
$X \to D^0 \bar D^0 \pi^0$, which would require at least two units of relative
orbital angular momentum in the three-body state, very near threshold.

Setting aside the $X(3872) \to \pi^+ \pi^- \pi^0 J/\psi$ observation for the
sake of argument, among conventional $c\bar{c}$ states only the $1D$ and $2P$
multiplets are nearby in mass.  Taking into account the angular distribution
analysis, only the $J^{PC}=1^{++}$ $2^3P_1$ and $2^{-+}$ $1^1D_2$ assignments
are possible.  The decay $X(3872)\to \gamma J/\psi$ is observed at a rate about
a quarter or less
of that for $X(3872)\to \pi^+\pi^-J/\psi$~\cite{Aubert:2006aj,Abe:2005ix}.
This would be an E1 transition for $2^3P_1$ but a more suppressed higher
multipole for $2^{-+}$, and therefore the $J^{PC} =
1^{++}$ interpretation appears more likely assuming $c\bar{c}$ content.
For a $1^{++}$ state the only surviving candidate is the $2^3P_1$.  However, we
will see that the identification of the $Z(3931)$ with the $2^3P_2$ implies a
$2^3P_2$ mass of $\sim 3930$~MeV, which is inconsistent with the $2^3P_1$
interpretation of $X(3872)$ if the $2^3P_2 - 2^3P_1$ mass splittings
are decidedly lower than 50~MeV~\cite{Eichten:2005ga,Barnes:2005pb}.
This favors the conclusion that the $X(3872)$ may be a $D^0\bar{D}^{0*}$
molecule or ``tetraquark''~\cite{Maiani:2004vq,Ebert:2005nc} state.
A prediction of the tetraquark interpretation is the existence of a second $X$
particle decaying to $D^0\bar{D}^0\pi^0$ \cite{Maiani:2004vq}, which has been
reported by the Belle Collaboration \cite{Abe:2006fj}.
However, the $X(3872)$ also has many
features in common with an $S$-wave bound state of $(D^0 \bar D^{*0} + \bar D^0
D^{*0})/ \sqrt{2} \sim c \bar c u \bar u$ with $J^{PC} = 1^{++}$
\cite{Close:2003sg,Tornqvist:2003na,Swanson:2003tb,Swanson:2004pp}.  Its simultaneous decay to $\rho J/\psi$ and $\omega
J/\psi$ with roughly equal branching ratios is a consequence of this
``molecular'' assignment.
A new measurement of $m(D^0) = 1864.847 \pm 0.150 \pm 0.095$ MeV/$c^2$
\cite{Cawlfield:2007dw} implies
$m(D^0 \overline{D^{*0}}) = 3871.81 \pm 0.36$ MeV/$c^2$
and hence a binding energy of $0.6 \pm 0.6$ MeV
(see also Fig.~\ref{fig:x3872_mass}).
Irrespective of its eventual interpretation, the evidence is mounting that the
$X(3872)$ is not a conventional $c\bar{c}$ state.


\subsubsection{$Z(3930)$}

\begin{figure}
\begin{center}
\includegraphics[width=0.95\textwidth]{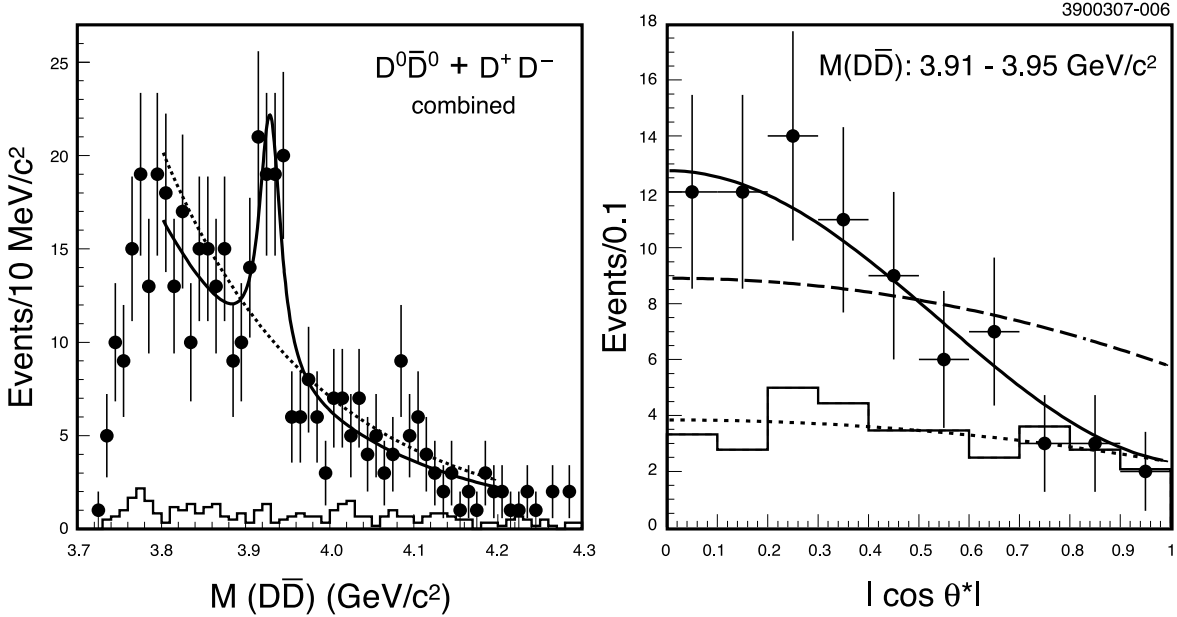}
\caption{Belle's $\chi_{c2}(2P)$ candidate~\cite{Uehara:2005qd}:
Left: The invariant mass $m(D \bar D)$ distribution in two-photon
production of the $Z(3930)$, $D^+D^-$ and $D^0 \bar D^0$ combined.
The signal yield is $64 \pm 18$ events. The two curves are fits
with and without a resonance component.
Right: $\cos \theta^*$, the angle of the $D$~meson relative to the
beam axis in the $\gamma\gamma$ center-of-mass frame for
events with $3.91 < m(D \bar D) < 3.95\,\mbox{GeV}$; the data (circles) are
compared with predictions for $J=2$ (solid) and $J=0$ (dashed).
The background level can be judged from the solid histogram
or the interpolated smooth dotted curve.
\label{fig:z3930}}
\end{center}
\end{figure}

Belle has reported a candidate for a $2^3P_2(\chi_{c2}(2P))$ state in $\gamma
\gamma$ collisions \cite{Uehara:2005qd}, decaying to $D \bar D$.
The state appears as an enhancement in the $m(D \bar D)$ distribution
at a statistical significance of $5.3\sigma$.  The relative $D^+D^-$ and $D^0
\bar D^0$ rates are consistent with expectations based on isospin invariance
and the $D^+ - D^0$ mass difference.  Combining charged and neutral modes, a
fit shown in the left-hand panel of Fig.\ \ref{fig:z3930} yields mass and width
$M=3929\pm 5\pm 2$~MeV and $\Gamma=29\pm 10 \pm 2$
MeV.  Although in principle the $D$-pair could be produced from $D^* \bar D$,
the observed transverse momentum spectrum of the $D \bar D$ pair is consistent
with no contribution from $D^* \bar D$.

The observation of decay to $D \bar D$ makes it impossible
for $Z(3930)$ to be the $\eta_c(3S)$ state.
Both $\chi_{c0}(2P)$ and $\chi_{c2}(2P)$ are
expected to decay to $D \bar D$ ($\chi_{c1}(2P)$ is not; it only decays to
$D^* \bar D$). To distinguish between the two remaining hypotheses,
the distribution in $\theta^*$, which is the angle of the $D$-meson relative to
the beam axis in the $\gamma\gamma$ center-of-mass frame, is examined.  This
distribution is consistent with $\sin^4 \theta^*$ as expected for a state with
$J=2, \lambda = \pm2$ (right-hand panel of Fig.\ \ref{fig:z3930}).
The two-photon width is, under the assumption of a tensor state, measured to be
$\Gamma_{\gamma\gamma}\cdot {\cal B}_{D\bar{D}}=0.18\pm 0.05\pm 0.03$~keV.

BaBar has searched for $Z(3930)$ decay into $\gamma
J/\psi$\cite{Aubert:2006aj}, and set an upper limit ${\cal B}(B \to Z(3930) +
K) \times {\cal B}(Z(3930) \to \gamma J/\psi) < 2.5 \times 10^{-6}$.

The predicted mass of the $\chi_{c2}(2P)$ is 3972~MeV and the
predicted partial widths and total width assuming
$ M[2^3P_2(c\bar{c})]=3930$~MeV 
are~\cite{Eichten:2005ga}\footnote{Barnes, Godfrey and Swanson 
\cite{Barnes:2005pb} obtain similar 
results when the $2^3P_2$ mass is rescaled to 3930~MeV. 
See \citet{Swanson:2006ap}.}
\newline
\mbox{\qquad} $\Gamma(\chi_{c2}(2P)\to D\bar{D})=21.5$~MeV, \newline
\mbox{\qquad} $\Gamma(\chi_{c2}(2P)\to D\bar{D}^*)=7.1$~MeV and \newline
\mbox{\qquad} $\Gamma_{\rm total}(\chi_{c2}(2P))=28.6$~MeV,\newline
in good agreement with the experimental measurement.  Furthermore, using
$\Gamma(\chi_{c2}(2P)\to \gamma\gamma)= 0.67$~keV \cite{Barnes:1992sg}
times $ {\cal B}(\chi_{c2}(2P)\to D\bar{D})=70\%$ implies
$\Gamma_{\gamma\gamma}\cdot {\cal B}_{D\bar{D}}=0.47$~keV, which is
within a factor of 2 of the observed number, fairly good agreement
considering the typical reliability of two-photon partial width predictions.

The observed $Z(3930)$ properties are consistent with those predicted
for the $\chi_{c2}(2P)$ $2^3P_2(c\bar{c})$ state.
So far, the only mild surprise is the observed mass, which is
$40-50\,\mbox{MeV}$ below expectations. Adjusting that, all other
properties observed so far can be accomodated within the framework of
\cite{Eichten:2005ga,Swanson:2006ap}.
The $\chi_{c2}(2P)$ interpretation could be confirmed by observation of the
$D\bar{D}^*$ final state.  We also note that the $\chi_{c2}(2P)$
is predicted to undergo radiative transitions to $\psi(2S)$
with a partial width of ${\cal O}(100\,\mathrm{keV})$
\cite{Eichten:2005ga,Barnes:2005pb}.


\subsubsection{$Y(3940)$}

\begin{figure}
\begin{center}
\includegraphics[width=0.95\textwidth]{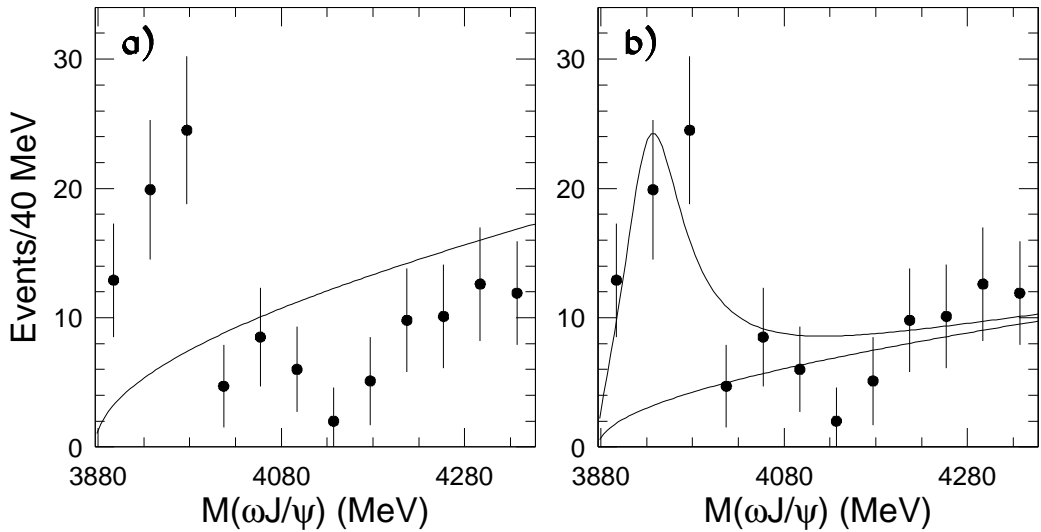}
\caption{Belle's $\chi_{c1}(2P)$ candidate~\cite{Abe:2004zs}:
The invariant mass $m(\omega J/\psi)$ distribution in
$m(B \to K\omega J/\psi)$ decay.
The signal yield is $58 \pm 11$ events. The two curves are fits
without (left) and including (right) a resonance component.
\label{fig:y3943}}
\end{center}
\end{figure}

The $Y(3940)$ was first seen by Belle in the $\omega J/\psi$ subsystem in the
decay $B\to K\omega(\to \pi^+\pi^-\pi^0) J/\psi$ \cite{Abe:2004zs}.  The final
state is selected by kinematic constraints that incorporate the parent particle
mass $m(B)$ and the fact that the $B$-meson pair is produced with no additional
particles.  Background from decays such as $K_1(1270) \to \omega K$
is reduced by requiring $m(\omega J/\psi) > 1.6\,\mbox{GeV}$.
The $K \omega J/\psi$ final state yield is then further examined
in bins of $m(\omega J/\psi)$. A threshold enhancement is observed,
shown in Figure~\ref{fig:y3943}, which is fit with a threshold function
suitable for phase-space production of this final state and an $S$-wave
Breit-Wigner shape. The reported mass and width of the enhancement are
$M=3943\pm 11\pm 13$~MeV and $\Gamma=87\pm 22 \pm 26$~MeV. A fit without
a resonance contribution gives no good description of the data.
BaBar confirmed the existence of the state~\cite{Aubert:2007vj}, also
for charged and neutral $B$~decays;
the values are $M=3914.6^{+3.8}_{-3.4} \pm 1.9$~MeV and
$\Gamma=33^{+12}_{-8} \pm 5$~MeV, somewhat different from Belle's.

The mass and width of $Y(3940)$ suggest a
radially excited $P$-wave charmonium state.  The combined branching ratio is
${\cal B}(B\to KY)\cdot {\cal B}(Y\to \omega J/\psi)= (7.1 \pm 1.3 \pm 3.1)
\times 10^{-5}$.  One expects that ${\cal B}(B\to K \chi_{cJ}(2P)) < {\cal B}
(B\to K \chi_{cJ}) = 4\times 10^{-4}$.  This implies that ${\cal B}(Y\to \omega
J/\psi) > 12\%$, which is unusual for a $c\bar{c}$ state above open charm
threshold.

For the $\chi_{c1}(2P)$
we expect $D\bar{D}^*$ to be the dominant decay mode with a predicted
width of 140~MeV \cite{Barnes:2006xq}, which is consistent with that of the
$Y(3940)$ within the theoretical and experimental uncertainties.  Furthermore,
the $\chi_{c1}$ is also seen in $B$-decays.  Although the decay $1^{++} \to
\omega J/\psi$ is unusual, the corresponding decay $\chi_{b1}(2P)\to \omega
\Upsilon (1S)$ has also been seen \cite{Severini:2003qw}.  One possible
explanation for this unusual decay mode is that rescattering through
$D\bar{D}^*$ is responsible: $1^{++} \to D\bar{D}^*\to \omega J/\psi$.
Another contributing factor might be mixing with the possible
molecular state tentatively identified with the $X(3872)$.

BaBar has searched for $Y(3940)$ decay into
$\gamma J/\psi$ \cite{Aubert:2006aj}, and set an upper limit ${\cal B}(B \to
Y(3940) + K) \times {\cal B}(Y(3940) \to \gamma J/\psi) < 1.4 \times 10^{-5}$.

The $\chi_{c1}(2P)$ assignment can be tested by searching for the $D\bar{D}$
and $D\bar{D}^*$ final states and by studying their angular distributions. With
the present experimental data, a $\chi_{c0}(2P)$ assignment
cannot be ruled out.


\subsubsection{Charmonium in $e^+e^- \to J/\psi + X$: $X(3940)$ and $X(4160)$}

Belle studied double-charmonium production and $e^+ e^- \to J/\psi + X$ near
the $\Upsilon(4S)$\ \cite{Abe:2005hd} and observed enhancements for the
well-known charmonium states $\eta_c$, $\chi_{c0}$, and $\eta_c(2S)$, at rates
and masses consistent with other determinations. In addition, a
peak at a higher energy was found. The mass and width were measured to
be $M=3936 \pm 14 \pm 6$~MeV and $\Gamma =39 \pm 26$ (stat) MeV.

\begin{figure}
\begin{center}
\includegraphics[width=0.57\textwidth]{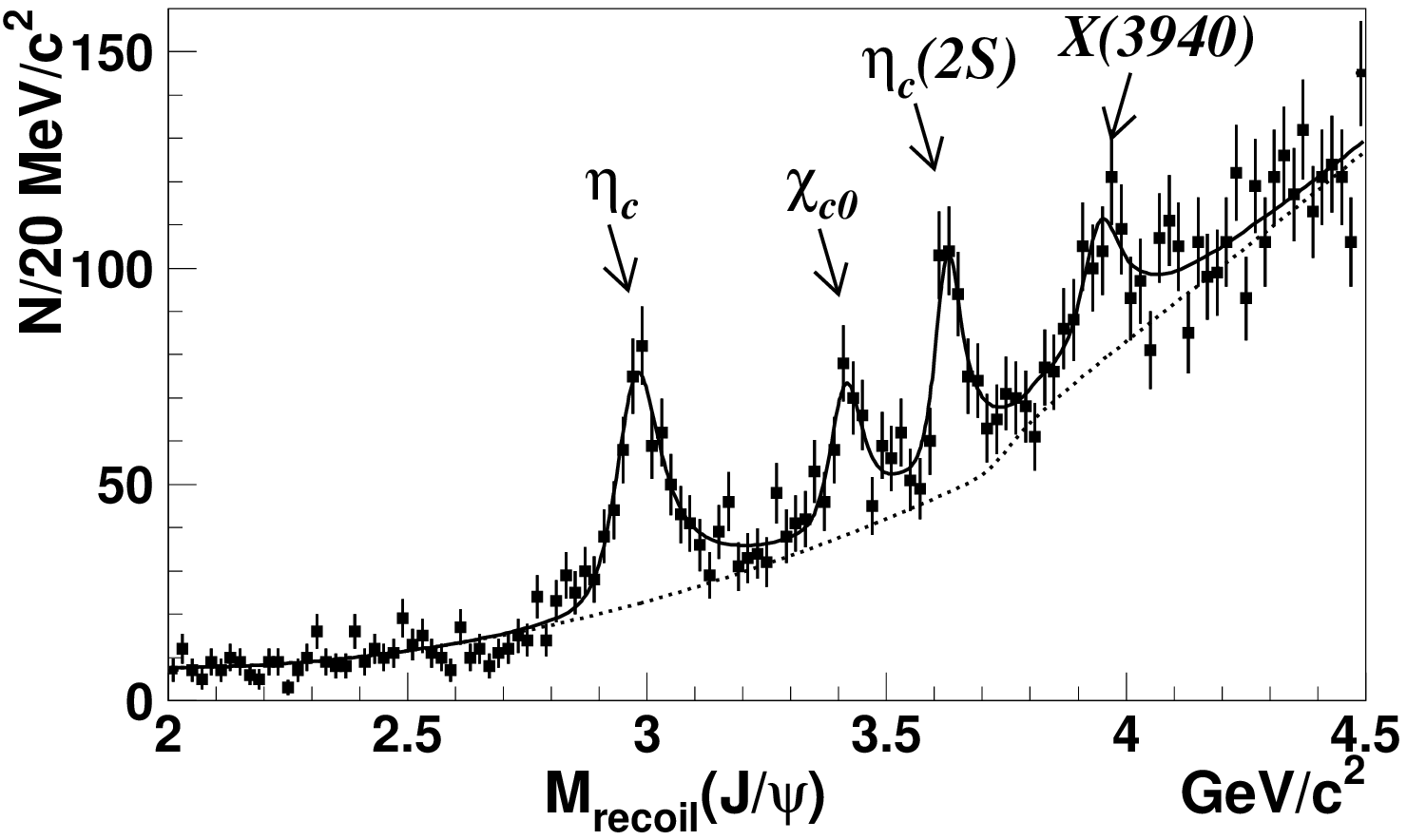}
\includegraphics[width=0.42\textwidth]{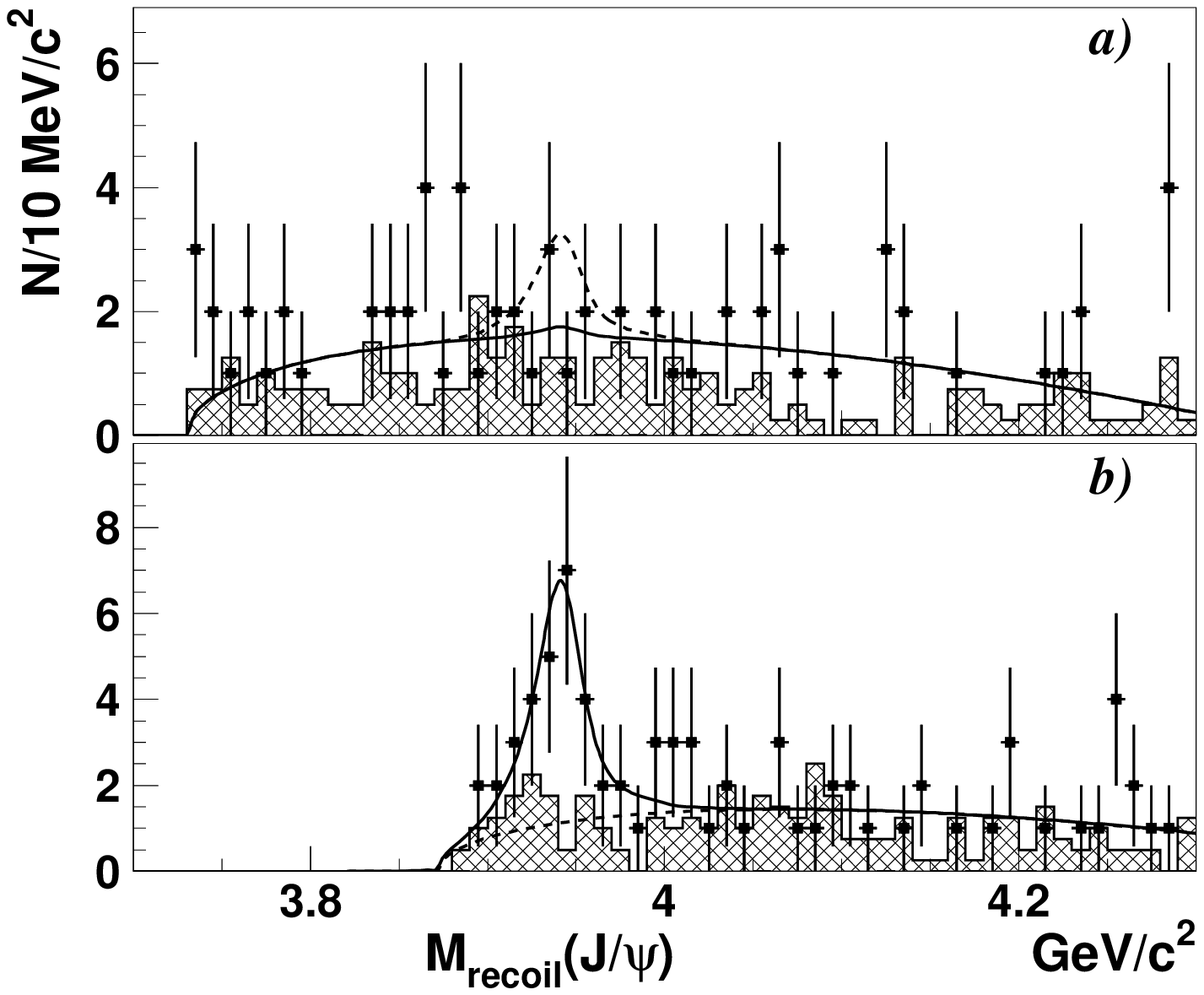}
\caption{Belle's $X(3940)$~\cite{Abe:2005hd}, sighted in
$e^+e^- \to J/\psi + X$:
Left: The mass of the system recoiling against the $J/\psi$.  The excess at
$X(3940)$ contains $266 \pm 63$ events and has a statistical significance of
$5.0\sigma$.  Right: Study of $X(3940)$ decay into $D$~mesons, $e^+e^- \to
J/\psi + D \bar D^{(*)}$.
Top: $D \bar D$, no signal is seen at $3940\,\mathrm{MeV}$.
Bottom: $D \bar D^*$, the signal amounts to $24.5 \pm 6.9$ events ($5.0\sigma$).
\label{fig:x3940}}
\end{center}
\end{figure}

To further examine the properties of this enhancement, Belle searched for
exclusive decays $J/\psi \to D \bar D^{(*)}$, given that these decays are
kinematically accessible. The $J/\psi$ recoil mass for the cases $D \bar D$ and
$D \bar D^{*}$ are also shown in Figure~\ref{fig:x3940}. An enhancement at the
$X(3940)$ mass is seen for $D \bar D^*$, but not for $D \bar D$. The mass and
width determined in this study are $M = (3943 \pm 6 \pm 6)\,\mbox{MeV}$,
$\Gamma < 52\,\mbox{MeV}$ (90\% c.l.).  Note that the inclusive and
exclusive samples have some overlap, and thus the two mass measurements are not
statistically independent.  The overlap has been eliminated for the branching
fraction determination.  A signal of $5.0\sigma$ significance was seen for
$D \bar D^*$, but none for $D \bar D$. In addition, the
$X(3940)$ did not show a signal for a decay $\omega J/\psi$,
unlike the $Y(3940)$. These findings are summarized in Table~\ref{tab:x3940}.

Belle updated their study with slightly higher luminosity
\cite{Abe:2007sy}. The study confirmed the observation in the exclusive
decay with comparable parameters of the $X(3940)$ but higher significance,
and added the following pieces of information: (1) There is no indication
of an $X(3940)$ signal in the invariant mass spectrum of $D \bar D$, but
there is a statistically significant population spread out over a wide range.
It is mandatory to understand this before it is possible to quantify
an upper limit on $X(3940) \to D \bar D$.
(2) In the final state $D^* \bar D^*$, a peak of $5.1\sigma$ statistical
significance is fit with a Breit-Wigner shape and claimed as a resonance
of mass
$M=(4156^{+25}_{-20} \pm 15) \,\mathrm{MeV}$,
and width $\Gamma = (139^{+111}_{-61} \pm 21) \,\mathrm{MeV}$,
distinct from the $X(3940)$ (preliminary).

If confirmed, the decay to $D\bar{D}^*$ but not $D\bar{D}$ suggests the
$X(3940)$ has unnatural parity.  The lower mass $\eta_c(1S)$ and $\eta_c(2S)$
are also produced in double charm production.  One is therefore led to try an
$\eta_c(3S)$ assignment, although this state is expected to have a somewhat
higher mass~\cite{Barnes:2005pb}.  The predicted width for a $3^1S_0$ state
with a mass of 3943~MeV is $\sim 50$~MeV~\cite{Eichten:2005ga}, which is in not
too bad agreement with the measured $X(3940)$ width.

\begin{table}
\caption{Properties of the $X(3940)$~\cite{Abe:2005hd}.
\label{tab:x3940}}
\begin{center}
\begin{tabular}{c c } \hline \hline
Mass                     & $3936 \pm 14 \pm 6 \,\mbox{MeV}$ (incl.)\\
& $3943 \pm  6 \pm 6 \,\mbox{MeV}$ ($D \bar D^*$)\\
Total width                    & $< 52\,\mbox{MeV}$ \\
\hline
${\cal B}(X(3940)\to D\bar{D}^*)$    &  $(96^{+45}_{-32} \pm 22)\%$, \\
&  $>45\%$ (90\% C.L.) \\
${\cal B}(X(3940)\to D\bar{D})$      &  $ <  41\%$ (90\% CL)     \\
${\cal B}(X(3940)\to \omega J/\psi)$ &  $ <  26\%$ (90\% CL)     \\
\hline\hline
\end{tabular}
\end{center}
\end{table}

Another possibility due to the dominant $D\bar{D}^*$ final states is that the
$X(3940)$ is the $2^3P_1(c\bar{c}) \; \chi_1(2P)$ state.
It is natural to consider
the $2P(c\bar{c})$ since the $2^3P_J$ states are predicted to lie in the
3920--3980~MeV mass region and the widths are predicted to be in the range
$\Gamma(2^3P_J)=30$--165 MeV \cite{Barnes:2005pb}.  The dominant $D\bar{D}^*$
mode would then suggest that the $X(3940)$ is the $2^3P_1(c\bar{c})$ state.
The problems with this interpretation are
(1) there is no evidence for the $1^3P_1(c\bar{c})$ state in the same data,
(2) the predicted width of the $2^3P_1(c\bar{c})$ is 140~MeV
(assuming $m(2^3P_1(c\bar{c}))=3943$~MeV) \cite{Barnes:2006xq}, and
(3) there is another candidate for the $2^3P_1(c\bar{c})$ state, the $Y(3940)$.

The most likely interpretation of the $X(3940)$ is that it is the
$3^1S_0(c\bar{c})$ $\eta_c(3S)$ state.  Tests of this assignment are to study
the angular distribution of the $D\bar{D}^*$ final state and to
observe it in $\gamma\gamma \to D\bar{D}^*$.


\subsubsection{$\pi^+\pi^- J/\psi$ in Initial State Radiation: $Y(4260)$ and
$X(4008)$}

\begin{figure}
\begin{center}
\includegraphics[width=0.8\textwidth, bb = 50 0 530 451]{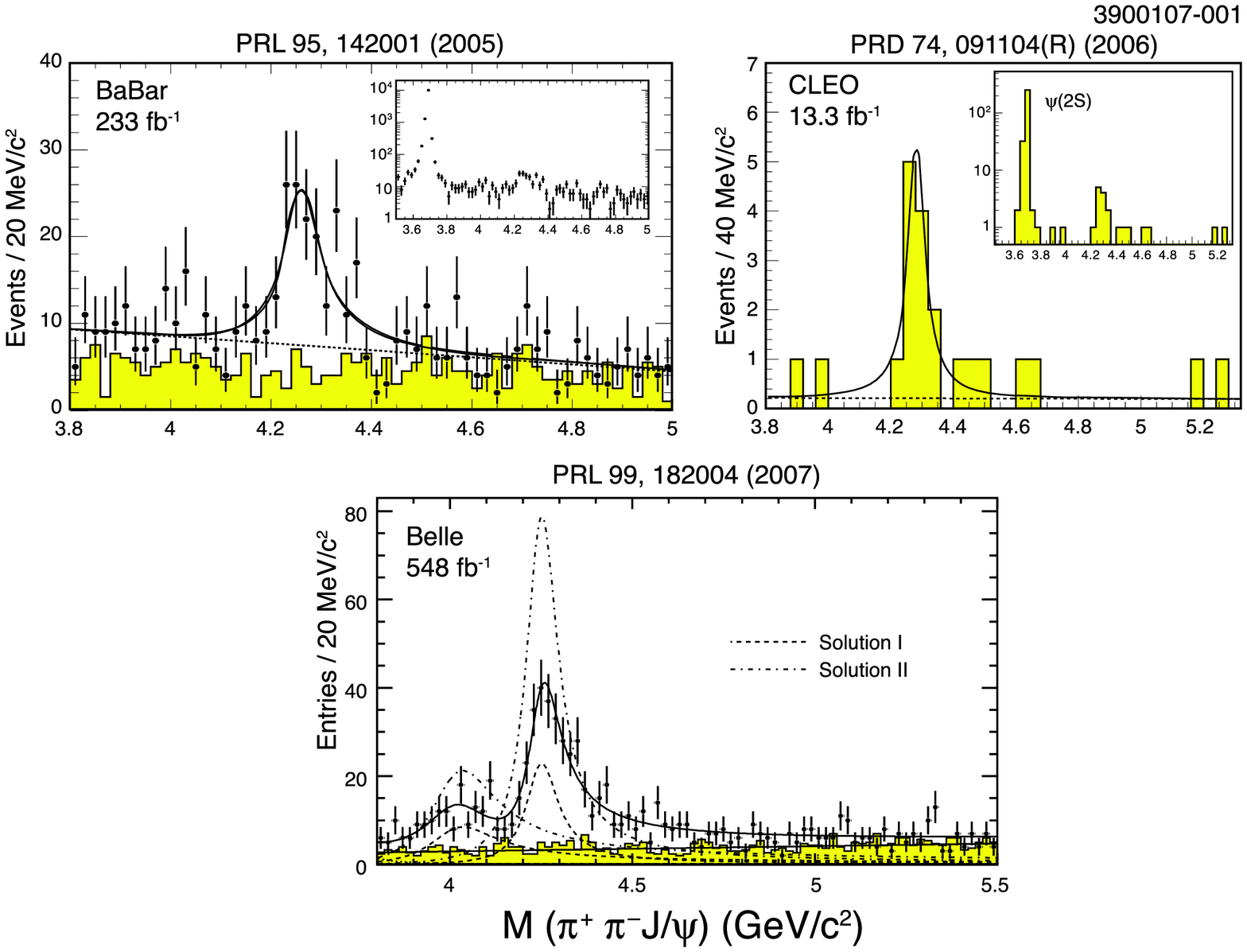}
\caption{$Y(4260)$ signal in ISR from the $\Upsilon(4S)$
by BaBar~\cite{Aubert:2005rm}, CLEO~\cite{He:2006kg},
and Belle~\cite{Yuan:2007sj}.
The fit parameters are given in Table~\ref{tab:Y4260}.
\label{fig:y4260isr}}
\end{center}
\end{figure}

Perhaps the most intriguing of the recently discovered states is the $Y(4260)$
reported by BaBar as an enhancement in the $\pi\pi J/\psi$ subsystem in
the radiative return reaction $e^+e^-\to \gamma_{\rm ISR} J/\psi\pi\pi$
\cite{Aubert:2005rm}, where ``ISR'' stands for ``initial state radiation.''
This and subsequent independent confirmation signals
\cite{He:2006kg,Yuan:2007sj}
are shown in Fig.~\ref{fig:y4260isr}.
The measured mass, width, and leptonic width times ${\cal B}(Y\to J/\psi
\pi^+\pi^-)$ are summarized in the first row of Table~\ref{tab:Y4260}.
Further evidence was seen by BaBar in $B\to K (\pi^+\pi^- J/\psi)$
\cite{Aubert:2005zh}.

\begin{figure}[h]
\begin{center}
\includegraphics[height=0.6\textwidth]{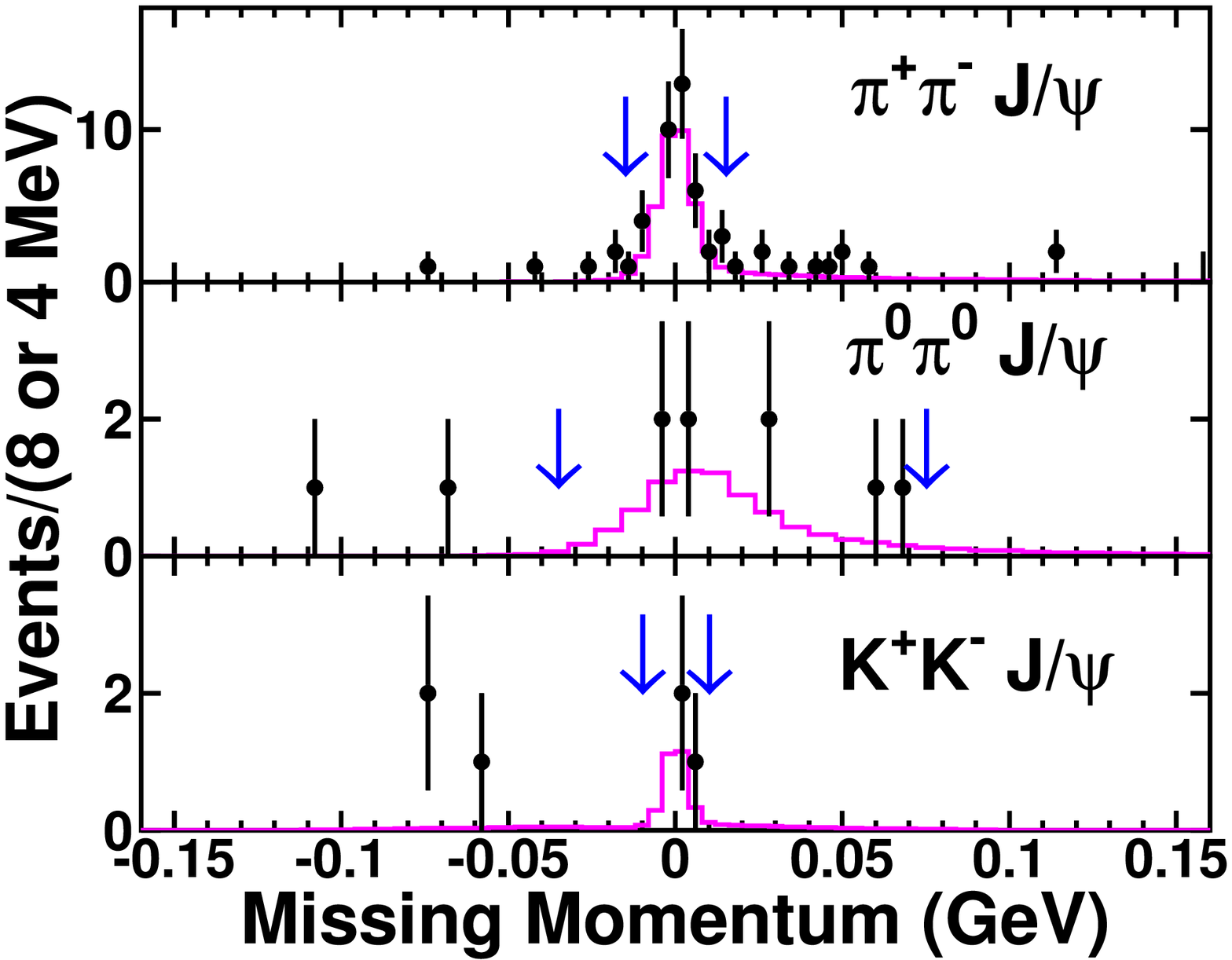}
\caption{Evidence for $Y(4260)$ from a direct scan by CLEO \cite{Coan:2006rv}.
\label{fig:cleo4260}}
\end{center}
\end{figure}

The CLEO Collaboration has confirmed the $Y(4260)$, both in a direct scan
\cite{Coan:2006rv} and in radiative return \cite{He:2006kg}.  Results from the
scan are shown in Fig.\ \ref{fig:cleo4260}, including
cross-section increases at $E_{\rm cm} = 4260$ MeV consistent with
$Y(4260) \to \pi^+ \pi^- J/\psi$ (11$\sigma$),
$\pi^0 \pi^0 J/\psi$ (5.1$\sigma$), and
$K^+ K^- J/\psi$ (3.7$\sigma$).  There are also weak signals for $\psi(4160)
\to \pi^+ \pi^- J/\psi$ (3.6$\sigma$) and $\pi^0 \pi^0 J/\psi$ (2.6$\sigma$),
consistent with the $Y(4260)$ tail, and for $\psi(4040) \to \pi^+ \pi^- J/\psi$
(3.3$\sigma$).  \citet{He:2006kg} determines the resonance parameters
shown in the second row of Table~\ref{tab:Y4260}.

Belle~\cite{Yuan:2007sj}, also in ISR events, fit the $\pi^+\pi^- J/\psi$
enhancement with an additional component, two coherent Breit-Wigner functions
in total, in order to achieve a better description of the low-side tail of the
$Y(4260)$.  The fit results in mass and width of $M = (4008 \pm
40^{+72}_{-28})$ MeV and $\Gamma = (226 \pm 44^{+87}_{-79})$ MeV
for the lower resonance. The values for the upper (the $Y(4260)$) are listed
in Table~\ref{tab:Y4260}.
Interference leads to a two-fold ambiguity in the rate, corresponding to
constructive and destructive interference.  Both solutions arrive at the same
fit function.  The two solutions differ markedly. The lower-lying state is not
associated with any presently known charmonium state.

The invariant mass distribution $m(\pi^+\pi^-)$ looks quite different
for events at $\sim 4260$ MeV than above and below; the distribution is shifted
towards higher values, not consistent with
phase space~\cite{Yuan:2007sj}.

\begin{table}
\caption{Comparison of parameters of $Y(4260)$ as measured by the
BaBar \cite{Aubert:2005rm}, CLEO \cite{He:2006kg},
and Belle \cite{Yuan:2007sj} Collaborations.
\label{tab:Y4260}}
\begin{center}
\begin{tabular}{c c c c} \hline \hline
Collab.\ &    Mass     &  $\Gamma$   & $\Gamma_{ee} \times {\cal B}(Y(4260)
\to \pi^+ \pi^- J/\psi)$ \\
& (MeV/$c^2$) & (MeV/$c^2$) & (eV) \\ \hline
\rule[0mm]{0mm}{4.2mm}
BaBar   & $4259\pm 8^{+2}_{-6} $
& $ 88 \pm 23^{+6}_{-4}$
& $5.5\pm 1.0^{+0.8}_{-0.7}$ \\
CLEO    & $4284^{+17}_{-16}\pm4$
& $ 73^{+39}_{-25}\pm5$
& $8.9^{+3.9}_{-3.1}\pm1.8$  \\
Belle   & \multicolumn{3}{l}{\qquad\qquad Two-resonance fit:} \\
& $4247\pm12^{+17}_{-32}~~$
& $~~108\pm 19 \pm 10~~$
& $6.0\pm1.2^{+4.7}_{-0.5}$ or
$20.6\pm 2.3 ^{+9.1}_{-1.7}$  \\
& \multicolumn{3}{l}{\qquad\qquad Single-resonance fit:} \\
& $4263 \pm 6$
& $126 \pm 18$
& $9.1 \pm 1.1$ \\
\hline \hline
\end{tabular}
\end{center}
\end{table}

A variety of ratios between channels have been measured now
\cite{Coan:2006rv,Heltsley:2006QWG,Aubert:2005cb,Gowdy:2006be,Ye:2006QWG},
which should
help narrow down the possible explanations of $Y(4260)$.  They are listed in
Table~\ref{tab:4260ratios}.  The preliminary upper limit for the
ratio of $D \bar D$ to $\pi^+\pi^-J/\psi$ of 7.6 may not seem particularly
tight at first glance, but is to be compared, for example, with the same ratio
for the $\psi(3770)$, where it is about 500.

{\small
\begin{table}[t]
\caption{Experimental results on $Y(4260)$ decay.  
The last column gives the relative rate compared 
to $\pi^+\pi^-J/\psi$ for each channel. 
Data are from \citet{Coan:2006rv} and
\citet{Heltsley:2006QWG}, except 
(a)~\citet{Aubert:2005cb},
(b)~\citet{Gowdy:2006be},
and
(c)~\citet{Ye:2006QWG}.
Unless indicated otherwise, upper limits are at 90\% CL.
\label{tab:4260ratios}}
\begin{center}
\begin{tabular}{|c|c|c|} \hline \hline
Channel & cross-section (pb) & ${\cal B}/{\cal B}_{\pi^+\pi^-J/\psi}$ \\
\hline
$\pi^+\pi^- J/\psi$         & $58^{+12}_{-10}\pm 4$ &
$1$ \\
& $51\pm 12$~(a)&
$1$ \\
$\pi^0\pi^0 J/\psi$         & $23^{+12}_{-8}\pm 1$  &
$0.39^{+0.20}_{-0.15}\pm 0.02$ \\
$K^+K^- J/\psi$             & $ 9^{+9}_{-5}\pm 1$   &
$0.15^{+0.10}_{-0.08}\pm 0.02$ \\
$\eta J/\psi$               & $< 32$ &
$<0.6$ \\
$\pi^0 J/\psi$              & $< 32$ &
$<0.2$ \\
$\eta' J/\psi$              & $< 19$ &
$<0.3$ \\
$\pi^+\pi^-\pi^0 J/\psi$    & $<  7$ &
$<0.1$ \\
$\eta \eta J/\psi$          & $< 44$ &
$<0.8$ \\
$\pi^+\pi^- \psi(2S)$       & $< 20$ &
$<0.3$ \\
$\eta \psi(2S)$             & $< 25$ &
$<0.4$ \\
$\omega \chi_{c0}$          & $<234$ &
$<4.0$ \\
$\gamma \chi_{c1}$          & $< 30$ &
$<0.5$ \\
$\gamma \chi_{c2}$          & $< 90$ &
$<1.6$ \\
$\pi^+\pi^-\pi^0 \chi_{c1}$ & $< 46$ &
$<0.8$ \\
$\pi^+\pi^-\pi^0 \chi_{c2}$ & $< 96$ &
$<1.7$ \\
$\pi^+\pi^-\phi$            & $<  5$ &
$<0.1$ (also see (b))\\
$D \bar D$            & & $<7.6$ (95\%CL)~(c)  \\
$p \bar p$            & & $<0.13$~(a)  \\
\hline \hline
\end{tabular}
\end{center}
\end{table}
}

A number of explanations have appeared in the literature: $\psi(4S)$
\cite{LlanesEstrada:2005hz}, $c s \bar c \bar s$ tetraquark
\cite{Maiani:2005pe}, and
$c\bar{c}$ hybrid \cite{Zhu:2005hp,Close:2005iz,Kou:2005gt}.
In some models the mass of the $Y(4260)$ is consistent with the $4S
(c\bar{c})$ level \cite{LlanesEstrada:2005hz}. Indeed, a $4S$ charmonium level
at 4260 MeV/$c^2$ was anticipated on exactly this basis \cite{Quigg:1977dd}.
With this assignment, the $nS$ levels of charmonium and bottomonium are
remarkably congruent to one another.  However, other calculations using a
linear plus Coulomb potential identify the $4^3S_1 (c\bar{c})$ level with the
$\psi(4415)$ state (e.g., \citet{Barnes:2005pb}).  If this is the case
the first unaccounted-for $1^{--}(c\bar{c})$ state is the $\psi(3^3 D_1)$.
Quark models estimate its mass to be $m(3^3 D_1)\simeq 4500$~MeV which is much
too heavy to be the $Y(4260)$.  The $Y(4260)$ therefore represents an
overpopulation of the expected $1^{--}$ states.  The absence of open charm
production also argues against it being a conventional $c\bar{c}$ state.

The hybrid interpretation of $Y(4260)$ is appealing.  The flux
tube model predicts that the lowest $c\bar{c}$ hybrid mass is $\sim 4200$~MeV
\cite{Barnes:1995hc} with lattice gauge theory having similar expectations
\cite{Lacock:1996ny}.  Models of hybrids typically expect the wavefunction at
the origin to vanish implying a small $e^+e^-$ width in agreement with the
observed value.  Lattice gauge theory found that the $b\bar{b}$ hybrids have
large couplings to closed flavor
channels \cite{McNeile:2002az} which is similar
to the BaBar observation of $Y\to J/\psi \pi^+\pi^-$; the branching ratio of
${\cal B}(Y\to J/\psi \pi^+\pi^-) > 8.8\%$ combined with the observed
width implies that $\Gamma (Y\to J/\psi \pi^+\pi^-) > 7.7\pm 2.1 $~MeV.
This is much larger than the typical charmonium transition widths of, for
example, $\Gamma(\psi(3770)\to J/\psi \pi^+\pi^-)\sim 80$~keV.  And
the $Y$ is seen in this mode while the conventional states
$\psi(4040)$, $\psi(4160)$, and $\psi(4415)$ are not.

One predicted consequence of the hybrid hypothesis is that the dominant hybrid
charmonium open-charm decay modes are expected to be a meson pair with an
$S$-wave ($D$, $D^*$, $D_s$, $D_s^*$) and a $P$-wave ($D_J$, $D_{sJ}$) in the
final state~\cite{Close:2005iz}.  The dominant decay mode is expected to be
$D \bar D_1 + \mathrm{c.c.}$.  (Subsequently we shall omit
``$+ \mathrm{c.c.}$'' in cases where it is to be understood.)
Evidence for a large $D \bar D_1$ signal would be strong evidence for the
hybrid interpretation.  A complication is that $D \bar D_1$ threshold is 4287
MeV/$c^2$ if we consider the lightest $D_1$ to be the narrow state noted in
\citet{Yao:2006px} at 2422 MeV/$c^2$.  The possibility also exists that
the $Y(4260)$ could be a $D \bar D_1$ {\it bound state}. It would decay
to $D \pi \bar D^*$, where the $D$ and $\pi$ are not in a $D^*$. Note that the
dip in $R_{e^+ e^-}$ occurs just below $D \bar D_1$ threshold,
which may be the first $S$-wave meson pair accessible in $c \bar c$
fragmentation \cite{Close:2005iz,Rosner:2006vc}.  In addition to the hybrid
decay modes given above, lattice gauge theory suggests that we search for other
closed charm modes with
$J^{PC}=1^{--}$: $J/\psi \eta$, $J/\psi \eta'$, $\chi_{cJ} \omega$ and more.
Distinguishing among the interpretations of the $Y(4260)$ will likely
require careful measurement of several decay modes.

If the $Y(4260)$ is a hybrid it is expected to be a member of a multiplet
consisting of eight states with masses in the 4.0 to 4.5~GeV mass range with
lattice gauge theory preferring the higher end of the range \cite{Liao:2002rj}.
It would be most convincing if some of these partners were found, especially
the $J^{PC}$ exotics.  In the flux-tube model the exotic states have
$J^{PC}=0^{+-}$, $1^{-+}$, and $2^{+-}$ while the non-exotic low-lying hybrids
have $0^{-+}$, $1^{+-}$, $2^{-+}$, $1^{++}$, and $1^{--}$.


\subsubsection{States decaying to $\pi^+ \pi^- \psi(2S)$}
\label{sec:pipipsi(2S)}

In the radiative return process $e^+ e^- \to \gamma + X$, the BaBar
Collaboration \cite{Aubert:2006ge} reports a broad structure
decaying to $\pi^+ \pi^- \psi(2S)$, where
$\psi(2S) \to \pi^+ \pi^- J/\psi$.
A single-resonance hypothesis with
$m(X)=(4324 \pm 24\,\mathrm{(stat)})$~MeV and
$\Gamma(X) = (172 \pm 33 \,\mathrm{(stat)})$~MeV
is adequate to fit the observed mass spectrum.

Belle, with more than twice the sample size used in the BaBar analysis,
observed two enhancements in the same reaction~\cite{Wang:2007ea}:
One that confirms BaBar's measurement, at
$M=(4361\pm 9\pm 9)$ MeV with a width $\Gamma= (74\pm 15\pm 10)$ MeV
(statistical signficance $8\sigma$),
and a second, $M=(4664\pm 11\pm 5)$ MeV with a width of
$\Gamma=(48\pm 15\pm 3)$ MeV ($5.8\sigma$). The existence of the
higher-energy peak is not excluded in the BaBar data.

Given the uncertainty in the masses and widths of the lower state
decaying into $\pi^+\pi^- \psi(2S)$ and the $Y(4260)$, the possibility
that they are different manifestations of the same state cannot be
excluded.

\section{Bottomonium}
\label{sec:bottomonium}

\subsection{Overview}

Some properties and decays of the $\Upsilon$ ($b \bar b$) levels are
summarized in Fig.~\ref{fig:ups}.
The measured masses of the $\Upsilon$ states below
open flavor threshold have accuracies comparable to those in charmonium
since similar techniques are used. Experimentally, the situation is
more difficult due to the larger multiplicities involved and due to
the increased continuum background compared to the charmonium region.

Modern data samples are CLEO's 22M, 9M, 6M $\Upsilon(1,2,3S)$ decays
(with smaller off-resonance samples in addition) and Belle's
$\Upsilon(3S)$ sample of 11M  $\Upsilon(3S)$.

\begin{figure}
\begin{center}
\includegraphics[width=0.48\textwidth]{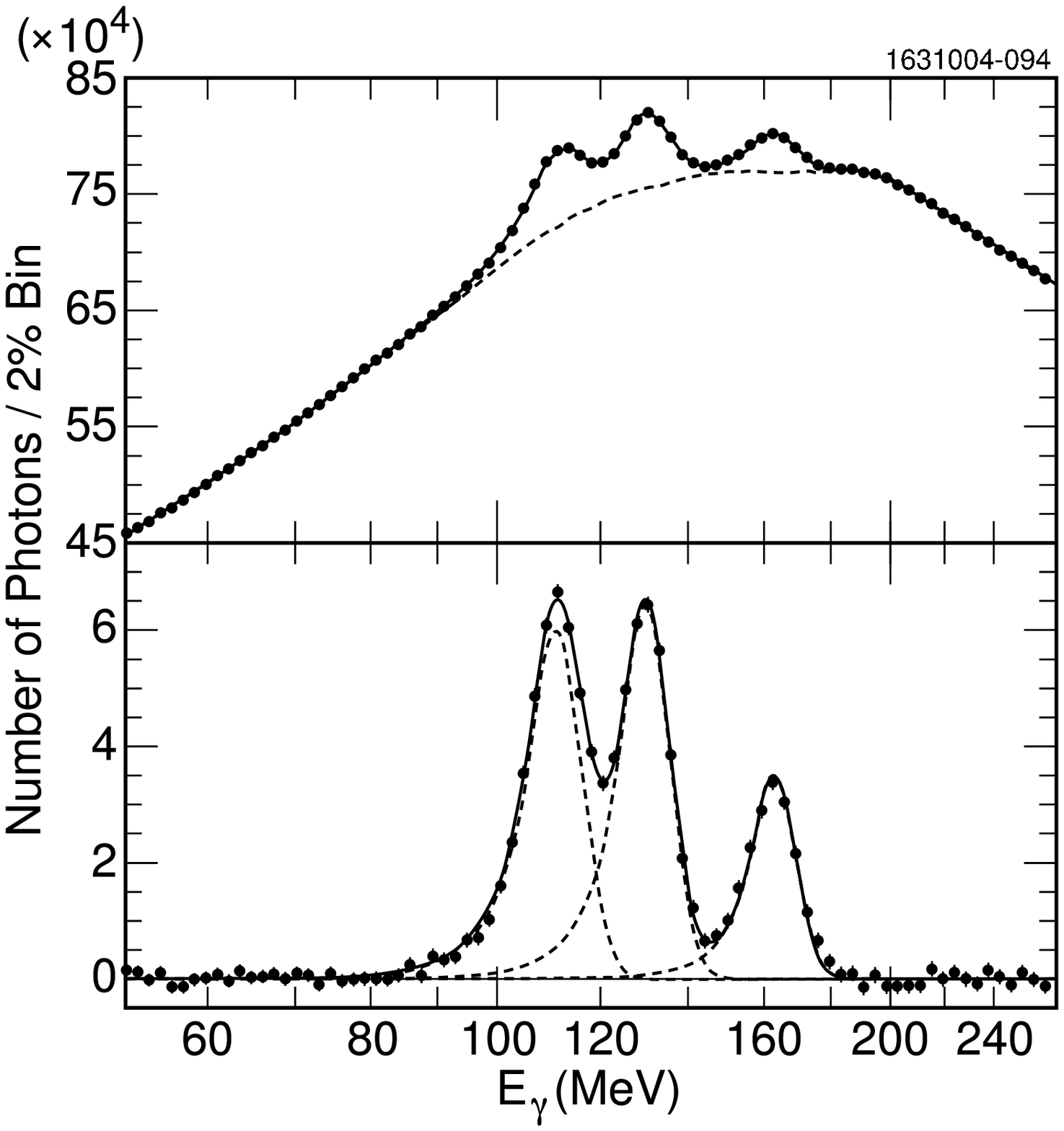}
\includegraphics[width=0.48\textwidth]{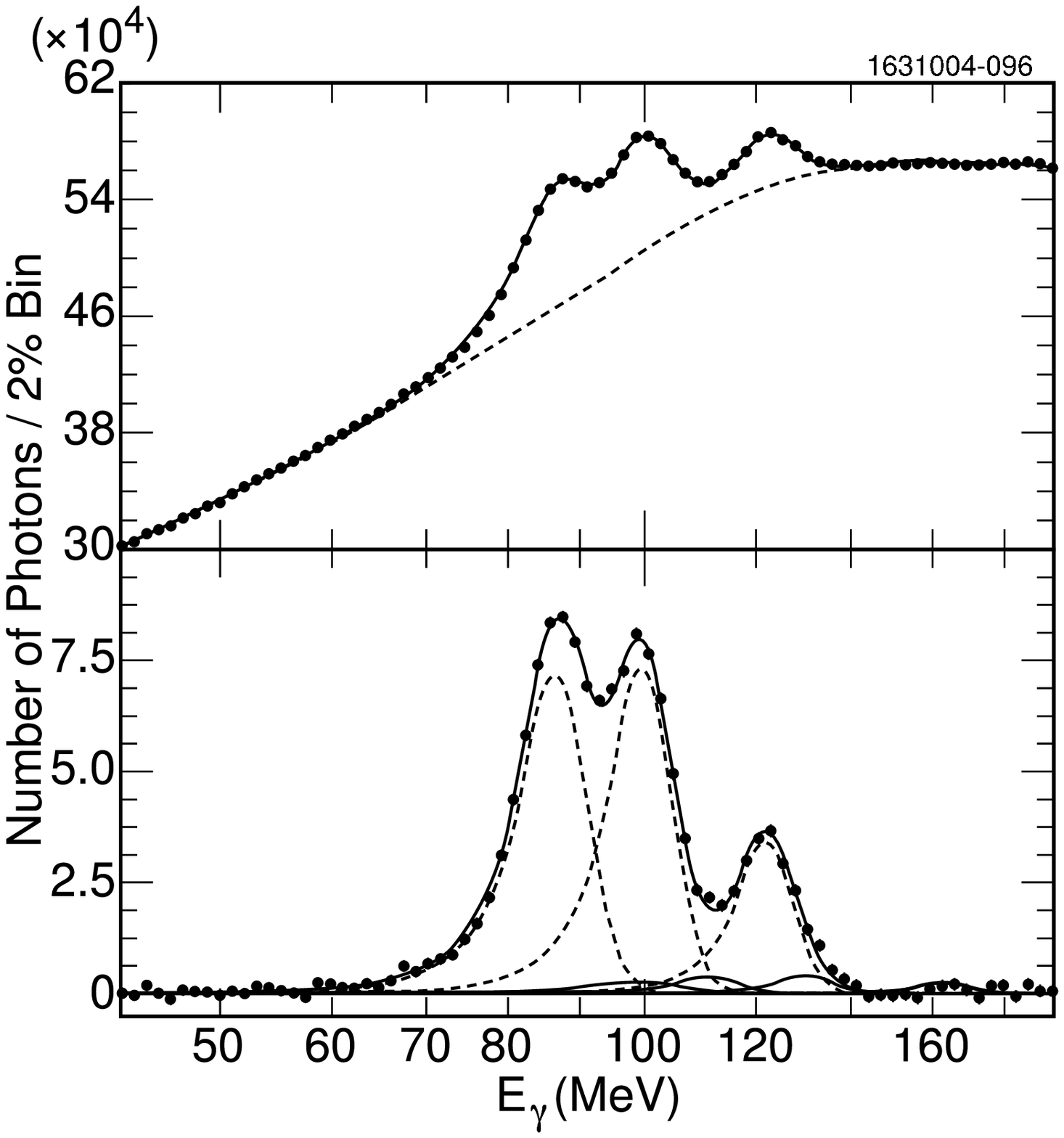}
\caption{From \citet{Artuso:2004fp}:
Inclusive photon spectrum in $\Upsilon(nS) \to \gamma X$,
for $n=2$ (left) and $n=3$ (right), before and after background
subtraction. In the upper plots, the dashed line indicates
the background level; in the lower plots, the fit contribution
for each resonance is delineated. The low-lying solid curves
in the lower right plot show two background contributions.
The three peaks corresponding to the $\chi_{bJ}(1,2P)$
are clearly visible. The peak position determines the
$\chi_{bJ}(1,2P)$ masses. The signal area is used to determine the
$\Upsilon(nS) \to \gamma \chi_{bJ}(1,2P)$ branching fraction.
\label{fig:chibJ}}
\end{center}
\end{figure}

The $\chi_{bJ}(1,2P)$ states are reached through E1 transitions;
branching fractions for $n \to n-1$ range from 4-14\%.
Their masses are determined from the transition photon energies.
Their intrinsic widths are not known. Examples of fits to the
inclusive photon spectrum that led to $\chi_{bJ}(1,2P)$ mass
determinations \cite{Artuso:2004fp} are shown in Fig.~\ref{fig:chibJ}.
Exclusive hadronic decays of the $\chi_{bJ}(1,2P)$ states
have not been reported;
information exists only on transitions within the bottomonium spectrum.
An $\Upsilon(1D)$ candidate has been observed; the singlets
$\eta_b(1,2S)$ and $h_b(1,2P)$ have thus far escaped detection.

Mass differences within the bottomonium spectrum are in agreement
with unquenched lattice QCD calculations~\cite{Lepage:2005eg}.  Direct photons
have been observed in 1S, 2S, and 3S decays, leading to estimates
of the strong fine-structure constant~$\alpha_S$
consistent with others \cite{Besson:2005jv}.  The transitions
$\chi_b(2P) \to \pi \pi \chi_b(1P)$ have been seen \cite{Cawlfield:2005ra}.
BaBar has observed $\Upsilon(4S) \to \pi^+ \pi^- \Upsilon(1S,2S)$ transitions
\cite{Aubert:2006bm}, while Belle has seen $\Upsilon(4S) \to \pi^+ \pi^- \Upsilon(1S)$
\cite{Sokolov:2006sd}.

Decays to light hadrons proceed, as in the case of the charmonium
states, via annihilation of the heavy quarks into $ggg$, $gg\gamma$
or $\gamma^*$, which subsequently hadronize.
At higher energies, fragmentation into low-multiplicity states
is suppressed, and so the second step makes it difficult to
arrive at a simple scaling prediction to translate bottomonium
and charmonium results into each other. Comparing the $\Upsilon$
states with each other, for example by constructing a prescription
akin to the 12\%~rule in charmonium, is possible, but to date only
a few exclusive radiative decays to light mesons, but no exclusive
non-radiative decays to light mesons, have been observed.

\subsection{$\Upsilon(1S,2S,3S)$}

\subsubsection{Masses and total widths}

The best measurements of the narrow $\Upsilon(nS)$ states,
as was the case for the $J^{PC}=1^{--}$ states in charmonium,
come from fits to the cross-section $\Upsilon(nS) \to \mbox{hadrons}$
around the resonance together with a very precise beam energy
calibration using resonant depolarization.
This leads to precision mass determinations with uncertainties
of order 100~keV.

The $\Upsilon(1S)$ mass measurements from
CUSB ~\cite{MacKay:1984kv} and MD-1~\cite{Artamonov:2000cz}
each have a relative precision of one part in $10^5$,
but are about 0.5~MeV apart.
The $\Upsilon(2S)$ determinations by MD-1~\cite{Artamonov:2000cz} and
DORIS experiments~\cite{Barber:1983im} agree well.
There is only one measurement of $m(\Upsilon(3S))$, again
by MD-1~\cite{Artamonov:2000cz}.

The below-flavor $\Upsilon(nS)$ states are narrow, some ten keV,
whereas the $\Upsilon(4S)$, for which the decay to $B \bar B$ is
kinematically possible, has a full width three orders of magnitude higher.
The intrinsic widths of the $\Upsilon(1,2,3S)$ cannot be
determined directly in $e^+e^-$ collisions as they lie
well below the beam energy spread.
They can be determined indirectly, by using the relation
\begin{equation}
\Gamma = \frac{\Gamma_{\ell\ell} }{  {\cal B}_{\ell\ell} }
= \frac{ \Gamma_{ee} }{ {\cal B}_{\mu\mu} },
\end{equation}
where the last step assumes lepton universality.
Expanding by the hadronic partial width,
$\Gamma_{\mathrm{had}} = (1 - 3 {\cal B}_{\mu\mu})/\Gamma$,
the equation reads:
\begin{equation}
\Gamma =  \frac{ \Gamma_{ee} \Gamma_{\mathrm{had}} / \Gamma }{
{\cal B}_{\mu\mu} (1 - 3 {\cal B}_{\mu\mu} ) }.
\end{equation}

The expression in the numerator is directly accessible in the
reaction $e^+ e^- \to \Upsilon(nS) \to \mathrm{hadrons}$; the
integral of the hadronic cross-section over the resonance is
proportional to the product of widths. The muonic branching
fraction can be determined from a measurement of
$\xi = \Gamma_{\mu\mu} / \Gamma_{\mathrm{hadrons}}$, which is
independent of the total width; ${\cal B}_{\mu\mu} =
\Gamma_{\mu\mu}/\Gamma
= \Gamma_{\mu\mu}/(\Gamma_{\mathrm{had}} + 3\Gamma_{\mu\mu})
= \xi / ( 1+3\xi)$.  The current status of experimental precision
is below 2\% for $ \Gamma_{ee} \Gamma_{\mathrm{had}} / \Gamma$ and
3-4\% for ${\cal B}_{\mu\mu}$. The corresponding measurements
are discussed in Section~\ref{sec:ups_dilep}.

\subsubsection{Leptonic branching ratios and partial widths}
\label{sec:ups_dilep}

New values of ${\cal B}[\Upsilon(1S,2S,3S) \to \mu^+ \mu^-] = (2.49 \pm 0.02
\pm 0.07,~2.03\pm0.03\pm0.08,~2.39\pm0.07\pm0.10)\%$ \cite{Adams:2004xa},
when combined with new measurements
$\Gamma_{ee}(1S,2S,3S) = (1.354\pm0.004 \pm0.020,
~0.619\pm0.004\pm0.010,
~0.446\pm0.004\pm0.007)$ keV \cite{Rosner:2005eu},
imply total widths 
$\Gamma_{\rm tot}(1S,2S,3S)=
(54.4\pm 1.8, 
~30.5\pm 1.4, 
~18.6\pm 1.0)$ keV. 
The values of $\Gamma_{\rm tot}(2S,3S)$ changed considerably with respect
to previous world averages.  Combining with previous data, the Particle Data
Group \cite{Yao:2006px} now quotes $\Gamma_{\rm tot}(1S,2S,3S)=(54.02\pm1.25,%
~31.98\pm2.63,~20.32\pm1.85)$ keV, which we shall use in what follows.  This
will lead to changes in comparisons of predicted and observed transition rates.
As one example, the study of $\Upsilon(2S,3S) \to \gamma X$ decays
\cite{Artuso:2004fp} has provided new branching ratios for E1 transitions
to $\chi_{bJ}(1P),~\chi_{bJ}(2P)$ states.  These may be combined with the
new total widths to obtain updated partial decay widths [line (a) in
Table~\ref{tab:E1}], which may be compared with one set of nonrelativistic
predictions \cite{Kwong:1988ae} [line (b)].  The suppression of transitions to $J=0$
states by 10--20\% with respect to nonrelativistic expectations agrees
with relativistic predictions \cite{Moxhay:1983vu,McClary:1983xw,Skwarnicki:2005pq}.
The partial width for $\Upsilon(3S) \to \gamma 1^3P_0$ is found to be $61 \pm
23$ eV, about nine times the highly-suppressed value predicted in 
\citet{Kwong:1988ae}.  
That prediction is very sensitive to details of wave functions;
the discrepancy indicates the importance of relativistic distortions.

\begin{table}
\caption{Comparison of observed (a) and predicted (b) partial widths
for $2S \to 1 P_J$ and $3S \to 2 P_J$ transitions in $b \bar b$ systems.
\label{tab:E1}}
\begin{center}
\begin{tabular}{|c|c c c|c c c|} \hline \hline
& \multicolumn{3}{c|}{$\Gamma$ (keV), $2S \to 1P_J$ transitions}
& \multicolumn{3}{c|}{$\Gamma$ (keV), $3S \to 2P_J$ transitions} \\
& $J=0$ & $J=1$ & $J=2$ & $J=0$ & $J=1$ & $J=2$ \\ \hline
(a) & 1.20$\pm$0.18 & 2.22$\pm$0.23 & 2.32$\pm$0.23 &
1.38$\pm$0.19 & 2.95$\pm$0.30 & 3.21$\pm$0.33 \\
(b)& 1.39 & 2.18 & 2.14 & 1.65 & 2.52 & 2.78 \\ \hline \hline
\end{tabular}
\end{center}
\end{table}

The branching ratios ${\cal B}[\Upsilon(1S,2S,3S) \to \tau^+ \tau^-]$ have
been measured by the CLEO Collaboration \cite{Besson:2006gj}, and are shown in
Table \ref{tab:tautau}.  They are consistent with lepton universality and
represent the first measurement of the $\Upsilon(3S) \to \tau \tau$ branching
ratio.

\begin{table}
\caption{Ratio $R_{\tau \tau} \equiv {\cal B}[\Upsilon(nS) \to \tau \tau]/
{\cal B}[\Upsilon(nS) \to \mu \mu]$ and ${\cal B}[\Upsilon(nS) \to \tau \tau]$
\cite{Besson:2006gj}.
\label{tab:tautau}}
\begin{center}
\begin{tabular}{c c c} \hline \hline
& $R_{\tau \tau}$ & ${\cal B}[\Upsilon(nS) \to \tau \tau]~(\%)$ \\ \hline
$\Upsilon(1S)$ & $1.02\pm0.02\pm0.05$ & $2.54\pm0.04\pm0.12$ \\
$\Upsilon(2S)$ & $1.04\pm0.04\pm0.05$ & $2.11\pm0.07\pm0.13$ \\
$\Upsilon(3S)$ & $1.05\pm0.08\pm0.05$ & $2.52\pm0.19\pm0.15$ \\ \hline \hline
\end{tabular}
\end{center}
\end{table}

\subsubsection{$\gamma gg/ggg$ ratios}

The direct photon spectrum in $1S,2S,3S$ decays has been measured using CLEO
III data \cite{Besson:2005jv}.  The ratios $R_\gamma \equiv {\cal B}(g g
\gamma)/ {\cal B}(g g g)$ are found to be
$R_\gamma(1S) = (2.70\pm0.01\pm0.13\pm0.24)\%$,
$R_\gamma(2S) = (3.18\pm0.04\pm0.22\pm0.41)\%$,
$R_\gamma(3S) = (2.72\pm0.06\pm0.32\pm0.37)\%$.
$R_\gamma(1S)$ is consistent with an earlier CLEO value
of $(2.54\pm0.18\pm0.14)\%$.

\subsection{E1 transitions between $\chi_{bJ}(nP)$ and $S$ states}

We have already discussed the inclusive branching ratios for the transitions
$\Upsilon(2S) \to \gamma \chi_{bJ}(1P)$,
$\Upsilon(3S) \to \gamma \chi_{bJ}(1P)$, and
$\Upsilon(3S) \to \gamma \chi_{bJ}(2P)$.
When these are combined with branching ratios for exclusive transitions
where the photons from $\chi_{bJ} \to \gamma \Upsilon(1S)$ and
$\chi_{bJ}(2P) \to \gamma \Upsilon(1S,2S)$ and the subsequent
decays $\Upsilon(1S,2S) \to \ell^+ \ell^-$ also are observed, one can obtain
branching ratios for the radiative E1 decays of the $\chi_{bJ}(1P)$ and
$\chi_{bJ}(2P)$ states.
The $\chi_{bJ}(1P)$ branching ratios have not changed since
the treatment of \citet{Kwong:1988ae}, and are consistent with the predictions
quoted there.  There has been some improvement in knowledge of the
$\chi_{bJ}(2P)$ branching ratios, as summarized in Table~\ref{tab:chibrs}.

\begin{table}
\caption{Predicted \cite{Kwong:1988ae} and measured \cite{Yao:2006px}
branching ratios for $\chi_{bJ}(2P) = 2^3P_J$ radiative E1 decays.
\label{tab:chibrs}}
\begin{center}
\begin{tabular}{c c c c} \hline \hline
      & Final & Predicted ${\cal B} (\%)$ & Measured ${\cal B}$ (\%) \\
Level & state & \cite{Kwong:1988ae}       & \cite{Yao:2006px}        \\ \hline
$2^3P_0$ & $\gamma + 1S$ & 0.96 & $0.9\pm0.6$ \\
& $\gamma + 2S$ & 1.27 & $4.6\pm2.1$ \\
$2^3P_1$ & $\gamma + 1S$ & 11.8 & $8.5\pm1.3$ \\
& $\gamma + 2S$ & 20.2 & $21\pm4$    \\
$2^3P_2$ & $\gamma + 1S$ &  5.3 & $7.1\pm1.0$ \\
& $\gamma + 2S$ & 18.9 & $16.2\pm2.4$ \\ \hline \hline
\end{tabular}
\end{center}
\end{table}

\begin{figure}
\begin{center}
\includegraphics[width=0.98\textwidth]{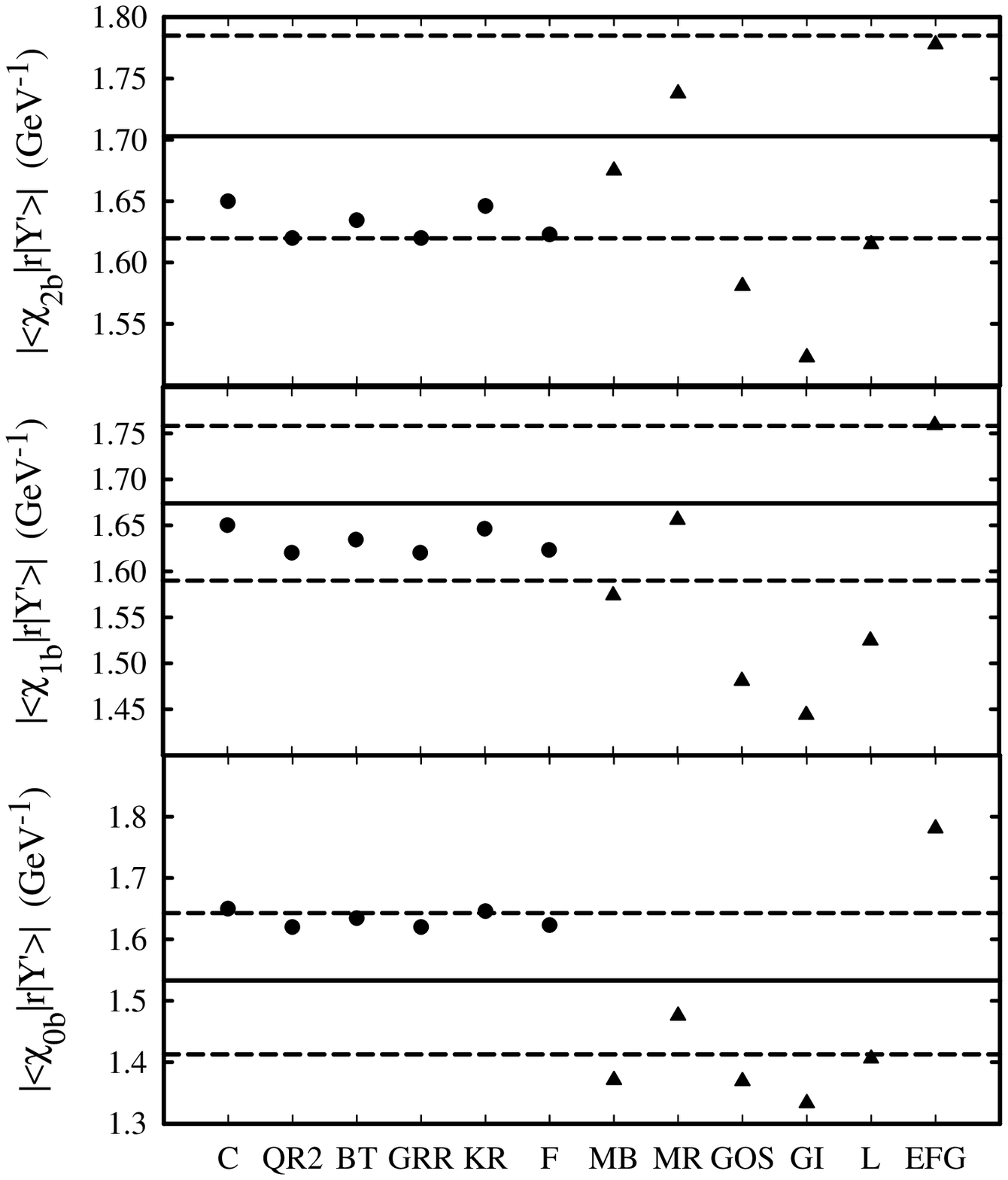}
\caption{E1 dipole transition matrix elements for the bottomonium decays
$2^3S_1\to 1^3P_J$.  The labels are the same as in Fig.~\ref{fig:cc2sto1p}
with the addition of two sets of predictions:
KR-Kwong Rosner \cite{Kwong:1988ae}, F-Fulcher \cite{Fulcher:1990kx}.
\label{fig:bb2sto1p}}
\end{center}
\end{figure}

\begin{figure}
\begin{center}
\includegraphics[width=0.98\textwidth]{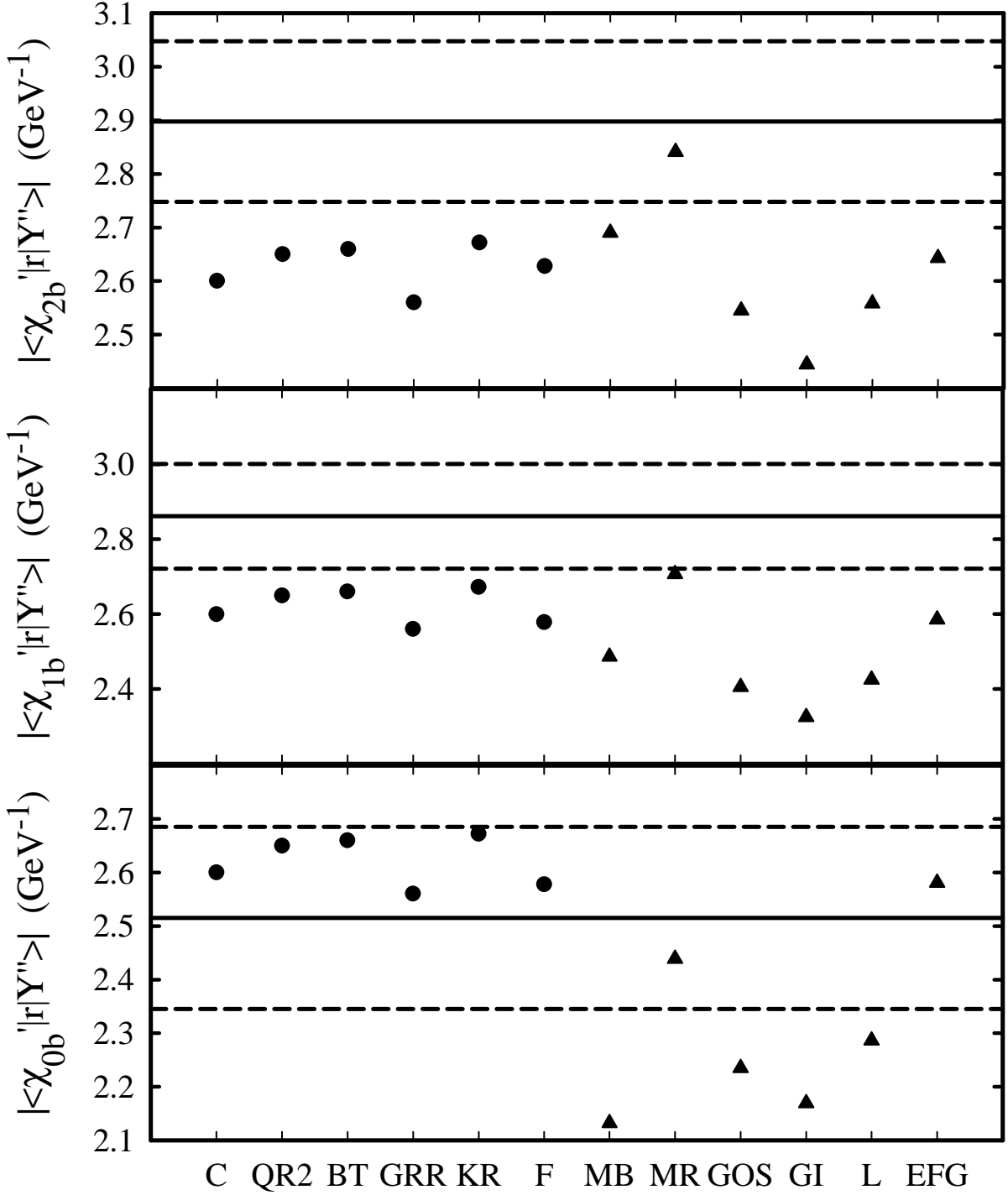}
\caption{E1 dipole transition matrix elements for the bottomonium decays
$3^3S_1\to 2^3P_J$.  The labels are the same as in Fig.~\ref{fig:bb2sto1p}.
\label{fig:bb3sto2p}}
\end{center}
\end{figure}

The dipole matrix elements for $\Upsilon(2S) \to \gamma \chi_{bJ}(1P)$ and
$\Upsilon(3S) \to \gamma \chi_{bJ}(2P)$ are
shown in Figs.\ \ref{fig:bb2sto1p} and \ref{fig:bb3sto2p},
along with predictions of various models.  The dipole matrix element
predictions are in generally good agreement with the observed values.

As already pointed out, the most notable exceptions are the matrix elements
$\langle 3^3S_1 | r | 1^3P_J \rangle$.  In the NR limit this overlap is less
than 5\% of any other $S-P$ overlap, and its suppression occurs for a broad
range of potential shapes \cite{Grant:1992fi}. This dynamical accident makes
these transition rates very sensitive to the details of wave functions and
relativistic corrections which are not known to this level of precision.  This
sensitivity is shown most clearly looking at the signs of the matrix elements
as well as their magnitudes.  The average experimental value for this matrix
element is $\langle 3^3S_1 | r | 1^3P_J \rangle = 0.050 \pm 0.006$ GeV$^{-1}$
\cite{Cinabro:2002ji}.  Taking the predictions of \citet{Godfrey:1985xj} for
comparison, the average over $J$ values gives 0.052 GeV$^{-1}$ which is in good
agreement with the observed value.  However, more detailed scrutiny gives
0.097, 0.045, and --0.015 GeV$^{-1}$ for $J=2$, 1, and 0 matrix elements
respectively.  Not only is there a large variation in the magnitudes but the
sign also changes, highlighting how sensitive the results for this particular
transition are to details of the model due to delicate cancellations in the
integral.

The branching ratios can also be used to measure the ratios of various E1
matrix elements which can then be compared to potential model predictions.
CLEO \cite{Cinabro:2002ji} obtained the following values for ratios:
\begin{eqnarray}
\frac{|\langle 2^3P_2 | r | 1^3S_1 \rangle |}{|\langle 2^3P_2 | r |
2^3S_1 \rangle |} & = & 0.105 \pm 0.004 \pm 0.006, \nonumber \\
\frac{|\langle 2^3P_1 | r | 1^3S_1 \rangle |}{|\langle 2^3P_1 | r |
2^3S_1 \rangle |} & = & 0.087 \pm 0.002 \pm 0.005, \nonumber \\
\frac{|\langle 2^3P_{1,2} | r | 1^3S_1 \rangle |}{|\langle 2^3P_{1,2} | r |
2^3S_1 \rangle |} & = & 0.096 \pm 0.002 \pm 0.005, \nonumber
\end{eqnarray}
where the final ratio averages the results for $J=1$ and $J=2$.
In nonrelativistic calculations the E1 matrix elements do not depend on $J$.
The deviation of the results for $J=1$ and $J=2$ from each other
suggests relativistic contributions to the matrix elements.

\subsection{$D$-wave states}

The precise information on the masses of $S$-wave and $P$-wave $b \bar b$ levels
leads to highly constrained predictions for the masses and production rates
for the $D$-wave levels \cite{Kwong:1988ae,Godfrey:2001vc}.  The CLEO Collaboration
\cite{Bonvicini:2004yj} has presented evidence for at least one of these levels
in the four-photon cascade $\Upsilon(3S) \to \gamma \chi_b (2P)$,
$\chi_b (2P) \to \gamma \Upsilon (1D)$, $\Upsilon (1D) \to \gamma
\chi_b(1P)$, $\chi_b(1P)\to \gamma  \Upsilon(1S)$,
followed by the $\Upsilon(1S)$ annihilation into $e^+e^-$ or
$\mu^+\mu^-$.  CLEO III \cite{Bonvicini:2004yj} finds their data are
dominated by the production of one $\Upsilon(1D)$ state consistent
with the $J=2$ assignment and a mass $(10161.1\pm 0.6 \pm 1.6)$~MeV,
which is consistent with predictions from potential models and lattice
QCD calculations.  The signal product branching ratio obtained is
${\cal B}(\gamma\gamma\gamma\gamma
\ell^+\ell^-)_{\Upsilon(1D)}=(2.5\pm 0.5\pm 0.5 ) \cdot 10^{-5}$ where
the first error is statistical and the second one is systematic.  The
branching ratio is consistent with the theoretical estimate of $2.6\times
10^{-5}$ \cite{Kwong:1988ae,Godfrey:2001vc} for the $\Upsilon (1^3D_2)$
intermediate state.

\subsection{New hadronic transitions}

\subsubsection{$\chi_{b1,2}(2P) \to \omega \Upsilon(1S)$}

The first transition of one heavy quarkonium state to another involving
$\omega$ emission was reported by the CLEO Collaboration
\cite{Severini:2003qw}: $\Upsilon(2^3P_{1,2})\to \omega \Upsilon(1S)$, which we
have already mentioned in connection with the corresponding
transition for the $\chi_{c1}(2P)\ (2^3P_1)$ charmonium state.

\subsubsection{$\chi_{b1,2}(2P) \to \chi_{b1,2}$}

The transitions $\chi_b(2P) \to \chi_b(1P) \pi \pi$ have been observed for the
first time \cite{Cawlfield:2005ra}.  One looks for $\Upsilon(3S) \to \gamma
\chi_b(2P) \to \gamma \pi \pi \chi_b(1P)
\to \gamma \pi \pi \gamma \Upsilon(1S)$ in
CLEO data consisting of 5.8 million 3S events.  Both charged and neutral pions
are detected.  Assuming that
$\Gamma(\chi_{b1}(2P) \to \pi \pi \chi_{b1}(1P)) =
\Gamma(\chi_{b2}(2P) \to \pi \pi \chi_{b2}(1P))$,
both are found equal to $(0.83 \pm
0.22 \pm 0.08 \pm 0.19)$ keV, with the uncertainties being statistical,
internal CLEO systematics, and common systematics from outside sources.  This
value is in satisfactory agreement with theoretical expectations \cite{Kuang:1981se}.

\subsubsection{Searches for $\Upsilon(2S,3S) \to \eta \Upsilon(1S)$}

The decay $\psi(2S) \to \eta J/\psi(1S)$ has been known to occur since the
early decays of charmonium spectroscopy.  The world average for its branching
ratio is $\b[\psi(2S) \to \eta J/\psi(1S)] = (3.09 \pm 0.08)\%$
\cite{Yao:2006px}. The corresponding $\Upsilon(2S) \to \eta \Upsilon(1S)$
process
is represented by the published upper limit
$\b < 2 \times 10^{-3}$ \cite{Fonseca:1984fg}.  The corresponding upper limit for
$\Upsilon(3S) \to \eta \Upsilon(1S)$ is $\b < 2.2 \times 10^{-3}$
\cite{Brock:1990pj}.  However, because these transitions involve a quark spin
flip, they are expected to be highly suppressed in the $b \bar b$ system.
Defining the ratios
\beq
R' \equiv \frac{\Gamma[\Upsilon(2S) \to \eta \Upsilon(1S)]}
{\Gamma[\psi(2S) \to \eta J/\psi(1S)]}~~,~~~
R'' \equiv \frac{\Gamma[\Upsilon(3S) \to \eta \Upsilon(1S)]}
{\Gamma[\psi(2S) \to \eta J/\psi(1S)]}~~,~~~
\eeq
Yan \cite{Yan:1980uh} estimates $R' \simeq 1/400$, while
Kuang \cite{Kuang:2006me} finds in one model $R' = 0.0025$, $R'' = 0.0013$.

Combining these results with the latest total widths \cite{Yao:2006px},
one predicts
\begin{eqnarray}
\b[\Upsilon(2S) \to \eta \Upsilon(1S)] &=&(8.1 \pm 0.8) \times 10^{-4}~~,~~~
\label{eqn:2Spred} \\
\b[\Upsilon(3S) \to \eta \Upsilon(1S)] &=& (6.7 \pm 0.7) \times 10^{-4}~~.
\label{eqn:3Spred}
\end{eqnarray}
The present CLEO III samples of 9~million $\Upsilon(2S)$ and 6~million
$\Upsilon(3S)$ decays are being used to test these predictions.
Preliminary results \cite{Kreinick:2007gh} indicate $\b[\Upsilon(2S) \to \eta
\Upsilon(1S)] = (2.5 \pm 0.7 \pm 0.5) \times 10^{-4}$ and $\b[\Upsilon(2S) \to
\pi^0 \Upsilon(1S)] < 2.1 \times 10^{-4}~(90\%$ c.l.)

\subsection{Searches for spin-singlets}

Decays of the $\Upsilon(1S,2S,3S)$ states can yield $b \bar b$ spin-singlets,
but none has been seen yet.  One expects $1S$, $2S$, and $3S$ hyperfine
splittings to be approximately 60, 30, 20 MeV/$c^2$ \cite{Godfrey:2001eb}. The
lowest $P$-wave singlet state (``$h_b$'') is expected to be near $\langle
m(1^3P_J) \rangle \simeq 9900$ MeV/$c^2$ \cite{Godfrey:2002rp}.

The CDF Collaboration has identified events of the form $B_c \to J/\psi
\pi^\pm$, allowing for the first time a precise determination of the mass.
The value quoted in \citet{Aaltonen:2007gv},
$m(B_c)$=(6275.6$\pm$2.9$\pm$2.5) MeV/$c^2$,
is in reasonable accord with the latest lattice prediction of
6304$\pm$12$^{+18}_{-0}$ MeV \cite{Allison:2004be}.

The mass of the observed $b \bar c$ state can be used to distinguish among
various theoretical approaches to $c \bar c$, $b \bar c$, and $b \bar b$
spectra.  In this manner, in principle, one can obtain a more reliable
prediction of the masses of unseen $b \bar b$ states such as
$\eta_b(1S,2S,3S)$.  For example, by comparing predictions of potential
models to the measured values of the $J/\psi$, $\eta_c$, $\Upsilon$,
and $B_c$ states one could use the prediction of the most reliable models
\cite{Godfrey:1985xj,Ebert:2002pp,Eichten:1994gt,Fulcher:1991dm,Fulcher:1998ka}
to estimate the mass of the $\eta_b(1S)=9400-9410$~MeV.

Several searches have been performed or are under way in $1S,~2S$, and $3S$
CLEO data.  The allowed M1 transition in $\Upsilon(1S) \to \gamma \eta_b(1S)$
can be studied by reconstructing exclusive final states in $\eta_b(1S)$ decays.
One may be able to dispense with the soft photon, which could be swallowed up
in background.  Final states are likely to be of high multiplicity.

One can search for higher-energy but suppressed M1 photons in $\Upsilon(n'S)
\to \gamma \eta_b(nS)$ $(n \ne n')$ decays.  Inclusive searches already exclude
many models. The strongest upper limit obtained is for $n'=3$, $n=1$: ${\cal B}
\le 4.3 \times 10^{-4}$ (90\% c.l.)~\cite{Artuso:2004fp}.  Exclusive searches
(in which $\eta_b$ decay products are reconstructed) also hold some promise.
Searches for $\eta_b$ using the sequential processes
$\Upsilon(3S) \to \pi^0 h_b(1^1P_1) \to \pi^0 \gamma \eta_b(1S)$ and
$\Upsilon(3S) \to \gamma \chi_{b0}(2P) \to \gamma \eta \eta_b(1S)$
(suggested in \citet{Voloshin:2004hs}) are being conducted.
Additional searches for $h_b$ involve the transition
$\Upsilon(3S) \to \pi^+ \pi^- h_b$ [for which a typical experimental upper
bound based on earlier CLEO data \cite{Brock:1990pj,Butler:1993rq} is
${\cal O}(10^{-3})$].  The $h_b \to \gamma \eta_b$ transition is
expected to have a 40\% branching ratio \cite{Godfrey:2002rp}, much like $h_c
\to \gamma \eta_c$.

\subsection{$\Upsilon(4S)$}

The $\Upsilon(4S)$ is the lowest-lying bound bottomonium state above
open-flavor threshold. Its mass and total width as well as electronic width
have been determined in scans, most recently by BaBar~\cite{Aubert:2004pwa}:
$M = (10579.3 \pm 0.4 \pm 1.2)\,\mathrm{MeV}/c^2$,
$\Gamma_{ee} = (0.321 \pm 0.017 \pm 0.029)\,\mathrm{keV}$,
$\Gamma = (20.7 \pm 1.6 \pm 2.5)\,\mathrm{MeV}$.
Although the $\Upsilon(4S)$ has primarily been regarded as a $B \bar B$
``factory,''  its decays to bound $b \bar b$ states are beginning to be
observed in the large data samples accumulated by BaBar and Belle.  This is
not surprising, as the corresponding first charmonium state above flavor
threshold, the $\psi(3770)$, does decay -- rarely -- to
charmonium~\cite{Yao:2006px}.

The BaBar Collaboration \cite{Aubert:2006bm} measures the product branching
fractions \linebreak
${\cal B}[\Upsilon(4S) \to \pi^+ \pi^- \Upsilon(1S)] \times
{\cal B}(\Upsilon(1S) \to \mu^+ \mu^-) =
(2.23 \pm 0.25 \pm 0.27)\times 10^{-6}$ and
${\cal B}[\Upsilon(4S) \to \pi^+ \pi^- \Upsilon(2S)] \times
{\cal B}(\Upsilon(2S) \to \mu^+ \mu^-) =
(1.69 \pm 0.26 \pm 0.20)\times 10^{-6}$, while
the Belle Collaboration \cite{Sokolov:2006sd} finds
${\cal B}[\Upsilon(4S) \to \pi^+ \pi^- \Upsilon(1S)] \times
{\cal B}(\Upsilon(1S) \to \mu^+ \mu^-) =
(4.4 \pm 0.8 \pm 0.6)\times 10^{-6}$.
These product branching fractions, when combined with
${\cal B}(\Upsilon(1S) [\Upsilon(2S)] \to \mu^+ \mu^-)
= (2.48 \pm 0.05)\% [ (1.93 \pm 0.17)\% ]$~\cite{Yao:2006px}
result in branching fractions of the order of $10^{-4}$ and
partial widths of a few keV, comparable with other
partial widths for dipion transitions in the $\Upsilon$ system
of the same order of magnitude.
An interesting feature is that the distribution
of $m(\pi^+\pi^-)$ in $\Upsilon(4S) \to \Upsilon(2S)$
looks markedly different from the $\Upsilon$ dipion transitions with $\Delta
n = 1$ [$\Upsilon(3S) \to \Upsilon(2S)$, $\Upsilon(2S) \to \Upsilon(1S)$]
and more resembles that of $\Upsilon(3S) \to \Upsilon(1S)$; however,
the $\Upsilon(4S) \to \Upsilon(1S)$ dipion spectrum
($\Delta n = 3$) can be described by a model that
suits the $\Delta n = 1$ bottomonium transitions and also
the shape in $\psi(2S) \to \pi^+\pi^- J/\psi$~\cite{Kuang:1981se}.

\begin{figure}
\begin{center}
\includegraphics[width=\textwidth]{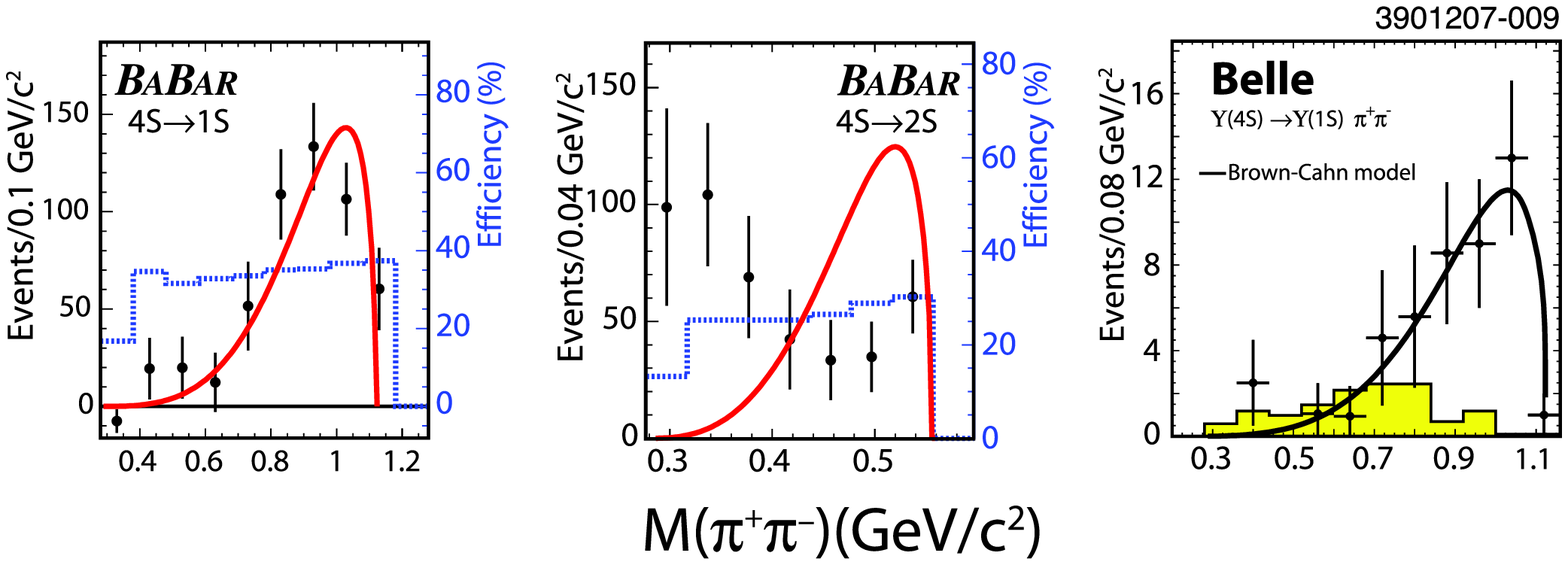}
\caption{Invariant mass of the dipion system in $\Upsilon(4S)
\to \pi^+\pi^- \Upsilon(1,2S)$ as measured in data from
BaBar~\cite{Aubert:2006bm} and Belle~\cite{Sokolov:2006sd} (points), after
efficiency correction.
For the BaBar figures, the dotted line is the selection
efficiency, and the solid line is the prediction of
\citet{Kuang:1981se}.
In the Belle plot, the shaded histogram is a background
estimate, and the curve is based on the model detailed
in \citet{Brown:1975dz,Voloshin:1975yb,Yan:1980uh}.
\label{fig:mpipi}}
\end{center}
\end{figure}

The measured dipion invariant mass distributions
for $\Upsilon(4S) \to \pi^+\pi^-\Upsilon(1S,2S)$
are shown in Figure~\ref{fig:mpipi}.

\subsection{States above open flavor threshold}
Two states have been seen in $e^+e^-$ scattering~\cite{Yao:2006px},
establishing quantum numbers $J^{PC} = 1^{--}$: $\Upsilon(10860)$
(mass $10.865 \pm 0.008$~GeV, total width $110 \pm 13$~MeV) and
$\Upsilon(11020)$ (mass $11.019 \pm 0.008$~GeV, total width $79 \pm 16$~MeV).
These states are often identified as $5S$ and $6S$ bottomonium levels.

\section{Summary}
\label{sec:summary}

In the presence of much more accurate data, multipole expansions for both
electromagnetic and hadronic transitions hold up well.  The coefficients
appearing in these expansions have been described in the past by a combination
of potential models and perturbative QCD.  As expected there are significant
relativistic corrections for the charmonium system.  The overall scales of
these corrections are reduced for the $\bar bb$ system and are consistent with
expectations from the NRQCD velocity expansion.  Relativistic corrections are
determined in the same framework as leading order terms.  However, relativistic
corrections have not improved markedly upon the nonrelativistic treatments,
though some qualitiative patterns (such as hierarchies in electric dipole
matrix elements) are reproduced.

Electromagnetic transitions for which the leading-order expansion coefficient
is dynamically suppressed are particularly sensitive to relativistic
corrections.  For the $\Upsilon (3S) \rightarrow \chi_b(1P)$ E1 transitions
there is a large cancellation in overlap amplitude because of the node in the
$3S$ radial wavefunction.  The result is a wide scatter of theoretical
predictions.  For the $\Upsilon (3S) \rightarrow \eta_b(1S)$  M1 transition,
the overlap coefficient vanishes in leading order (a hindered transition).
Here the experimental upper bound on the rate is smaller than expected in
potential models for relativistic corrections.
Modern theoretical tools (effective theories and nonperturbative lattice QCD)
combined with more detailed high-statistics experimental data will help pin
down the various relativistic corrections.

Decays described by perturbative QED or QCD, such as $\chi_{c0,2} \to (\gamma
\gamma,~gg$), appear to behave as expected, yielding values of $\alpha_S$ for
the most part consistent with other determinations.  Exceptions (as in the case
of the anomalously small $J/\psi$ hadronic width) can be ascribed to large
QCD or relativistic corrections
or to neglected color-octet components of the wave function
which are not yet fully under control.

Recent experiments have also observed a number of new hadronic transitions.
Many details remain to be understood.  The two-pion invariant mass
distributions in both the $\Upsilon (3S) \rightarrow \Upsilon (1S) + 2\pi$ and
$\Upsilon (4S) \rightarrow \Upsilon (2S) + 2\pi$ transitions do not show
typically strong $S$-wave behavior.  Perhaps some dynamical suppression plays a
role in these transitions.  To further complicate the situation, the $\Upsilon
(4S) \rightarrow \Upsilon(1S) + 2\pi$ decay seems to show the usual $S$-wave
behavior with the dipion spectrum peaked toward the highest effective masses.

Coupled-channel effects appear to be important in understanding quarkonium
behavior, especially in such cases as the $X(3872)$
which lies right at the $D^0 \bar D^{*0}$ threshold.  It seems that
long-awaited states such as ``molecular charmonium'' [with $X(3872)$ the
leading candidate] and hybrids [perhaps such as $Y(4260)$] are making their
appearance, and the study of their transitions will shed much light on their
nature.  Now that we are entering the era of precise lattice QCD predictions
for low-lying quarkonium states, it is time for lattice theorists to grapple with
these issues as well.

\section*{Acknowledgments}

Input from Frank Close, Richard Galik, Brian Heltsley, and Kamal Seth is
gratefully acknowledged.  Fermilab is operated by Fermi Research Alliance, LLC
under Contract No.\ DE-AC02-07CH11359 with the United States Department of
Energy.  This work was supported in part by the United States Department of
Energy under Grant No.\ DE FG02 90ER40560, the Natural Sciences and Engineering
Research Council of Canada, and the US National Science Foundation under
cooperative agreement PHY-0202078.  J. L. R. thanks the Laboratory for
Elementary-Particle Physics (Cornell) and the Aspen Center for Physics for
hospitality during part of this research.

\bibliography{OniaReview}

\begin{thebibliography}{260}
\expandafter\ifx\csname natexlab\endcsname\relax\def\natexlab#1{#1}\fi
\expandafter\ifx\csname bibnamefont\endcsname\relax
  \def\bibnamefont#1{#1}\fi
\expandafter\ifx\csname bibfnamefont\endcsname\relax
  \def\bibfnamefont#1{#1}\fi
\expandafter\ifx\csname citenamefont\endcsname\relax
  \def\citenamefont#1{#1}\fi
\expandafter\ifx\csname url\endcsname\relax
  \def\url#1{\texttt{#1}}\fi
\expandafter\ifx\csname urlprefix\endcsname\relax\def\urlprefix{URL }\fi
\providecommand{\bibinfo}[2]{#2}
\providecommand{\eprint}[2][]{\url{#2}}

\bibitem[{Aaltonen \emph{et~al.}(2007)\citenamefont{Aaltonen}
  \emph{et~al.}}]{Aaltonen:2007gv}
\bibinfo{author}{\bibnamefont{Aaltonen}, \bibfnamefont{T.}}, \emph{et~al.}
  (\bibinfo{collaboration}{CDF}), \bibinfo{year}{2007}, \eprint{arXiv:0712.1506
  [hep-ex]}.

\bibitem[{Abazov \emph{et~al.}(2004)\citenamefont{Abazov}
  \emph{et~al.}}]{Abazov:2004kp}
\bibinfo{author}{\bibnamefont{Abazov}, \bibfnamefont{V.~M.}}, \emph{et~al.}
  (\bibinfo{collaboration}{D0}), \bibinfo{year}{2004}, \bibinfo{journal}{Phys.
  Rev. Lett.} \textbf{\bibinfo{volume}{93}}, \bibinfo{pages}{162002}.

\bibitem[{Abe \emph{et~al.}(2002{\natexlab{a}})\citenamefont{Abe}
  \emph{et~al.}}]{Abe:2002va}
\bibinfo{author}{\bibnamefont{Abe}, \bibfnamefont{K.}}, \emph{et~al.}
  (\bibinfo{collaboration}{Belle}), \bibinfo{year}{2002}{\natexlab{a}},
  \bibinfo{journal}{Phys. Lett.} \textbf{\bibinfo{volume}{B540}},
  \bibinfo{pages}{33}.

\bibitem[{Abe \emph{et~al.}(2002{\natexlab{b}})\citenamefont{Abe}
  \emph{et~al.}}]{Abe:2002rb}
\bibinfo{author}{\bibnamefont{Abe}, \bibfnamefont{K.}}, \emph{et~al.}
  (\bibinfo{collaboration}{Belle}), \bibinfo{year}{2002}{\natexlab{b}},
  \bibinfo{journal}{Phys. Rev. Lett.} \textbf{\bibinfo{volume}{89}},
  \bibinfo{pages}{142001}.

\bibitem[{Abe \emph{et~al.}(2004)\citenamefont{Abe} \emph{et~al.}}]{Abe:2003zv}
\bibinfo{author}{\bibnamefont{Abe}, \bibfnamefont{K.}}, \emph{et~al.}
  (\bibinfo{collaboration}{Belle}), \bibinfo{year}{2004},
  \bibinfo{journal}{Phys. Rev. Lett.} \textbf{\bibinfo{volume}{93}},
  \bibinfo{pages}{051803}.

\bibitem[{Abe \emph{et~al.}(2005{\natexlab{a}})\citenamefont{Abe}
  \emph{et~al.}}]{Abe:2005ix}
\bibinfo{author}{\bibnamefont{Abe}, \bibfnamefont{K.}}, \emph{et~al.},
  \bibinfo{year}{2005}{\natexlab{a}}, in \emph{\bibinfo{booktitle}{22th
  International Symposium on Lepton-Photon Interactions at High Energy}}
  (\bibinfo{address}{Uppsala, Sweden, June 30 - July 5, 2005}),
  \eprint{hep-ex/0505037}.

\bibitem[{Abe \emph{et~al.}(2005{\natexlab{b}})\citenamefont{Abe}
  \emph{et~al.}}]{Abe:2005iya}
\bibinfo{author}{\bibnamefont{Abe}, \bibfnamefont{K.}}, \emph{et~al.},
  \bibinfo{year}{2005}{\natexlab{b}}, in \emph{\bibinfo{booktitle}{22th
  International Symposium on Lepton-Photon Interactions at High Energy}}
  (\bibinfo{address}{Uppsala, Sweden, June 30 - July 5, 2005}),
  \eprint{hep-ex/0505038}.

\bibitem[{Abe \emph{et~al.}(2005{\natexlab{c}})\citenamefont{Abe}
  \emph{et~al.}}]{Abe:2004zs}
\bibinfo{author}{\bibnamefont{Abe}, \bibfnamefont{K.}}, \emph{et~al.}
  (\bibinfo{collaboration}{Belle}), \bibinfo{year}{2005}{\natexlab{c}},
  \bibinfo{journal}{Phys. Rev. Lett.} \textbf{\bibinfo{volume}{94}},
  \bibinfo{pages}{182002}.

\bibitem[{Abe \emph{et~al.}(2007{\natexlab{a}})\citenamefont{Abe}
  \emph{et~al.}}]{Abe:2006fj}
\bibinfo{author}{\bibnamefont{Abe}, \bibfnamefont{K.}}, \emph{et~al.}
  (\bibinfo{collaboration}{Belle}), \bibinfo{year}{2007}{\natexlab{a}},
  \bibinfo{journal}{Phys. Rev. Lett.} \textbf{\bibinfo{volume}{98}},
  \bibinfo{pages}{092001}.

\bibitem[{Abe \emph{et~al.}(2007{\natexlab{b}})\citenamefont{Abe}
  \emph{et~al.}}]{Abe:2005hd}
\bibinfo{author}{\bibnamefont{Abe}, \bibfnamefont{K.}}, \emph{et~al.},
  \bibinfo{year}{2007}{\natexlab{b}}, \bibinfo{journal}{Phys. Rev. Lett.}
  \textbf{\bibinfo{volume}{98}}, \bibinfo{pages}{082001}.

\bibitem[{Abe \emph{et~al.}(2007{\natexlab{c}})\citenamefont{Abe}
  \emph{et~al.}}]{Abe:2007sy}
\bibinfo{author}{\bibnamefont{Abe}, \bibfnamefont{K.}}, \emph{et~al.}
  (\bibinfo{collaboration}{Belle}), \bibinfo{year}{2007}{\natexlab{c}}, in
  \emph{\bibinfo{booktitle}{International Europhysics Conference on High Energy
  Physics (EPS-HEP2007)}} (\bibinfo{address}{Manchester, England, July 19-25,
  2007}), \eprint{arXiv:0708.3812 [hep-ex]}.

\bibitem[{Ablikim \emph{et~al.}(2004{\natexlab{a}})\citenamefont{Ablikim}
  \emph{et~al.}}]{Ablikim:2004ck}
\bibinfo{author}{\bibnamefont{Ablikim}, \bibfnamefont{M.}}, \emph{et~al.}
  (\bibinfo{collaboration}{BES}), \bibinfo{year}{2004}{\natexlab{a}},
  \bibinfo{journal}{Phys. Lett.} \textbf{\bibinfo{volume}{B603}},
  \bibinfo{pages}{130}.

\bibitem[{Ablikim \emph{et~al.}(2004{\natexlab{b}})\citenamefont{Ablikim}
  \emph{et~al.}}]{Ablikim:2004kv}
\bibinfo{author}{\bibnamefont{Ablikim}, \bibfnamefont{M.}}, \emph{et~al.}
  (\bibinfo{collaboration}{BES}), \bibinfo{year}{2004}{\natexlab{b}},
  \bibinfo{journal}{Phys. Rev.} \textbf{\bibinfo{volume}{D70}},
  \bibinfo{pages}{112007}, \bibinfo{note}{[Erratum-ibid.\ {\bf D71}, 019901
  (2005)]}.

\bibitem[{Ablikim \emph{et~al.}(2004{\natexlab{c}})\citenamefont{Ablikim}
  \emph{et~al.}}]{Ablikim:2004sf}
\bibinfo{author}{\bibnamefont{Ablikim}, \bibfnamefont{M.}}, \emph{et~al.}
  (\bibinfo{collaboration}{BES}), \bibinfo{year}{2004}{\natexlab{c}},
  \bibinfo{journal}{Phys. Rev.} \textbf{\bibinfo{volume}{D70}},
  \bibinfo{pages}{112003}.

\bibitem[{Ablikim \emph{et~al.}(2004{\natexlab{d}})\citenamefont{Ablikim}
  \emph{et~al.}}]{Ablikim:2004bc}
\bibinfo{author}{\bibnamefont{Ablikim}, \bibfnamefont{M.}}, \emph{et~al.}
  (\bibinfo{collaboration}{BES}), \bibinfo{year}{2004}{\natexlab{d}},
  \bibinfo{journal}{Phys. Rev.} \textbf{\bibinfo{volume}{D70}},
  \bibinfo{pages}{077101}.

\bibitem[{Ablikim \emph{et~al.}(2004{\natexlab{e}})\citenamefont{Ablikim}
  \emph{et~al.}}]{Ablikim:2004mv}
\bibinfo{author}{\bibnamefont{Ablikim}, \bibfnamefont{M.}}, \emph{et~al.}
  (\bibinfo{collaboration}{BES}), \bibinfo{year}{2004}{\natexlab{e}},
  \bibinfo{journal}{Phys. Rev.} \textbf{\bibinfo{volume}{D70}},
  \bibinfo{pages}{012003}.

\bibitem[{Ablikim \emph{et~al.}(2005{\natexlab{a}})\citenamefont{Ablikim}
  \emph{et~al.}}]{Ablikim:2004ky}
\bibinfo{author}{\bibnamefont{Ablikim}, \bibfnamefont{M.}}, \emph{et~al.}
  (\bibinfo{collaboration}{BES}), \bibinfo{year}{2005}{\natexlab{a}},
  \bibinfo{journal}{Phys. Lett.} \textbf{\bibinfo{volume}{B614}},
  \bibinfo{pages}{37}.

\bibitem[{Ablikim \emph{et~al.}(2005{\natexlab{b}})\citenamefont{Ablikim}
  \emph{et~al.}}]{Ablikim:2005jy}
\bibinfo{author}{\bibnamefont{Ablikim}, \bibfnamefont{M.}}, \emph{et~al.}
  (\bibinfo{collaboration}{BES}), \bibinfo{year}{2005}{\natexlab{b}},
  \bibinfo{journal}{Phys. Lett.} \textbf{\bibinfo{volume}{B619}},
  \bibinfo{pages}{247}.

\bibitem[{Ablikim \emph{et~al.}(2005{\natexlab{c}})\citenamefont{Ablikim}
  \emph{et~al.}}]{Ablikim:2005yd}
\bibinfo{author}{\bibnamefont{Ablikim}, \bibfnamefont{M.}}, \emph{et~al.}
  (\bibinfo{collaboration}{BES}), \bibinfo{year}{2005}{\natexlab{c}},
  \bibinfo{journal}{Phys. Rev.} \textbf{\bibinfo{volume}{D71}},
  \bibinfo{pages}{092002}.

\bibitem[{Ablikim \emph{et~al.}(2006{\natexlab{a}})\citenamefont{Ablikim}
  \emph{et~al.}}]{Ablikim:2006zq}
\bibinfo{author}{\bibnamefont{Ablikim}, \bibfnamefont{M.}}, \emph{et~al.}
  (\bibinfo{collaboration}{BES}), \bibinfo{year}{2006}{\natexlab{a}},
  \bibinfo{journal}{Phys. Rev. Lett.} \textbf{\bibinfo{volume}{97}},
  \bibinfo{pages}{121801}.

\bibitem[{Ablikim \emph{et~al.}(2006{\natexlab{b}})\citenamefont{Ablikim}
  \emph{et~al.}}]{Ablikim:2006aj}
\bibinfo{author}{\bibnamefont{Ablikim}, \bibfnamefont{M.}}, \emph{et~al.}
  (\bibinfo{collaboration}{BES}), \bibinfo{year}{2006}{\natexlab{b}},
  \bibinfo{journal}{Phys. Lett.} \textbf{\bibinfo{volume}{B641}},
  \bibinfo{pages}{145}.

\bibitem[{Ablikim \emph{et~al.}(2007{\natexlab{a}})\citenamefont{Ablikim}
  \emph{et~al.}}]{Ablikim:2006aw}
\bibinfo{author}{\bibnamefont{Ablikim}, \bibfnamefont{M.}}, \emph{et~al.}
  (\bibinfo{collaboration}{BES}), \bibinfo{year}{2007}{\natexlab{a}},
  \bibinfo{journal}{Phys. Lett.} \textbf{\bibinfo{volume}{B648}},
  \bibinfo{pages}{149}.

\bibitem[{Ablikim \emph{et~al.}(2007{\natexlab{b}})\citenamefont{Ablikim}
  \emph{et~al.}}]{Ablikim:2007ss}
\bibinfo{author}{\bibnamefont{Ablikim}, \bibfnamefont{M.}}, \emph{et~al.}
  (\bibinfo{collaboration}{BES}), \bibinfo{year}{2007}{\natexlab{b}},
  \bibinfo{journal}{Phys. Lett.} \textbf{\bibinfo{volume}{B650}},
  \bibinfo{pages}{111}.

\bibitem[{Ablikim \emph{et~al.}(2007{\natexlab{c}})\citenamefont{Ablikim}
  \emph{et~al.}}]{Ablikim:2006md}
\bibinfo{author}{\bibnamefont{Ablikim}, \bibfnamefont{M.}}, \emph{et~al.}
  (\bibinfo{collaboration}{BES}), \bibinfo{year}{2007}{\natexlab{c}},
  \bibinfo{journal}{Phys. Lett.} \textbf{\bibinfo{volume}{B652}},
  \bibinfo{pages}{238}.

\bibitem[{Ablikim \emph{et~al.}(2008)\citenamefont{Ablikim}
  \emph{et~al.}}]{Ablikim:2007gd}
\bibinfo{author}{\bibnamefont{Ablikim}, \bibfnamefont{M.}}, \emph{et~al.}
  (\bibinfo{collaboration}{BES}), \bibinfo{year}{2008}, \bibinfo{journal}{Phys.
  Lett.} \textbf{\bibinfo{volume}{B660}}, \bibinfo{pages}{315}.

\bibitem[{Abrams \emph{et~al.}(1974)\citenamefont{Abrams}
  \emph{et~al.}}]{Abrams:1974yy}
\bibinfo{author}{\bibnamefont{Abrams}, \bibfnamefont{G.~S.}}, \emph{et~al.},
  \bibinfo{year}{1974}, \bibinfo{journal}{Phys. Rev. Lett.}
  \textbf{\bibinfo{volume}{33}}, \bibinfo{pages}{1453}.

\bibitem[{Abulencia \emph{et~al.}(2007)\citenamefont{Abulencia}
  \emph{et~al.}}]{Abulencia:2006ma}
\bibinfo{author}{\bibnamefont{Abulencia}, \bibfnamefont{A.}}, \emph{et~al.}
  (\bibinfo{collaboration}{CDF}), \bibinfo{year}{2007}, \bibinfo{journal}{Phys.
  Rev. Lett.} \textbf{\bibinfo{volume}{98}}, \bibinfo{pages}{132002}.

\bibitem[{Acosta \emph{et~al.}(2004)\citenamefont{Acosta}
  \emph{et~al.}}]{Acosta:2003zx}
\bibinfo{author}{\bibnamefont{Acosta}, \bibfnamefont{D.~E.}}, \emph{et~al.}
  (\bibinfo{collaboration}{CDF}), \bibinfo{year}{2004}, \bibinfo{journal}{Phys.
  Rev. Lett.} \textbf{\bibinfo{volume}{93}}, \bibinfo{pages}{072001}.

\bibitem[{Adam \emph{et~al.}(2005{\natexlab{a}})\citenamefont{Adam}
  \emph{et~al.}}]{Adam:2005uh}
\bibinfo{author}{\bibnamefont{Adam}, \bibfnamefont{N.~E.}}, \emph{et~al.}
  (\bibinfo{collaboration}{CLEO}), \bibinfo{year}{2005}{\natexlab{a}},
  \bibinfo{journal}{Phys. Rev. Lett.} \textbf{\bibinfo{volume}{94}},
  \bibinfo{pages}{232002}.

\bibitem[{Adam \emph{et~al.}(2005{\natexlab{b}})\citenamefont{Adam}
  \emph{et~al.}}]{Adam:2004pr}
\bibinfo{author}{\bibnamefont{Adam}, \bibfnamefont{N.~E.}}, \emph{et~al.}
  (\bibinfo{collaboration}{CLEO}), \bibinfo{year}{2005}{\natexlab{b}},
  \bibinfo{journal}{Phys. Rev. Lett.} \textbf{\bibinfo{volume}{94}},
  \bibinfo{pages}{012005}.

\bibitem[{Adam \emph{et~al.}(2006)\citenamefont{Adam}
  \emph{et~al.}}]{Adam:2005mr}
\bibinfo{author}{\bibnamefont{Adam}, \bibfnamefont{N.~E.}}, \emph{et~al.}
  (\bibinfo{collaboration}{CLEO}), \bibinfo{year}{2006},
  \bibinfo{journal}{Phys. Rev. Lett.} \textbf{\bibinfo{volume}{96}},
  \bibinfo{pages}{082004}.

\bibitem[{Adams \emph{et~al.}(2005)\citenamefont{Adams}
  \emph{et~al.}}]{Adams:2004xa}
\bibinfo{author}{\bibnamefont{Adams}, \bibfnamefont{G.~S.}}, \emph{et~al.}
  (\bibinfo{collaboration}{CLEO}), \bibinfo{year}{2005},
  \bibinfo{journal}{Phys. Rev. Lett.} \textbf{\bibinfo{volume}{94}},
  \bibinfo{pages}{012001}.

\bibitem[{Adams \emph{et~al.}(2006{\natexlab{a}})\citenamefont{Adams}
  \emph{et~al.}}]{Adams:2005ks}
\bibinfo{author}{\bibnamefont{Adams}, \bibfnamefont{G.~S.}}, \emph{et~al.}
  (\bibinfo{collaboration}{CLEO}), \bibinfo{year}{2006}{\natexlab{a}},
  \bibinfo{journal}{Phys. Rev.} \textbf{\bibinfo{volume}{D73}},
  \bibinfo{pages}{012002}.

\bibitem[{Adams \emph{et~al.}(2006{\natexlab{b}})\citenamefont{Adams}
  \emph{et~al.}}]{Adams:2005mp}
\bibinfo{author}{\bibnamefont{Adams}, \bibfnamefont{G.~S.}}, \emph{et~al.}
  (\bibinfo{collaboration}{CLEO}), \bibinfo{year}{2006}{\natexlab{b}},
  \bibinfo{journal}{Phys. Rev.} \textbf{\bibinfo{volume}{D73}},
  \bibinfo{pages}{051103}.

\bibitem[{Allison \emph{et~al.}(2005)\citenamefont{Allison}
  \emph{et~al.}}]{Allison:2004be}
\bibinfo{author}{\bibnamefont{Allison}, \bibfnamefont{I.~F.}}, \emph{et~al.}
  (\bibinfo{collaboration}{HPQCD}), \bibinfo{year}{2005},
  \bibinfo{journal}{Phys. Rev. Lett.} \textbf{\bibinfo{volume}{94}},
  \bibinfo{pages}{172001}.

\bibitem[{Ambrogiani \emph{et~al.}(2000)\citenamefont{Ambrogiani}
  \emph{et~al.}}]{Ambrogiani:2000vc}
\bibinfo{author}{\bibnamefont{Ambrogiani}, \bibfnamefont{M.}}, \emph{et~al.}
  (\bibinfo{collaboration}{Fermilab E835}), \bibinfo{year}{2000},
  \bibinfo{journal}{Phys. Rev.} \textbf{\bibinfo{volume}{D62}},
  \bibinfo{pages}{052002}.

\bibitem[{Ambrogiani \emph{et~al.}(2002)\citenamefont{Ambrogiani}
  \emph{et~al.}}]{Ambrogiani:2001jw}
\bibinfo{author}{\bibnamefont{Ambrogiani}, \bibfnamefont{M.}}, \emph{et~al.}
  (\bibinfo{collaboration}{Fermilab E835}), \bibinfo{year}{2002},
  \bibinfo{journal}{Phys. Rev.} \textbf{\bibinfo{volume}{D65}},
  \bibinfo{pages}{052002}.

\bibitem[{Ambrogiani \emph{et~al.}(2003)\citenamefont{Ambrogiani}
  \emph{et~al.}}]{Ambrogiani:2003md}
\bibinfo{author}{\bibnamefont{Ambrogiani}, \bibfnamefont{M.}}, \emph{et~al.}
  (\bibinfo{collaboration}{Fermilab E835}), \bibinfo{year}{2003},
  \bibinfo{journal}{Phys. Lett.} \textbf{\bibinfo{volume}{B566}},
  \bibinfo{pages}{45}.

\bibitem[{Andreotti \emph{et~al.}(2003)\citenamefont{Andreotti}
  \emph{et~al.}}]{Andreotti:2003sk}
\bibinfo{author}{\bibnamefont{Andreotti}, \bibfnamefont{M.}}, \emph{et~al.}
  (\bibinfo{collaboration}{Fermilab E835}), \bibinfo{year}{2003},
  \bibinfo{journal}{Phys. Rev. Lett.} \textbf{\bibinfo{volume}{91}},
  \bibinfo{pages}{091801}.

\bibitem[{Andreotti \emph{et~al.}(2004)\citenamefont{Andreotti}
  \emph{et~al.}}]{Andreotti:2004ru}
\bibinfo{author}{\bibnamefont{Andreotti}, \bibfnamefont{M.}}, \emph{et~al.}
  (\bibinfo{collaboration}{Fermilab E835}), \bibinfo{year}{2004},
  \bibinfo{journal}{Phys. Lett.} \textbf{\bibinfo{volume}{B584}},
  \bibinfo{pages}{16}.

\bibitem[{Andreotti \emph{et~al.}(2005{\natexlab{a}})\citenamefont{Andreotti}
  \emph{et~al.}}]{Andreotti:2005pf}
\bibinfo{author}{\bibnamefont{Andreotti}, \bibfnamefont{M.}}, \emph{et~al.}
  (\bibinfo{collaboration}{Fermilab E835}), \bibinfo{year}{2005}{\natexlab{a}},
  \bibinfo{journal}{Phys. Rev.} \textbf{\bibinfo{volume}{D71}},
  \bibinfo{pages}{032006}.

\bibitem[{Andreotti \emph{et~al.}(2005{\natexlab{b}})\citenamefont{Andreotti}
  \emph{et~al.}}]{Andreotti:2005ts}
\bibinfo{author}{\bibnamefont{Andreotti}, \bibfnamefont{M.}}, \emph{et~al.}
  (\bibinfo{collaboration}{Fermilab E835}), \bibinfo{year}{2005}{\natexlab{b}},
  \bibinfo{journal}{Nucl. Phys.} \textbf{\bibinfo{volume}{B717}},
  \bibinfo{pages}{34}.

\bibitem[{Andreotti \emph{et~al.}(2005{\natexlab{c}})\citenamefont{Andreotti}
  \emph{et~al.}}]{Andreotti:2005vu}
\bibinfo{author}{\bibnamefont{Andreotti}, \bibfnamefont{M.}}, \emph{et~al.}
  (\bibinfo{collaboration}{Fermilab E835}), \bibinfo{year}{2005}{\natexlab{c}},
  \bibinfo{journal}{Phys. Rev.} \textbf{\bibinfo{volume}{D72}},
  \bibinfo{pages}{032001}.

\bibitem[{Andreotti \emph{et~al.}(2007)\citenamefont{Andreotti}
  \emph{et~al.}}]{Andreotti:2007ur}
\bibinfo{author}{\bibnamefont{Andreotti}, \bibfnamefont{M.}}, \emph{et~al.}
  (\bibinfo{collaboration}{Fermilab E835}), \bibinfo{year}{2007},
  \bibinfo{journal}{Phys. Lett.} \textbf{\bibinfo{volume}{B654}},
  \bibinfo{pages}{74}.

\bibitem[{\citenamefont{Appelquist}
  \emph{et~al.}(1978)\citenamefont{Appelquist, Barnett, and
  Lane}}]{Appelquist:1978aq}
\bibinfo{author}{\bibnamefont{Appelquist}, \bibfnamefont{T.}},
  \bibinfo{author}{\bibfnamefont{R.~M.} \bibnamefont{Barnett}}, and
  \bibinfo{author}{\bibfnamefont{K.~D.} \bibnamefont{Lane}},
  \bibinfo{year}{1978}, \bibinfo{journal}{Ann. Rev. Nucl. Part. Sci.}
  \textbf{\bibinfo{volume}{28}}, \bibinfo{pages}{387}.

\bibitem[{\citenamefont{Appelquist}
  \emph{et~al.}(1975)\citenamefont{Appelquist, De~Rujula, Politzer, and
  Glashow}}]{Appelquist:1974yr}
\bibinfo{author}{\bibnamefont{Appelquist}, \bibfnamefont{T.}},
  \bibinfo{author}{\bibfnamefont{A.}~\bibnamefont{De~Rujula}},
  \bibinfo{author}{\bibfnamefont{H.~D.} \bibnamefont{Politzer}}, and
  \bibinfo{author}{\bibfnamefont{S.~L.} \bibnamefont{Glashow}},
  \bibinfo{year}{1975}, \bibinfo{journal}{Phys. Rev. Lett.}
  \textbf{\bibinfo{volume}{34}}, \bibinfo{pages}{365}.

\bibitem[{Armstrong \emph{et~al.}(1992)\citenamefont{Armstrong}
  \emph{et~al.}}]{Armstrong:1991he}
\bibinfo{author}{\bibnamefont{Armstrong}, \bibfnamefont{T.~A.}}, \emph{et~al.}
  (\bibinfo{collaboration}{Fermilab E760}), \bibinfo{year}{1992},
  \bibinfo{journal}{Phys. Rev. Lett.} \textbf{\bibinfo{volume}{68}},
  \bibinfo{pages}{1468}.

\bibitem[{Armstrong \emph{et~al.}(1993)\citenamefont{Armstrong}
  \emph{et~al.}}]{Armstrong:1992wu}
\bibinfo{author}{\bibnamefont{Armstrong}, \bibfnamefont{T.~A.}}, \emph{et~al.}
  (\bibinfo{collaboration}{Fermilab E760}), \bibinfo{year}{1993},
  \bibinfo{journal}{Phys. Rev.} \textbf{\bibinfo{volume}{D47}},
  \bibinfo{pages}{772}.

\bibitem[{Artamonov \emph{et~al.}(2000)\citenamefont{Artamonov}
  \emph{et~al.}}]{Artamonov:2000cz}
\bibinfo{author}{\bibnamefont{Artamonov}, \bibfnamefont{A.~S.}}, \emph{et~al.}
  (\bibinfo{collaboration}{OLYA}), \bibinfo{year}{2000},
  \bibinfo{journal}{Phys. Lett.} \textbf{\bibinfo{volume}{B474}},
  \bibinfo{pages}{427}.

\bibitem[{Artuso \emph{et~al.}(2005)\citenamefont{Artuso}
  \emph{et~al.}}]{Artuso:2004fp}
\bibinfo{author}{\bibnamefont{Artuso}, \bibfnamefont{M.}}, \emph{et~al.}
  (\bibinfo{collaboration}{CLEO}), \bibinfo{year}{2005},
  \bibinfo{journal}{Phys. Rev. Lett.} \textbf{\bibinfo{volume}{94}},
  \bibinfo{pages}{032001}.

\bibitem[{Asner \emph{et~al.}(2004)\citenamefont{Asner}
  \emph{et~al.}}]{Asner:2003wv}
\bibinfo{author}{\bibnamefont{Asner}, \bibfnamefont{D.~M.}}, \emph{et~al.}
  (\bibinfo{collaboration}{CLEO}), \bibinfo{year}{2004},
  \bibinfo{journal}{Phys. Rev. Lett.} \textbf{\bibinfo{volume}{92}},
  \bibinfo{pages}{142001}.

\bibitem[{Athar \emph{et~al.}(2004)\citenamefont{Athar}
  \emph{et~al.}}]{Athar:2004dn}
\bibinfo{author}{\bibnamefont{Athar}, \bibfnamefont{S.~B.}}, \emph{et~al.}
  (\bibinfo{collaboration}{CLEO}), \bibinfo{year}{2004},
  \bibinfo{journal}{Phys. Rev.} \textbf{\bibinfo{volume}{D70}},
  \bibinfo{pages}{112002}.

\bibitem[{Aubert \emph{et~al.}(2004{\natexlab{a}})\citenamefont{Aubert}
  \emph{et~al.}}]{Aubert:2003sv}
\bibinfo{author}{\bibnamefont{Aubert}, \bibfnamefont{B.}}, \emph{et~al.}
  (\bibinfo{collaboration}{BABAR}), \bibinfo{year}{2004}{\natexlab{a}},
  \bibinfo{journal}{Phys. Rev.} \textbf{\bibinfo{volume}{D69}},
  \bibinfo{pages}{011103}.

\bibitem[{Aubert \emph{et~al.}(2004{\natexlab{b}})\citenamefont{Aubert}
  \emph{et~al.}}]{Aubert:2003pt}
\bibinfo{author}{\bibnamefont{Aubert}, \bibfnamefont{B.}}, \emph{et~al.}
  (\bibinfo{collaboration}{BABAR}), \bibinfo{year}{2004}{\natexlab{b}},
  \bibinfo{journal}{Phys. Rev. Lett.} \textbf{\bibinfo{volume}{92}},
  \bibinfo{pages}{142002}.

\bibitem[{Aubert \emph{et~al.}(2004{\natexlab{c}})\citenamefont{Aubert}
  \emph{et~al.}}]{Aubert:2004fc}
\bibinfo{author}{\bibnamefont{Aubert}, \bibfnamefont{B.}}, \emph{et~al.}
  (\bibinfo{collaboration}{BABAR}), \bibinfo{year}{2004}{\natexlab{c}},
  \bibinfo{journal}{Phys. Rev. Lett.} \textbf{\bibinfo{volume}{93}},
  \bibinfo{pages}{041801}.

\bibitem[{Aubert \emph{et~al.}(2005{\natexlab{a}})\citenamefont{Aubert}
  \emph{et~al.}}]{Aubert:2004pwa}
\bibinfo{author}{\bibnamefont{Aubert}, \bibfnamefont{B.}}, \emph{et~al.}
  (\bibinfo{collaboration}{BABAR}), \bibinfo{year}{2005}{\natexlab{a}},
  \bibinfo{journal}{Phys. Rev.} \textbf{\bibinfo{volume}{D72}},
  \bibinfo{pages}{032005}.

\bibitem[{Aubert \emph{et~al.}(2005{\natexlab{b}})\citenamefont{Aubert}
  \emph{et~al.}}]{Aubert:2005rm}
\bibinfo{author}{\bibnamefont{Aubert}, \bibfnamefont{B.}}, \emph{et~al.}
  (\bibinfo{collaboration}{BABAR}), \bibinfo{year}{2005}{\natexlab{b}},
  \bibinfo{journal}{Phys. Rev. Lett.} \textbf{\bibinfo{volume}{95}},
  \bibinfo{pages}{142001}.

\bibitem[{Aubert \emph{et~al.}(2005{\natexlab{c}})\citenamefont{Aubert}
  \emph{et~al.}}]{Aubert:2004zr}
\bibinfo{author}{\bibnamefont{Aubert}, \bibfnamefont{B.}}, \emph{et~al.}
  (\bibinfo{collaboration}{BABAR}), \bibinfo{year}{2005}{\natexlab{c}},
  \bibinfo{journal}{Phys. Rev.} \textbf{\bibinfo{volume}{D71}},
  \bibinfo{pages}{031501}.

\bibitem[{Aubert \emph{et~al.}(2005{\natexlab{d}})\citenamefont{Aubert}
  \emph{et~al.}}]{Aubert:2004ns}
\bibinfo{author}{\bibnamefont{Aubert}, \bibfnamefont{B.}}, \emph{et~al.}
  (\bibinfo{collaboration}{BABAR}), \bibinfo{year}{2005}{\natexlab{d}},
  \bibinfo{journal}{Phys. Rev.} \textbf{\bibinfo{volume}{D71}},
  \bibinfo{pages}{071103}.

\bibitem[{Aubert \emph{et~al.}(2006{\natexlab{a}})\citenamefont{Aubert}
  \emph{et~al.}}]{Aubert:2005vi}
\bibinfo{author}{\bibnamefont{Aubert}, \bibfnamefont{B.}}, \emph{et~al.}
  (\bibinfo{collaboration}{BABAR}), \bibinfo{year}{2006}{\natexlab{a}},
  \bibinfo{journal}{Phys. Rev. Lett.} \textbf{\bibinfo{volume}{96}},
  \bibinfo{pages}{052002}.

\bibitem[{Aubert \emph{et~al.}(2006{\natexlab{b}})\citenamefont{Aubert}
  \emph{et~al.}}]{Aubert:2006bm}
\bibinfo{author}{\bibnamefont{Aubert}, \bibfnamefont{B.}}, \emph{et~al.}
  (\bibinfo{collaboration}{BABAR}), \bibinfo{year}{2006}{\natexlab{b}},
  \bibinfo{journal}{Phys. Rev. Lett.} \textbf{\bibinfo{volume}{96}},
  \bibinfo{pages}{232001}.

\bibitem[{Aubert \emph{et~al.}(2006{\natexlab{c}})\citenamefont{Aubert}
  \emph{et~al.}}]{Aubert:2006aj}
\bibinfo{author}{\bibnamefont{Aubert}, \bibfnamefont{B.}}, \emph{et~al.}
  (\bibinfo{collaboration}{BABAR}), \bibinfo{year}{2006}{\natexlab{c}},
  \bibinfo{journal}{Phys. Rev.} \textbf{\bibinfo{volume}{D74}},
  \bibinfo{pages}{071101}.

\bibitem[{Aubert \emph{et~al.}(2006{\natexlab{d}})\citenamefont{Aubert}
  \emph{et~al.}}]{Aubert:2005cb}
\bibinfo{author}{\bibnamefont{Aubert}, \bibfnamefont{B.}}, \emph{et~al.}
  (\bibinfo{collaboration}{BABAR}), \bibinfo{year}{2006}{\natexlab{d}},
  \bibinfo{journal}{Phys. Rev.} \textbf{\bibinfo{volume}{D73}},
  \bibinfo{pages}{012005}.

\bibitem[{Aubert \emph{et~al.}(2006{\natexlab{e}})\citenamefont{Aubert}
  \emph{et~al.}}]{Aubert:2005zh}
\bibinfo{author}{\bibnamefont{Aubert}, \bibfnamefont{B.}}, \emph{et~al.}
  (\bibinfo{collaboration}{BABAR}), \bibinfo{year}{2006}{\natexlab{e}},
  \bibinfo{journal}{Phys. Rev.} \textbf{\bibinfo{volume}{D73}},
  \bibinfo{pages}{011101}.

\bibitem[{Aubert \emph{et~al.}(2007{\natexlab{a}})\citenamefont{Aubert}
  \emph{et~al.}}]{Aubert:2006ge}
\bibinfo{author}{\bibnamefont{Aubert}, \bibfnamefont{B.}}, \emph{et~al.}
  (\bibinfo{collaboration}{BABAR}), \bibinfo{year}{2007}{\natexlab{a}},
  \bibinfo{journal}{Phys. Rev. Lett.} \textbf{\bibinfo{volume}{98}},
  \bibinfo{pages}{212001}.

\bibitem[{Aubert \emph{et~al.}(2007{\natexlab{b}})\citenamefont{Aubert}
  \emph{et~al.}}]{Aubert:2007vj}
\bibinfo{author}{\bibnamefont{Aubert}, \bibfnamefont{B.}}, \emph{et~al.}
  (\bibinfo{collaboration}{BaBar}), \bibinfo{year}{2007}{\natexlab{b}},
  \eprint{arXiv:0711.2047 [hep-ex]}.

\bibitem[{Aubert \emph{et~al.}(1974)\citenamefont{Aubert}
  \emph{et~al.}}]{Aubert:1974js}
\bibinfo{author}{\bibnamefont{Aubert}, \bibfnamefont{J.~J.}}, \emph{et~al.}
  (\bibinfo{collaboration}{BNL E598}), \bibinfo{year}{1974},
  \bibinfo{journal}{Phys. Rev. Lett.} \textbf{\bibinfo{volume}{33}},
  \bibinfo{pages}{1404}.

\bibitem[{Augustin \emph{et~al.}(1974)\citenamefont{Augustin}
  \emph{et~al.}}]{Augustin:1974xw}
\bibinfo{author}{\bibnamefont{Augustin}, \bibfnamefont{J.~E.}}, \emph{et~al.}
  (\bibinfo{collaboration}{SLAC-SP-017}), \bibinfo{year}{1974},
  \bibinfo{journal}{Phys. Rev. Lett.} \textbf{\bibinfo{volume}{33}},
  \bibinfo{pages}{1406}.

\bibitem[{Aulchenko \emph{et~al.}(2003)\citenamefont{Aulchenko}
  \emph{et~al.}}]{Aulchenko:2003qq}
\bibinfo{author}{\bibnamefont{Aulchenko}, \bibfnamefont{V.~M.}}, \emph{et~al.}
  (\bibinfo{collaboration}{KEDR}), \bibinfo{year}{2003},
  \bibinfo{journal}{Phys. Lett.} \textbf{\bibinfo{volume}{B573}},
  \bibinfo{pages}{63}.

\bibitem[{Bai \emph{et~al.}(1995)\citenamefont{Bai} \emph{et~al.}}]{Bai:1995ik}
\bibinfo{author}{\bibnamefont{Bai}, \bibfnamefont{J.~Z.}}, \emph{et~al.}
  (\bibinfo{collaboration}{BES}), \bibinfo{year}{1995}, \bibinfo{journal}{Phys.
  Lett.} \textbf{\bibinfo{volume}{B355}}, \bibinfo{pages}{374},
  \bibinfo{note}{[Erratum-ibid.\ {\bf B363}, 267 (1995)]}.

\bibitem[{Bai \emph{et~al.}(2000)\citenamefont{Bai} \emph{et~al.}}]{Bai:2000sr}
\bibinfo{author}{\bibnamefont{Bai}, \bibfnamefont{J.~Z.}}, \emph{et~al.}
  (\bibinfo{collaboration}{BES}), \bibinfo{year}{2000}, \bibinfo{journal}{Phys.
  Rev.} \textbf{\bibinfo{volume}{D62}}, \bibinfo{pages}{072001}.

\bibitem[{Bai \emph{et~al.}(2002)\citenamefont{Bai} \emph{et~al.}}]{Bai:2002zn}
\bibinfo{author}{\bibnamefont{Bai}, \bibfnamefont{J.~Z.}}, \emph{et~al.}
  (\bibinfo{collaboration}{BES}), \bibinfo{year}{2002}, \bibinfo{journal}{Phys.
  Lett.} \textbf{\bibinfo{volume}{B550}}, \bibinfo{pages}{24}.

\bibitem[{Bai \emph{et~al.}(2003)\citenamefont{Bai} \emph{et~al.}}]{Bai:2003et}
\bibinfo{author}{\bibnamefont{Bai}, \bibfnamefont{J.~Z.}}, \emph{et~al.}
  (\bibinfo{collaboration}{BES}), \bibinfo{year}{2003}, \bibinfo{journal}{Phys.
  Lett.} \textbf{\bibinfo{volume}{B555}}, \bibinfo{pages}{174}.

\bibitem[{Bai \emph{et~al.}(2004{\natexlab{a}})\citenamefont{Bai}
  \emph{et~al.}}]{Bai:2003vf}
\bibinfo{author}{\bibnamefont{Bai}, \bibfnamefont{J.~Z.}}, \emph{et~al.}
  (\bibinfo{collaboration}{BES}), \bibinfo{year}{2004}{\natexlab{a}},
  \bibinfo{journal}{Phys. Rev.} \textbf{\bibinfo{volume}{D69}},
  \bibinfo{pages}{072001}.

\bibitem[{Bai \emph{et~al.}(2004{\natexlab{b}})\citenamefont{Bai}
  \emph{et~al.}}]{Bai:2004cg}
\bibinfo{author}{\bibnamefont{Bai}, \bibfnamefont{J.~Z.}}, \emph{et~al.}
  (\bibinfo{collaboration}{BES}), \bibinfo{year}{2004}{\natexlab{b}},
  \bibinfo{journal}{Phys. Rev.} \textbf{\bibinfo{volume}{D70}},
  \bibinfo{pages}{012006}.

\bibitem[{Bai \emph{et~al.}(2005)\citenamefont{Bai} \emph{et~al.}}]{Bai:2003hv}
\bibinfo{author}{\bibnamefont{Bai}, \bibfnamefont{J.~Z.}}, \emph{et~al.}
  (\bibinfo{collaboration}{BES}), \bibinfo{year}{2005}, \bibinfo{journal}{Phys.
  Lett.} \textbf{\bibinfo{volume}{B605}}, \bibinfo{pages}{63}.

\bibitem[{\citenamefont{Bali}(2001)}]{Bali:2000gf}
\bibinfo{author}{\bibnamefont{Bali}, \bibfnamefont{G.~S.}},
  \bibinfo{year}{2001}, \bibinfo{journal}{Phys. Rept.}
  \textbf{\bibinfo{volume}{343}}, \bibinfo{pages}{1}.

\bibitem[{Barber \emph{et~al.}(1984)\citenamefont{Barber}
  \emph{et~al.}}]{Barber:1983im}
\bibinfo{author}{\bibnamefont{Barber}, \bibfnamefont{D.~P.}}, \emph{et~al.}
  (\bibinfo{collaboration}{ARGUS}), \bibinfo{year}{1984},
  \bibinfo{journal}{Phys. Lett.} \textbf{\bibinfo{volume}{B135}},
  \bibinfo{pages}{498}.

\bibitem[{\citenamefont{Barnes}(1992)}]{Barnes:1992sg}
\bibinfo{author}{\bibnamefont{Barnes}, \bibfnamefont{T.}},
  \bibinfo{year}{1992}, in \emph{\bibinfo{booktitle}{International Workshop on
  Photon-Photon Collisions}} (\bibinfo{address}{La Jolla, CA, March 22-26,
  1992}), \bibinfo{note}{[unpublished]}.

\bibitem[{\citenamefont{Barnes}(2006)}]{Barnes:2006xq}
\bibinfo{author}{\bibnamefont{Barnes}, \bibfnamefont{T.}},
  \bibinfo{year}{2006}, \bibinfo{journal}{Int. J. Mod. Phys.}
  \textbf{\bibinfo{volume}{A21}}, \bibinfo{pages}{5583}.

\bibitem[{\citenamefont{Barnes} \emph{et~al.}(1995)\citenamefont{Barnes, Close,
  and Swanson}}]{Barnes:1995hc}
\bibinfo{author}{\bibnamefont{Barnes}, \bibfnamefont{T.}},
  \bibinfo{author}{\bibfnamefont{F.~E.} \bibnamefont{Close}}, and
  \bibinfo{author}{\bibfnamefont{E.~S.} \bibnamefont{Swanson}},
  \bibinfo{year}{1995}, \bibinfo{journal}{Phys. Rev.}
  \textbf{\bibinfo{volume}{D52}}, \bibinfo{pages}{5242}.

\bibitem[{\citenamefont{Barnes and Godfrey}(2004)}]{Barnes:2003vb}
\bibinfo{author}{\bibnamefont{Barnes}, \bibfnamefont{T.}}, and
  \bibinfo{author}{\bibfnamefont{S.}~\bibnamefont{Godfrey}},
  \bibinfo{year}{2004}, \bibinfo{journal}{Phys. Rev.}
  \textbf{\bibinfo{volume}{D69}}, \bibinfo{pages}{054008}.

\bibitem[{\citenamefont{Barnes} \emph{et~al.}(2005)\citenamefont{Barnes,
  Godfrey, and Swanson}}]{Barnes:2005pb}
\bibinfo{author}{\bibnamefont{Barnes}, \bibfnamefont{T.}},
  \bibinfo{author}{\bibfnamefont{S.}~\bibnamefont{Godfrey}}, and
  \bibinfo{author}{\bibfnamefont{E.~S.} \bibnamefont{Swanson}},
  \bibinfo{year}{2005}, \bibinfo{journal}{Phys. Rev.}
  \textbf{\bibinfo{volume}{D72}}, \bibinfo{pages}{054026}.

\bibitem[{Besson \emph{et~al.}(2006{\natexlab{a}})\citenamefont{Besson}
  \emph{et~al.}}]{Besson:2005hm}
\bibinfo{author}{\bibnamefont{Besson}, \bibfnamefont{D.}}, \emph{et~al.}
  (\bibinfo{collaboration}{CLEO}), \bibinfo{year}{2006}{\natexlab{a}},
  \bibinfo{journal}{Phys. Rev. Lett.} \textbf{\bibinfo{volume}{96}},
  \bibinfo{pages}{092002}.

\bibitem[{Besson \emph{et~al.}(2006{\natexlab{b}})\citenamefont{Besson}
  \emph{et~al.}}]{Besson:2005jv}
\bibinfo{author}{\bibnamefont{Besson}, \bibfnamefont{D.}}, \emph{et~al.}
  (\bibinfo{collaboration}{CLEO}), \bibinfo{year}{2006}{\natexlab{b}},
  \bibinfo{journal}{Phys. Rev.} \textbf{\bibinfo{volume}{D74}},
  \bibinfo{pages}{012003}.

\bibitem[{Besson \emph{et~al.}(2007)\citenamefont{Besson}
  \emph{et~al.}}]{Besson:2006gj}
\bibinfo{author}{\bibnamefont{Besson}, \bibfnamefont{D.}}, \emph{et~al.}
  (\bibinfo{collaboration}{CLEO}), \bibinfo{year}{2007},
  \bibinfo{journal}{Phys. Rev. Lett.} \textbf{\bibinfo{volume}{98}},
  \bibinfo{pages}{052002}.

\bibitem[{\citenamefont{Bethke}(2007)}]{Bethke:2006ac}
\bibinfo{author}{\bibnamefont{Bethke}, \bibfnamefont{S.}},
  \bibinfo{year}{2007}, \bibinfo{journal}{Prog. Part. Nucl. Phys.}
  \textbf{\bibinfo{volume}{58}}, \bibinfo{pages}{351}.

\bibitem[{\citenamefont{Bhanot} \emph{et~al.}(1979)\citenamefont{Bhanot,
  Fischler, and Rudaz}}]{Bhanot:1979af}
\bibinfo{author}{\bibnamefont{Bhanot}, \bibfnamefont{G.}},
  \bibinfo{author}{\bibfnamefont{W.}~\bibnamefont{Fischler}}, and
  \bibinfo{author}{\bibfnamefont{S.}~\bibnamefont{Rudaz}},
  \bibinfo{year}{1979}, \bibinfo{journal}{Nucl. Phys.}
  \textbf{\bibinfo{volume}{B155}}, \bibinfo{pages}{208}.

\bibitem[{\citenamefont{Bhanot and Peskin}(1979)}]{Bhanot:1979vb}
\bibinfo{author}{\bibnamefont{Bhanot}, \bibfnamefont{G.}}, and
  \bibinfo{author}{\bibfnamefont{M.~E.} \bibnamefont{Peskin}},
  \bibinfo{year}{1979}, \bibinfo{journal}{Nucl. Phys.}
  \textbf{\bibinfo{volume}{B156}}, \bibinfo{pages}{391}.

\bibitem[{\citenamefont{Bodwin} \emph{et~al.}(1995)\citenamefont{Bodwin,
  Braaten, and Lepage}}]{Bodwin:1994jh}
\bibinfo{author}{\bibnamefont{Bodwin}, \bibfnamefont{G.~T.}},
  \bibinfo{author}{\bibfnamefont{E.}~\bibnamefont{Braaten}}, and
  \bibinfo{author}{\bibfnamefont{G.~P.} \bibnamefont{Lepage}},
  \bibinfo{year}{1995}, \bibinfo{journal}{Phys. Rev.}
  \textbf{\bibinfo{volume}{D51}}, \bibinfo{pages}{1125},
  \bibinfo{note}{[Erratum-ibid.\ {\bf D55}, 5853 (1997)]}.

\bibitem[{Bonvicini \emph{et~al.}(2004)\citenamefont{Bonvicini}
  \emph{et~al.}}]{Bonvicini:2004yj}
\bibinfo{author}{\bibnamefont{Bonvicini}, \bibfnamefont{G.}}, \emph{et~al.}
  (\bibinfo{collaboration}{CLEO}), \bibinfo{year}{2004},
  \bibinfo{journal}{Phys. Rev.} \textbf{\bibinfo{volume}{D70}},
  \bibinfo{pages}{032001}.

\bibitem[{\citenamefont{Brambilla} \emph{et~al.}(2006)\citenamefont{Brambilla,
  Jia, and Vairo}}]{Brambilla:2005zw}
\bibinfo{author}{\bibnamefont{Brambilla}, \bibfnamefont{N.}},
  \bibinfo{author}{\bibfnamefont{Y.}~\bibnamefont{Jia}}, and
  \bibinfo{author}{\bibfnamefont{A.}~\bibnamefont{Vairo}},
  \bibinfo{year}{2006}, \bibinfo{journal}{Phys. Rev.}
  \textbf{\bibinfo{volume}{D73}}, \bibinfo{pages}{054005}.

\bibitem[{Brambilla \emph{et~al.}(2004)\citenamefont{Brambilla}
  \emph{et~al.}}]{Brambilla:2004wf}
\bibinfo{author}{\bibnamefont{Brambilla}, \bibfnamefont{N.}}, \emph{et~al.}
  (\bibinfo{collaboration}{Quarkonium Working Group}), \bibinfo{year}{2004},
  \eprint{hep-ph/0412158}.

\bibitem[{Briere \emph{et~al.}(2001)\citenamefont{Briere}
  \emph{et~al.}}]{Briere:2001CLEOc}
\bibinfo{author}{\bibnamefont{Briere}, \bibfnamefont{R.~A.}}, \emph{et~al.}
  (\bibinfo{collaboration}{CESR-c Taskforce, CLEO-c Taskforce and CLEO-c
  Collaboration}), \bibinfo{year}{2001}, \bibinfo{journal}{CLNS}
  \textbf{\bibinfo{volume}{01/1742 {\rm [unpublished]}}}.

\bibitem[{Briere \emph{et~al.}(2005)\citenamefont{Briere}
  \emph{et~al.}}]{Briere:2005rc}
\bibinfo{author}{\bibnamefont{Briere}, \bibfnamefont{R.~A.}}, \emph{et~al.}
  (\bibinfo{collaboration}{CLEO}), \bibinfo{year}{2005},
  \bibinfo{journal}{Phys. Rev. Lett.} \textbf{\bibinfo{volume}{95}},
  \bibinfo{pages}{062001}.

\bibitem[{Briere \emph{et~al.}(2006)\citenamefont{Briere}
  \emph{et~al.}}]{Briere:2006ff}
\bibinfo{author}{\bibnamefont{Briere}, \bibfnamefont{R.~A.}}, \emph{et~al.}
  (\bibinfo{collaboration}{CLEO}), \bibinfo{year}{2006},
  \bibinfo{journal}{Phys. Rev.} \textbf{\bibinfo{volume}{D74}},
  \bibinfo{pages}{031106}.

\bibitem[{Brock \emph{et~al.}(1991)\citenamefont{Brock}
  \emph{et~al.}}]{Brock:1990pj}
\bibinfo{author}{\bibnamefont{Brock}, \bibfnamefont{I.~C.}}, \emph{et~al.},
  \bibinfo{year}{1991}, \bibinfo{journal}{Phys. Rev.}
  \textbf{\bibinfo{volume}{D43}}, \bibinfo{pages}{1448}.

\bibitem[{\citenamefont{Brodsky} \emph{et~al.}(1983)\citenamefont{Brodsky,
  Lepage, and Mackenzie}}]{Brodsky:1982gc}
\bibinfo{author}{\bibnamefont{Brodsky}, \bibfnamefont{S.~J.}},
  \bibinfo{author}{\bibfnamefont{G.~P.} \bibnamefont{Lepage}}, and
  \bibinfo{author}{\bibfnamefont{P.~B.} \bibnamefont{Mackenzie}},
  \bibinfo{year}{1983}, \bibinfo{journal}{Phys. Rev.}
  \textbf{\bibinfo{volume}{D28}}, \bibinfo{pages}{228}.

\bibitem[{\citenamefont{Brown and Cahn}(1975)}]{Brown:1975dz}
\bibinfo{author}{\bibnamefont{Brown}, \bibfnamefont{L.~S.}}, and
  \bibinfo{author}{\bibfnamefont{R.~N.} \bibnamefont{Cahn}},
  \bibinfo{year}{1975}, \bibinfo{journal}{Phys. Rev. Lett.}
  \textbf{\bibinfo{volume}{35}}, \bibinfo{pages}{1}.

\bibitem[{\citenamefont{Buchmuller and Tye}(1980)}]{Buchmuller:1979gy}
\bibinfo{author}{\bibnamefont{Buchmuller}, \bibfnamefont{W.}}, and
  \bibinfo{author}{\bibfnamefont{S.~H.~H.} \bibnamefont{Tye}},
  \bibinfo{year}{1980}, \bibinfo{journal}{Phys. Rev. Lett.}
  \textbf{\bibinfo{volume}{44}}, \bibinfo{pages}{850}.

\bibitem[{\citenamefont{Buchmuller and Tye}(1981)}]{Buchmuller:1980su}
\bibinfo{author}{\bibnamefont{Buchmuller}, \bibfnamefont{W.}}, and
  \bibinfo{author}{\bibfnamefont{S.~H.~H.} \bibnamefont{Tye}},
  \bibinfo{year}{1981}, \bibinfo{journal}{Phys. Rev.}
  \textbf{\bibinfo{volume}{D24}}, \bibinfo{pages}{132}.

\bibitem[{Bulava \emph{et~al.}(2007)\citenamefont{Bulava}
  \emph{et~al.}}]{Bulava:2007ka}
\bibinfo{author}{\bibnamefont{Bulava}, \bibfnamefont{J.}}, \emph{et~al.},
  \bibinfo{year}{2007}, \bibinfo{journal}{AIP Conf. Proc.}
  \textbf{\bibinfo{volume}{947}}, \bibinfo{pages}{77}, \bibinfo{note}{eprint
  arXiv:0708.2072 [hep-lat]}.

\bibitem[{Butler \emph{et~al.}(1994)\citenamefont{Butler}
  \emph{et~al.}}]{Butler:1993rq}
\bibinfo{author}{\bibnamefont{Butler}, \bibfnamefont{F.}}, \emph{et~al.}
  (\bibinfo{collaboration}{CLEO}), \bibinfo{year}{1994},
  \bibinfo{journal}{Phys. Rev.} \textbf{\bibinfo{volume}{D49}},
  \bibinfo{pages}{40}.

\bibitem[{\citenamefont{Cassel and Rosner}(2006)}]{Cassel:2006ccour}
\bibinfo{author}{\bibnamefont{Cassel}, \bibfnamefont{D.}}, and
  \bibinfo{author}{\bibfnamefont{J.~L.} \bibnamefont{Rosner}},
  \bibinfo{year}{2006}, \bibinfo{journal}{CERN Courier}
  \textbf{\bibinfo{volume}{46N5}}, \bibinfo{pages}{33}.

\bibitem[{\citenamefont{Caswell and Lepage}(1986)}]{Caswell:1985ui}
\bibinfo{author}{\bibnamefont{Caswell}, \bibfnamefont{W.~E.}}, and
  \bibinfo{author}{\bibfnamefont{G.~P.} \bibnamefont{Lepage}},
  \bibinfo{year}{1986}, \bibinfo{journal}{Phys. Lett.}
  \textbf{\bibinfo{volume}{B167}}, \bibinfo{pages}{437}.

\bibitem[{Cawlfield \emph{et~al.}(2006)\citenamefont{Cawlfield}
  \emph{et~al.}}]{Cawlfield:2005ra}
\bibinfo{author}{\bibnamefont{Cawlfield}, \bibfnamefont{C.}}, \emph{et~al.}
  (\bibinfo{collaboration}{CLEO}), \bibinfo{year}{2006},
  \bibinfo{journal}{Phys. Rev.} \textbf{\bibinfo{volume}{D73}},
  \bibinfo{pages}{012003}.

\bibitem[{Cawlfield \emph{et~al.}(2007)\citenamefont{Cawlfield}
  \emph{et~al.}}]{Cawlfield:2007dw}
\bibinfo{author}{\bibnamefont{Cawlfield}, \bibfnamefont{C.}}, \emph{et~al.}
  (\bibinfo{collaboration}{CLEO}), \bibinfo{year}{2007},
  \bibinfo{journal}{Phys. Rev. Lett.} \textbf{\bibinfo{volume}{98}},
  \bibinfo{pages}{092002}.

\bibitem[{Choi \emph{et~al.}(2002)\citenamefont{Choi}
  \emph{et~al.}}]{Choi:2002na}
\bibinfo{author}{\bibnamefont{Choi}, \bibfnamefont{S.~K.}}, \emph{et~al.}
  (\bibinfo{collaboration}{Belle}), \bibinfo{year}{2002},
  \bibinfo{journal}{Phys. Rev. Lett.} \textbf{\bibinfo{volume}{89}},
  \bibinfo{pages}{102001}.

\bibitem[{Choi \emph{et~al.}(2003)\citenamefont{Choi}
  \emph{et~al.}}]{Choi:2003ue}
\bibinfo{author}{\bibnamefont{Choi}, \bibfnamefont{S.~K.}}, \emph{et~al.}
  (\bibinfo{collaboration}{Belle}), \bibinfo{year}{2003},
  \bibinfo{journal}{Phys. Rev. Lett.} \textbf{\bibinfo{volume}{91}},
  \bibinfo{pages}{262001}.

\bibitem[{Cinabro \emph{et~al.}(2002)\citenamefont{Cinabro}
  \emph{et~al.}}]{Cinabro:2002ji}
\bibinfo{author}{\bibnamefont{Cinabro}, \bibfnamefont{D.}}, \emph{et~al.}
  (\bibinfo{collaboration}{CLEO}), \bibinfo{year}{2002}, in
  \emph{\bibinfo{booktitle}{31st International Conference on High Energy
  Physics (ICHEP 2002)}} (\bibinfo{address}{Amsterdam, The Netherlands, July
  24-31, 2002}), \eprint{hep-ex/0207062}.

\bibitem[{\citenamefont{Close and Page}(2004)}]{Close:2003sg}
\bibinfo{author}{\bibnamefont{Close}, \bibfnamefont{F.~E.}}, and
  \bibinfo{author}{\bibfnamefont{P.~R.} \bibnamefont{Page}},
  \bibinfo{year}{2004}, \bibinfo{journal}{Phys. Lett.}
  \textbf{\bibinfo{volume}{B578}}, \bibinfo{pages}{119}.

\bibitem[{\citenamefont{Close and Page}(2005)}]{Close:2005iz}
\bibinfo{author}{\bibnamefont{Close}, \bibfnamefont{F.~E.}}, and
  \bibinfo{author}{\bibfnamefont{P.~R.} \bibnamefont{Page}},
  \bibinfo{year}{2005}, \bibinfo{journal}{Phys. Lett.}
  \textbf{\bibinfo{volume}{B628}}, \bibinfo{pages}{215}.

\bibitem[{Coan \emph{et~al.}(2006{\natexlab{a}})\citenamefont{Coan}
  \emph{et~al.}}]{Coan:2006rv}
\bibinfo{author}{\bibnamefont{Coan}, \bibfnamefont{T.~E.}}, \emph{et~al.}
  (\bibinfo{collaboration}{CLEO}), \bibinfo{year}{2006}{\natexlab{a}},
  \bibinfo{journal}{Phys. Rev. Lett.} \textbf{\bibinfo{volume}{96}},
  \bibinfo{pages}{162003}.

\bibitem[{Coan \emph{et~al.}(2006{\natexlab{b}})\citenamefont{Coan}
  \emph{et~al.}}]{Coan:2005ps}
\bibinfo{author}{\bibnamefont{Coan}, \bibfnamefont{T.~E.}}, \emph{et~al.}
  (\bibinfo{collaboration}{CLEO}), \bibinfo{year}{2006}{\natexlab{b}},
  \bibinfo{journal}{Phys. Rev. Lett.} \textbf{\bibinfo{volume}{96}},
  \bibinfo{pages}{182002}.

\bibitem[{Cronin-Hennessy \emph{et~al.}(2006)\citenamefont{Cronin-Hennessy}
  \emph{et~al.}}]{CroninHennessy:2006su}
\bibinfo{author}{\bibnamefont{Cronin-Hennessy}, \bibfnamefont{D.}},
  \emph{et~al.} (\bibinfo{collaboration}{CLEO}), \bibinfo{year}{2006},
  \bibinfo{journal}{Phys. Rev.} \textbf{\bibinfo{volume}{D74}},
  \bibinfo{pages}{012005}.

\bibitem[{\citenamefont{Davier} \emph{et~al.}(2007)\citenamefont{Davier,
  Hocker, and Zhang}}]{Davier:2007ym}
\bibinfo{author}{\bibnamefont{Davier}, \bibfnamefont{M.}},
  \bibinfo{author}{\bibfnamefont{A.}~\bibnamefont{Hocker}}, and
  \bibinfo{author}{\bibfnamefont{Z.}~\bibnamefont{Zhang}},
  \bibinfo{year}{2007}, \bibinfo{journal}{Nucl. Phys. Proc. Suppl.}
  \textbf{\bibinfo{volume}{169}}, \bibinfo{pages}{22}.

\bibitem[{Davies \emph{et~al.}(2004)\citenamefont{Davies}
  \emph{et~al.}}]{Davies:2003ik}
\bibinfo{author}{\bibnamefont{Davies}, \bibfnamefont{C.~T.~H.}}, \emph{et~al.}
  (\bibinfo{collaboration}{HPQCD}), \bibinfo{year}{2004},
  \bibinfo{journal}{Phys. Rev. Lett.} \textbf{\bibinfo{volume}{92}},
  \bibinfo{pages}{022001}.

\bibitem[{Davies \emph{et~al.}(2006)\citenamefont{Davies}
  \emph{et~al.}}]{Davies:2006tx}
\bibinfo{author}{\bibnamefont{Davies}, \bibfnamefont{C.~T.~H.}}, \emph{et~al.}
  (\bibinfo{collaboration}{HPQCD}), \bibinfo{year}{2006},
  \bibinfo{journal}{PoS} \textbf{\bibinfo{volume}{LAT2006}},
  \bibinfo{pages}{082}.

\bibitem[{Dobbs \emph{et~al.}(2005)\citenamefont{Dobbs}
  \emph{et~al.}}]{Dobbs:2004di}
\bibinfo{author}{\bibnamefont{Dobbs}, \bibfnamefont{S.}}, \emph{et~al.}
  (\bibinfo{collaboration}{CLEO}), \bibinfo{year}{2005},
  \bibinfo{journal}{Phys. Rev. Lett.} \textbf{\bibinfo{volume}{94}},
  \bibinfo{pages}{032004}.

\bibitem[{Dobbs \emph{et~al.}(2006{\natexlab{a}})\citenamefont{Dobbs}
  \emph{et~al.}}]{Dobbs:2006fj}
\bibinfo{author}{\bibnamefont{Dobbs}, \bibfnamefont{S.}}, \emph{et~al.}
  (\bibinfo{collaboration}{CLEO}), \bibinfo{year}{2006}{\natexlab{a}},
  \bibinfo{journal}{Phys. Rev.} \textbf{\bibinfo{volume}{D74}},
  \bibinfo{pages}{011105}.

\bibitem[{Dobbs \emph{et~al.}(2006{\natexlab{b}})\citenamefont{Dobbs}
  \emph{et~al.}}]{Dobbs:2005yk}
\bibinfo{author}{\bibnamefont{Dobbs}, \bibfnamefont{S.}}, \emph{et~al.}
  (\bibinfo{collaboration}{CLEO}), \bibinfo{year}{2006}{\natexlab{b}},
  \bibinfo{journal}{Phys. Rev.} \textbf{\bibinfo{volume}{D73}},
  \bibinfo{pages}{071101}.

\bibitem[{\citenamefont{Dudek} \emph{et~al.}(2006)\citenamefont{Dudek, Edwards,
  and Richards}}]{Dudek:2006ej}
\bibinfo{author}{\bibnamefont{Dudek}, \bibfnamefont{J.~J.}},
  \bibinfo{author}{\bibfnamefont{R.~G.} \bibnamefont{Edwards}}, and
  \bibinfo{author}{\bibfnamefont{D.~G.} \bibnamefont{Richards}},
  \bibinfo{year}{2006}, \bibinfo{journal}{Phys. Rev.}
  \textbf{\bibinfo{volume}{D73}}, \bibinfo{pages}{074507}.

\bibitem[{\citenamefont{Ebert}
  \emph{et~al.}(2003{\natexlab{a}})\citenamefont{Ebert, Faustov, and
  Galkin}}]{Ebert:2002pp}
\bibinfo{author}{\bibnamefont{Ebert}, \bibfnamefont{D.}},
  \bibinfo{author}{\bibfnamefont{R.~N.} \bibnamefont{Faustov}}, and
  \bibinfo{author}{\bibfnamefont{V.~O.} \bibnamefont{Galkin}},
  \bibinfo{year}{2003}{\natexlab{a}}, \bibinfo{journal}{Phys. Rev.}
  \textbf{\bibinfo{volume}{D67}}, \bibinfo{pages}{014027}.

\bibitem[{\citenamefont{Ebert}
  \emph{et~al.}(2003{\natexlab{b}})\citenamefont{Ebert, Faustov, and
  Galkin}}]{Ebert:2003mu}
\bibinfo{author}{\bibnamefont{Ebert}, \bibfnamefont{D.}},
  \bibinfo{author}{\bibfnamefont{R.~N.} \bibnamefont{Faustov}}, and
  \bibinfo{author}{\bibfnamefont{V.~O.} \bibnamefont{Galkin}},
  \bibinfo{year}{2003}{\natexlab{b}}, \bibinfo{journal}{Mod. Phys. Lett.}
  \textbf{\bibinfo{volume}{A18}}, \bibinfo{pages}{601}.

\bibitem[{\citenamefont{Ebert} \emph{et~al.}(2006)\citenamefont{Ebert, Faustov,
  and Galkin}}]{Ebert:2005nc}
\bibinfo{author}{\bibnamefont{Ebert}, \bibfnamefont{D.}},
  \bibinfo{author}{\bibfnamefont{R.~N.} \bibnamefont{Faustov}}, and
  \bibinfo{author}{\bibfnamefont{V.~O.} \bibnamefont{Galkin}},
  \bibinfo{year}{2006}, \bibinfo{journal}{Phys. Lett.}
  \textbf{\bibinfo{volume}{B634}}, \bibinfo{pages}{214}.

\bibitem[{Edwards \emph{et~al.}(1982)\citenamefont{Edwards}
  \emph{et~al.}}]{Edwards:1981mq}
\bibinfo{author}{\bibnamefont{Edwards}, \bibfnamefont{C.}}, \emph{et~al.}
  (\bibinfo{collaboration}{Crystal Ball}), \bibinfo{year}{1982},
  \bibinfo{journal}{Phys. Rev. Lett.} \textbf{\bibinfo{volume}{48}},
  \bibinfo{pages}{70}.

\bibitem[{\citenamefont{Eichten} \emph{et~al.}(1975)\citenamefont{Eichten,
  Gottfried, Kinoshita, Kogut, Lane, and Yan}}]{Eichten:1974af}
\bibinfo{author}{\bibnamefont{Eichten}, \bibfnamefont{E.}},
  \bibinfo{author}{\bibfnamefont{K.}~\bibnamefont{Gottfried}},
  \bibinfo{author}{\bibfnamefont{T.}~\bibnamefont{Kinoshita}},
  \bibinfo{author}{\bibfnamefont{J.~B.} \bibnamefont{Kogut}},
  \bibinfo{author}{\bibfnamefont{K.~D.} \bibnamefont{Lane}}, and
  \bibinfo{author}{\bibfnamefont{T.-M.} \bibnamefont{Yan}},
  \bibinfo{year}{1975}, \bibinfo{journal}{Phys. Rev. Lett.}
  \textbf{\bibinfo{volume}{34}}, \bibinfo{pages}{369},
  \bibinfo{note}{[Erratum-ibid.\ {\bf 36}, 1276 (1976)]}.

\bibitem[{\citenamefont{Eichten} \emph{et~al.}(1976)\citenamefont{Eichten,
  Gottfried, Kinoshita, Lane, and Yan}}]{Eichten:1975ag}
\bibinfo{author}{\bibnamefont{Eichten}, \bibfnamefont{E.}},
  \bibinfo{author}{\bibfnamefont{K.}~\bibnamefont{Gottfried}},
  \bibinfo{author}{\bibfnamefont{T.}~\bibnamefont{Kinoshita}},
  \bibinfo{author}{\bibfnamefont{K.~D.} \bibnamefont{Lane}}, and
  \bibinfo{author}{\bibfnamefont{T.-M.} \bibnamefont{Yan}},
  \bibinfo{year}{1976}, \bibinfo{journal}{Phys. Rev. Lett.}
  \textbf{\bibinfo{volume}{36}}, \bibinfo{pages}{500}.

\bibitem[{\citenamefont{Eichten} \emph{et~al.}(1978)\citenamefont{Eichten,
  Gottfried, Kinoshita, Lane, and Yan}}]{Eichten:1978tg}
\bibinfo{author}{\bibnamefont{Eichten}, \bibfnamefont{E.}},
  \bibinfo{author}{\bibfnamefont{K.}~\bibnamefont{Gottfried}},
  \bibinfo{author}{\bibfnamefont{T.}~\bibnamefont{Kinoshita}},
  \bibinfo{author}{\bibfnamefont{K.~D.} \bibnamefont{Lane}}, and
  \bibinfo{author}{\bibfnamefont{T.-M.} \bibnamefont{Yan}},
  \bibinfo{year}{1978}, \bibinfo{journal}{Phys. Rev.}
  \textbf{\bibinfo{volume}{D17}}, \bibinfo{pages}{3090},
  \bibinfo{note}{[Erratum-ibid.\ {\bf D21}, 313 (1980)]}.

\bibitem[{\citenamefont{Eichten} \emph{et~al.}(1980)\citenamefont{Eichten,
  Gottfried, Kinoshita, Lane, and Yan}}]{Eichten:1979ms}
\bibinfo{author}{\bibnamefont{Eichten}, \bibfnamefont{E.}},
  \bibinfo{author}{\bibfnamefont{K.}~\bibnamefont{Gottfried}},
  \bibinfo{author}{\bibfnamefont{T.}~\bibnamefont{Kinoshita}},
  \bibinfo{author}{\bibfnamefont{K.~D.} \bibnamefont{Lane}}, and
  \bibinfo{author}{\bibfnamefont{T.-M.} \bibnamefont{Yan}},
  \bibinfo{year}{1980}, \bibinfo{journal}{Phys. Rev.}
  \textbf{\bibinfo{volume}{D21}}, \bibinfo{pages}{203}.

\bibitem[{\citenamefont{Eichten} \emph{et~al.}(2002)\citenamefont{Eichten,
  Lane, and Quigg}}]{Eichten:2002qv}
\bibinfo{author}{\bibnamefont{Eichten}, \bibfnamefont{E.~J.}},
  \bibinfo{author}{\bibfnamefont{K.}~\bibnamefont{Lane}}, and
  \bibinfo{author}{\bibfnamefont{C.}~\bibnamefont{Quigg}},
  \bibinfo{year}{2002}, \bibinfo{journal}{Phys. Rev. Lett.}
  \textbf{\bibinfo{volume}{89}}, \bibinfo{pages}{162002}.

\bibitem[{\citenamefont{Eichten} \emph{et~al.}(2004)\citenamefont{Eichten,
  Lane, and Quigg}}]{Eichten:2004uh}
\bibinfo{author}{\bibnamefont{Eichten}, \bibfnamefont{E.~J.}},
  \bibinfo{author}{\bibfnamefont{K.}~\bibnamefont{Lane}}, and
  \bibinfo{author}{\bibfnamefont{C.}~\bibnamefont{Quigg}},
  \bibinfo{year}{2004}, \bibinfo{journal}{Phys. Rev.}
  \textbf{\bibinfo{volume}{D69}}, \bibinfo{pages}{094019}.

\bibitem[{\citenamefont{Eichten} \emph{et~al.}(2006)\citenamefont{Eichten,
  Lane, and Quigg}}]{Eichten:2005ga}
\bibinfo{author}{\bibnamefont{Eichten}, \bibfnamefont{E.~J.}},
  \bibinfo{author}{\bibfnamefont{K.}~\bibnamefont{Lane}}, and
  \bibinfo{author}{\bibfnamefont{C.}~\bibnamefont{Quigg}},
  \bibinfo{year}{2006}, \bibinfo{journal}{Phys. Rev.}
  \textbf{\bibinfo{volume}{D73}}, \bibinfo{pages}{014014},
  \bibinfo{note}{[Erratum-ibid.\ {\bf D73}, 079903 (2006)]}.

\bibitem[{\citenamefont{Eichten and Quigg}(1994)}]{Eichten:1994gt}
\bibinfo{author}{\bibnamefont{Eichten}, \bibfnamefont{E.~J.}}, and
  \bibinfo{author}{\bibfnamefont{C.}~\bibnamefont{Quigg}},
  \bibinfo{year}{1994}, \bibinfo{journal}{Phys. Rev.}
  \textbf{\bibinfo{volume}{D49}}, \bibinfo{pages}{5845}.

\bibitem[{Fang \emph{et~al.}(2003)\citenamefont{Fang}
  \emph{et~al.}}]{Fang:2002gi}
\bibinfo{author}{\bibnamefont{Fang}, \bibfnamefont{F.}}, \emph{et~al.}
  (\bibinfo{collaboration}{Belle}), \bibinfo{year}{2003},
  \bibinfo{journal}{Phys. Rev. Lett.} \textbf{\bibinfo{volume}{90}},
  \bibinfo{pages}{071801}.

\bibitem[{Fang \emph{et~al.}(2006)\citenamefont{Fang}
  \emph{et~al.}}]{Fang:2006bz}
\bibinfo{author}{\bibnamefont{Fang}, \bibfnamefont{F.}}, \emph{et~al.}
  (\bibinfo{collaboration}{Belle}), \bibinfo{year}{2006},
  \bibinfo{journal}{Phys. Rev.} \textbf{\bibinfo{volume}{D74}},
  \bibinfo{pages}{012007}.

\bibitem[{\citenamefont{Feinberg and Sucher}(1975)}]{Feinberg:1975hk}
\bibinfo{author}{\bibnamefont{Feinberg}, \bibfnamefont{G.}}, and
  \bibinfo{author}{\bibfnamefont{J.}~\bibnamefont{Sucher}},
  \bibinfo{year}{1975}, \bibinfo{journal}{Phys. Rev. Lett.}
  \textbf{\bibinfo{volume}{35}}, \bibinfo{pages}{1740}.

\bibitem[{Follana \emph{et~al.}(2007)\citenamefont{Follana}
  \emph{et~al.}}]{Follana:2006rc}
\bibinfo{author}{\bibnamefont{Follana}, \bibfnamefont{E.}}, \emph{et~al.}
  (\bibinfo{collaboration}{HPQCD}), \bibinfo{year}{2007},
  \bibinfo{journal}{Phys. Rev.} \textbf{\bibinfo{volume}{D75}},
  \bibinfo{pages}{054502}.

\bibitem[{Fonseca \emph{et~al.}(1984)\citenamefont{Fonseca}
  \emph{et~al.}}]{Fonseca:1984fg}
\bibinfo{author}{\bibnamefont{Fonseca}, \bibfnamefont{V.}}, \emph{et~al.}
  (\bibinfo{collaboration}{CUSB}), \bibinfo{year}{1984},
  \bibinfo{journal}{Nucl. Phys.} \textbf{\bibinfo{volume}{B242}},
  \bibinfo{pages}{31}.

\bibitem[{\citenamefont{Fulcher}(1990)}]{Fulcher:1990kx}
\bibinfo{author}{\bibnamefont{Fulcher}, \bibfnamefont{L.~P.}},
  \bibinfo{year}{1990}, \bibinfo{journal}{Phys. Rev.}
  \textbf{\bibinfo{volume}{D42}}, \bibinfo{pages}{2337}.

\bibitem[{\citenamefont{Fulcher}(1991)}]{Fulcher:1991dm}
\bibinfo{author}{\bibnamefont{Fulcher}, \bibfnamefont{L.~P.}},
  \bibinfo{year}{1991}, \bibinfo{journal}{Phys. Rev.}
  \textbf{\bibinfo{volume}{D44}}, \bibinfo{pages}{2079}.

\bibitem[{\citenamefont{Fulcher}(1999)}]{Fulcher:1998ka}
\bibinfo{author}{\bibnamefont{Fulcher}, \bibfnamefont{L.~P.}},
  \bibinfo{year}{1999}, \bibinfo{journal}{Phys. Rev.}
  \textbf{\bibinfo{volume}{D60}}, \bibinfo{pages}{074006}.

\bibitem[{Gaiser \emph{et~al.}(1986)\citenamefont{Gaiser}
  \emph{et~al.}}]{Gaiser:1985ix}
\bibinfo{author}{\bibnamefont{Gaiser}, \bibfnamefont{J.}}, \emph{et~al.},
  \bibinfo{year}{1986}, \bibinfo{journal}{Phys. Rev.}
  \textbf{\bibinfo{volume}{D34}}, \bibinfo{pages}{711}.

\bibitem[{\citenamefont{Godfrey}(2006)}]{Godfrey:2006pd}
\bibinfo{author}{\bibnamefont{Godfrey}, \bibfnamefont{S.}},
  \bibinfo{year}{2006}, in \emph{\bibinfo{booktitle}{Proceedings of the Fourth
  International Conference on Flavor Physics and CP Violation (FPCP 2006)}}
  (\bibinfo{address}{Vancouver, B.C., Canada, April 9-12, 2006}),
  \bibinfo{note}{eConf {\bf C060409}, 221 (2006)}, \eprint{hep-ph/0605152}.

\bibitem[{\citenamefont{Godfrey and Isgur}(1985)}]{Godfrey:1985xj}
\bibinfo{author}{\bibnamefont{Godfrey}, \bibfnamefont{S.}}, and
  \bibinfo{author}{\bibfnamefont{N.}~\bibnamefont{Isgur}},
  \bibinfo{year}{1985}, \bibinfo{journal}{Phys. Rev.}
  \textbf{\bibinfo{volume}{D32}}, \bibinfo{pages}{189}.

\bibitem[{\citenamefont{Godfrey} \emph{et~al.}(1986)\citenamefont{Godfrey,
  Karl, and O'Donnell}}]{Godfrey:1985ei}
\bibinfo{author}{\bibnamefont{Godfrey}, \bibfnamefont{S.}},
  \bibinfo{author}{\bibfnamefont{G.}~\bibnamefont{Karl}}, and
  \bibinfo{author}{\bibfnamefont{P.~J.} \bibnamefont{O'Donnell}},
  \bibinfo{year}{1986}, \bibinfo{journal}{Z. Phys.}
  \textbf{\bibinfo{volume}{C31}}, \bibinfo{pages}{77}.

\bibitem[{\citenamefont{Godfrey and
  Rosner}(2001{\natexlab{a}})}]{Godfrey:2001vc}
\bibinfo{author}{\bibnamefont{Godfrey}, \bibfnamefont{S.}}, and
  \bibinfo{author}{\bibfnamefont{J.~L.} \bibnamefont{Rosner}},
  \bibinfo{year}{2001}{\natexlab{a}}, \bibinfo{journal}{Phys. Rev.}
  \textbf{\bibinfo{volume}{D64}}, \bibinfo{pages}{097501},
  \bibinfo{note}{[Erratum-ibid.\ {\bf D66}, 059902 (2002)]}.

\bibitem[{\citenamefont{Godfrey and
  Rosner}(2001{\natexlab{b}})}]{Godfrey:2001eb}
\bibinfo{author}{\bibnamefont{Godfrey}, \bibfnamefont{S.}}, and
  \bibinfo{author}{\bibfnamefont{J.~L.} \bibnamefont{Rosner}},
  \bibinfo{year}{2001}{\natexlab{b}}, \bibinfo{journal}{Phys. Rev.}
  \textbf{\bibinfo{volume}{D64}}, \bibinfo{pages}{074011},
  \bibinfo{note}{[Erratum-ibid.\ {\bf D65}, 039901 (2002)]}.

\bibitem[{\citenamefont{Godfrey and Rosner}(2002)}]{Godfrey:2002rp}
\bibinfo{author}{\bibnamefont{Godfrey}, \bibfnamefont{S.}}, and
  \bibinfo{author}{\bibfnamefont{J.~L.} \bibnamefont{Rosner}},
  \bibinfo{year}{2002}, \bibinfo{journal}{Phys. Rev.}
  \textbf{\bibinfo{volume}{D66}}, \bibinfo{pages}{014012}.

\bibitem[{Gokhroo \emph{et~al.}(2006)\citenamefont{Gokhroo}
  \emph{et~al.}}]{Gokhroo:2006bt}
\bibinfo{author}{\bibnamefont{Gokhroo}, \bibfnamefont{G.}}, \emph{et~al.}
  (\bibinfo{collaboration}{Belle}), \bibinfo{year}{2006},
  \bibinfo{journal}{Phys. Rev. Lett.} \textbf{\bibinfo{volume}{97}},
  \bibinfo{pages}{162002}.

\bibitem[{\citenamefont{Gottfried}(1978)}]{Gottfried:1977gp}
\bibinfo{author}{\bibnamefont{Gottfried}, \bibfnamefont{K.}},
  \bibinfo{year}{1978}, \bibinfo{journal}{Phys. Rev. Lett.}
  \textbf{\bibinfo{volume}{40}}, \bibinfo{pages}{598}.

\bibitem[{\citenamefont{Gowdy}(2006)}]{Gowdy:2006be}
\bibinfo{author}{\bibnamefont{Gowdy}, \bibfnamefont{S.~J.}},
  \bibinfo{year}{2006}, in \emph{\bibinfo{booktitle}{41th Rencontres de
  Moriond: QCD and Hadronic Interactions}} (\bibinfo{address}{La Thuile, Aosta
  Valley, March 18-25, 2006}), \eprint{hep-ex/0605086}.

\bibitem[{\citenamefont{Grant and Rosner}(1992)}]{Grant:1992fi}
\bibinfo{author}{\bibnamefont{Grant}, \bibfnamefont{A.}}, and
  \bibinfo{author}{\bibfnamefont{J.~L.} \bibnamefont{Rosner}},
  \bibinfo{year}{1992}, \bibinfo{journal}{Phys. Rev.}
  \textbf{\bibinfo{volume}{D46}}, \bibinfo{pages}{3862}.

\bibitem[{Gray \emph{et~al.}(2005)\citenamefont{Gray}
  \emph{et~al.}}]{Gray:2005ur}
\bibinfo{author}{\bibnamefont{Gray}, \bibfnamefont{A.}}, \emph{et~al.},
  \bibinfo{year}{2005}, \bibinfo{journal}{Phys. Rev.}
  \textbf{\bibinfo{volume}{D72}}, \bibinfo{pages}{094507}.

\bibitem[{\citenamefont{Grosse and Martin}(1980)}]{Grosse:1979xm}
\bibinfo{author}{\bibnamefont{Grosse}, \bibfnamefont{H.}}, and
  \bibinfo{author}{\bibfnamefont{A.}~\bibnamefont{Martin}},
  \bibinfo{year}{1980}, \bibinfo{journal}{Phys. Rept.}
  \textbf{\bibinfo{volume}{60}}, \bibinfo{pages}{341}.

\bibitem[{\citenamefont{Grotch} \emph{et~al.}(1984)\citenamefont{Grotch, Owen,
  and Sebastian}}]{Grotch:1984gf}
\bibinfo{author}{\bibnamefont{Grotch}, \bibfnamefont{H.}},
  \bibinfo{author}{\bibfnamefont{D.~A.} \bibnamefont{Owen}}, and
  \bibinfo{author}{\bibfnamefont{K.~J.} \bibnamefont{Sebastian}},
  \bibinfo{year}{1984}, \bibinfo{journal}{Phys. Rev.}
  \textbf{\bibinfo{volume}{D30}}, \bibinfo{pages}{1924}.

\bibitem[{\citenamefont{Grotch and Sebastian}(1982)}]{Grotch:1982bi}
\bibinfo{author}{\bibnamefont{Grotch}, \bibfnamefont{H.}}, and
  \bibinfo{author}{\bibfnamefont{K.~J.} \bibnamefont{Sebastian}},
  \bibinfo{year}{1982}, \bibinfo{journal}{Phys. Rev.}
  \textbf{\bibinfo{volume}{D25}}, \bibinfo{pages}{2944}.

\bibitem[{\citenamefont{Gupta and Johnson}(1996)}]{Gupta:1995ps}
\bibinfo{author}{\bibnamefont{Gupta}, \bibfnamefont{S.~N.}}, and
  \bibinfo{author}{\bibfnamefont{J.~M.} \bibnamefont{Johnson}},
  \bibinfo{year}{1996}, \bibinfo{journal}{Phys. Rev.}
  \textbf{\bibinfo{volume}{D53}}, \bibinfo{pages}{312}.

\bibitem[{\citenamefont{Gupta} \emph{et~al.}(1986)\citenamefont{Gupta, Radford,
  and Repko}}]{Gupta:1986xt}
\bibinfo{author}{\bibnamefont{Gupta}, \bibfnamefont{S.~N.}},
  \bibinfo{author}{\bibfnamefont{S.~F.} \bibnamefont{Radford}}, and
  \bibinfo{author}{\bibfnamefont{W.~W.} \bibnamefont{Repko}},
  \bibinfo{year}{1986}, \bibinfo{journal}{Phys. Rev.}
  \textbf{\bibinfo{volume}{D34}}, \bibinfo{pages}{201}.

\bibitem[{He \emph{et~al.}(2005)\citenamefont{He} \emph{et~al.}}]{He:2005bs}
\bibinfo{author}{\bibnamefont{He}, \bibfnamefont{Q.}}, \emph{et~al.}
  (\bibinfo{collaboration}{CLEO}), \bibinfo{year}{2005},
  \bibinfo{journal}{Phys. Rev. Lett.} \textbf{\bibinfo{volume}{95}},
  \bibinfo{pages}{121801}, \bibinfo{note}{[Erratum-ibid.\ {\bf 96}, 199903
  (2006)]}.

\bibitem[{He \emph{et~al.}(2006)\citenamefont{He} \emph{et~al.}}]{He:2006kg}
\bibinfo{author}{\bibnamefont{He}, \bibfnamefont{Q.}}, \emph{et~al.}
  (\bibinfo{collaboration}{CLEO}), \bibinfo{year}{2006},
  \bibinfo{journal}{Phys. Rev.} \textbf{\bibinfo{volume}{D74}},
  \bibinfo{pages}{091104}.

\bibitem[{\citenamefont{Heltsley}(2006)}]{Heltsley:2006QWG}
\bibinfo{author}{\bibnamefont{Heltsley}, \bibfnamefont{B.}},
  \bibinfo{year}{2006}, in \emph{\bibinfo{booktitle}{Quarkonium Working Group
  Meeting}} (\bibinfo{address}{Brookhaven Natl.\ Lab., June 27-30, 2006}),
  \bibinfo{note}{[unpublished]}.

\bibitem[{Huang \emph{et~al.}(2006)\citenamefont{Huang}
  \emph{et~al.}}]{Huang:2005fx}
\bibinfo{author}{\bibnamefont{Huang}, \bibfnamefont{G.~S.}}, \emph{et~al.}
  (\bibinfo{collaboration}{CLEO}), \bibinfo{year}{2006},
  \bibinfo{journal}{Phys. Rev. Lett.} \textbf{\bibinfo{volume}{96}},
  \bibinfo{pages}{032003}.

\bibitem[{\citenamefont{Ioffe and Shifman}(1980)}]{Ioffe:1980mx}
\bibinfo{author}{\bibnamefont{Ioffe}, \bibfnamefont{B.~L.}}, and
  \bibinfo{author}{\bibfnamefont{M.~A.} \bibnamefont{Shifman}},
  \bibinfo{year}{1980}, \bibinfo{journal}{Phys. Lett.}
  \textbf{\bibinfo{volume}{B95}}, \bibinfo{pages}{99}.

\bibitem[{\citenamefont{Ioffe and Shifman}(1981)}]{Ioffe:1981qa}
\bibinfo{author}{\bibnamefont{Ioffe}, \bibfnamefont{B.~L.}}, and
  \bibinfo{author}{\bibfnamefont{M.~A.} \bibnamefont{Shifman}},
  \bibinfo{year}{1981}, \bibinfo{journal}{Phys. Lett.}
  \textbf{\bibinfo{volume}{B107}}, \bibinfo{pages}{371}.

\bibitem[{\citenamefont{Kang and Sucher}(1978)}]{Kang:1978yw}
\bibinfo{author}{\bibnamefont{Kang}, \bibfnamefont{J.~S.}}, and
  \bibinfo{author}{\bibfnamefont{J.}~\bibnamefont{Sucher}},
  \bibinfo{year}{1978}, \bibinfo{journal}{Phys. Rev.}
  \textbf{\bibinfo{volume}{D18}}, \bibinfo{pages}{2698}.

\bibitem[{\citenamefont{Karl} \emph{et~al.}(1976)\citenamefont{Karl, Meshkov,
  and Rosner}}]{Karl:1975jp}
\bibinfo{author}{\bibnamefont{Karl}, \bibfnamefont{G.}},
  \bibinfo{author}{\bibfnamefont{S.}~\bibnamefont{Meshkov}}, and
  \bibinfo{author}{\bibfnamefont{J.~L.} \bibnamefont{Rosner}},
  \bibinfo{year}{1976}, \bibinfo{journal}{Phys. Rev.}
  \textbf{\bibinfo{volume}{D13}}, \bibinfo{pages}{1203}.

\bibitem[{\citenamefont{Karl} \emph{et~al.}(1980)\citenamefont{Karl, Meshkov,
  and Rosner}}]{Karl:1980wm}
\bibinfo{author}{\bibnamefont{Karl}, \bibfnamefont{G.}},
  \bibinfo{author}{\bibfnamefont{S.}~\bibnamefont{Meshkov}}, and
  \bibinfo{author}{\bibfnamefont{J.~L.} \bibnamefont{Rosner}},
  \bibinfo{year}{1980}, \bibinfo{journal}{Phys. Rev. Lett.}
  \textbf{\bibinfo{volume}{45}}, \bibinfo{pages}{215}.

\bibitem[{\citenamefont{Kluth}(2006)}]{Kluth:2006vf}
\bibinfo{author}{\bibnamefont{Kluth}, \bibfnamefont{S.}}, \bibinfo{year}{2006},
  in \emph{\bibinfo{booktitle}{33th International Conference on High Energy
  Physics (ICHEP06)}} (\bibinfo{address}{Moscow, Russia, July 26 - August 2,
  2006}), \eprint{hep-ex/0609020}.

\bibitem[{\citenamefont{Koma} \emph{et~al.}(2006)\citenamefont{Koma, Koma, and
  Wittig}}]{Koma:2005nq}
\bibinfo{author}{\bibnamefont{Koma}, \bibfnamefont{M.}},
  \bibinfo{author}{\bibfnamefont{Y.}~\bibnamefont{Koma}}, and
  \bibinfo{author}{\bibfnamefont{H.}~\bibnamefont{Wittig}},
  \bibinfo{year}{2006}, \bibinfo{journal}{PoS}
  \textbf{\bibinfo{volume}{LAT2005}}, \bibinfo{pages}{216}.

\bibitem[{\citenamefont{Koma and Koma}(2007)}]{Koma:2006fw}
\bibinfo{author}{\bibnamefont{Koma}, \bibfnamefont{Y.}}, and
  \bibinfo{author}{\bibfnamefont{M.}~\bibnamefont{Koma}}, \bibinfo{year}{2007},
  \bibinfo{journal}{Nucl. Phys.} \textbf{\bibinfo{volume}{B769}},
  \bibinfo{pages}{79}.

\bibitem[{\citenamefont{Kou and Pene}(2005)}]{Kou:2005gt}
\bibinfo{author}{\bibnamefont{Kou}, \bibfnamefont{E.}}, and
  \bibinfo{author}{\bibfnamefont{O.}~\bibnamefont{Pene}}, \bibinfo{year}{2005},
  \bibinfo{journal}{Phys. Lett.} \textbf{\bibinfo{volume}{B631}},
  \bibinfo{pages}{164}.

\bibitem[{\citenamefont{Kravchenko}(2006)}]{Kravchenko:2006qx}
\bibinfo{author}{\bibnamefont{Kravchenko}, \bibfnamefont{I.}}
  (\bibinfo{collaboration}{CDF}), \bibinfo{year}{2006}, in
  \emph{\bibinfo{booktitle}{Proceedings of the Fourth International Conference
  on Flavor Physics and CP Violation (FPCP 2006)}}
  (\bibinfo{address}{Vancouver, B.C., Canada, April 9-12, 2006}),
  \bibinfo{note}{eConf {\bf C060400},222 (2006)}, \eprint{hep-ex/0605076}.

\bibitem[{\citenamefont{Kreinick}(2007)}]{Kreinick:2007gh}
\bibinfo{author}{\bibnamefont{Kreinick}, \bibfnamefont{D.}},
  \bibinfo{year}{2007}, in \emph{\bibinfo{booktitle}{Proceedings of the Charm
  2007 Workshop}} (\bibinfo{address}{Cornell University, August 5-8, 2007}),
  \bibinfo{note}{eConf {\bf C070805}}, \eprint{arXiv:0710.5929 [hep-ex]}.

\bibitem[{\citenamefont{Kuang}(2002)}]{Kuang:2002hz}
\bibinfo{author}{\bibnamefont{Kuang}, \bibfnamefont{Y.-P.}},
  \bibinfo{year}{2002}, \bibinfo{journal}{Phys. Rev.}
  \textbf{\bibinfo{volume}{D65}}, \bibinfo{pages}{094024}.

\bibitem[{\citenamefont{Kuang}(2006)}]{Kuang:2006me}
\bibinfo{author}{\bibnamefont{Kuang}, \bibfnamefont{Y.-P.}},
  \bibinfo{year}{2006}, \bibinfo{journal}{Front. Phys. China}
  \textbf{\bibinfo{volume}{1}}, \bibinfo{pages}{19},
  \bibinfo{note}{[arXiv:hep-ph/0601044]}.

\bibitem[{\citenamefont{Kuang} \emph{et~al.}(1988)\citenamefont{Kuang, Tuan,
  and Yan}}]{Kuang:1988bz}
\bibinfo{author}{\bibnamefont{Kuang}, \bibfnamefont{Y.-P.}},
  \bibinfo{author}{\bibfnamefont{S.~F.} \bibnamefont{Tuan}}, and
  \bibinfo{author}{\bibfnamefont{T.-M.} \bibnamefont{Yan}},
  \bibinfo{year}{1988}, \bibinfo{journal}{Phys. Rev.}
  \textbf{\bibinfo{volume}{D37}}, \bibinfo{pages}{1210}.

\bibitem[{\citenamefont{Kuang and Yan}(1981)}]{Kuang:1981se}
\bibinfo{author}{\bibnamefont{Kuang}, \bibfnamefont{Y.-P.}}, and
  \bibinfo{author}{\bibfnamefont{T.-M.} \bibnamefont{Yan}},
  \bibinfo{year}{1981}, \bibinfo{journal}{Phys. Rev.}
  \textbf{\bibinfo{volume}{D24}}, \bibinfo{pages}{2874}.

\bibitem[{\citenamefont{Kuang and Yan}(1990)}]{Kuang:1989ub}
\bibinfo{author}{\bibnamefont{Kuang}, \bibfnamefont{Y.-P.}}, and
  \bibinfo{author}{\bibfnamefont{T.-M.} \bibnamefont{Yan}},
  \bibinfo{year}{1990}, \bibinfo{journal}{Phys. Rev.}
  \textbf{\bibinfo{volume}{D41}}, \bibinfo{pages}{155}.

\bibitem[{\citenamefont{Kwong} \emph{et~al.}(1988)\citenamefont{Kwong,
  Mackenzie, Rosenfeld, and Rosner}}]{Kwong:1987ak}
\bibinfo{author}{\bibnamefont{Kwong}, \bibfnamefont{W.}},
  \bibinfo{author}{\bibfnamefont{P.~B.} \bibnamefont{Mackenzie}},
  \bibinfo{author}{\bibfnamefont{R.}~\bibnamefont{Rosenfeld}}, and
  \bibinfo{author}{\bibfnamefont{J.~L.} \bibnamefont{Rosner}},
  \bibinfo{year}{1988}, \bibinfo{journal}{Phys. Rev.}
  \textbf{\bibinfo{volume}{D37}}, \bibinfo{pages}{3210}.

\bibitem[{\citenamefont{Kwong} \emph{et~al.}(1987)\citenamefont{Kwong, Quigg,
  and Rosner}}]{Kwong:1987mj}
\bibinfo{author}{\bibnamefont{Kwong}, \bibfnamefont{W.}},
  \bibinfo{author}{\bibfnamefont{C.}~\bibnamefont{Quigg}}, and
  \bibinfo{author}{\bibfnamefont{J.~L.} \bibnamefont{Rosner}},
  \bibinfo{year}{1987}, \bibinfo{journal}{Ann. Rev. Nucl. Part. Sci.}
  \textbf{\bibinfo{volume}{37}}, \bibinfo{pages}{325}.

\bibitem[{\citenamefont{Kwong and Rosner}(1986)}]{Kwong:1985ti}
\bibinfo{author}{\bibnamefont{Kwong}, \bibfnamefont{W.}}, and
  \bibinfo{author}{\bibfnamefont{J.~L.} \bibnamefont{Rosner}},
  \bibinfo{year}{1986}, \bibinfo{journal}{Prog. Theor. Phys. Suppl.}
  \textbf{\bibinfo{volume}{86}}, \bibinfo{pages}{366}.

\bibitem[{\citenamefont{Kwong and Rosner}(1988)}]{Kwong:1988ae}
\bibinfo{author}{\bibnamefont{Kwong}, \bibfnamefont{W.}}, and
  \bibinfo{author}{\bibfnamefont{J.~L.} \bibnamefont{Rosner}},
  \bibinfo{year}{1988}, \bibinfo{journal}{Phys. Rev.}
  \textbf{\bibinfo{volume}{D38}}, \bibinfo{pages}{279}.

\bibitem[{\citenamefont{Lacock} \emph{et~al.}(1997)\citenamefont{Lacock,
  Michael, Boyle, and Rowland}}]{Lacock:1996ny}
\bibinfo{author}{\bibnamefont{Lacock}, \bibfnamefont{P.}},
  \bibinfo{author}{\bibfnamefont{C.}~\bibnamefont{Michael}},
  \bibinfo{author}{\bibfnamefont{P.}~\bibnamefont{Boyle}}, and
  \bibinfo{author}{\bibfnamefont{P.}~\bibnamefont{Rowland}}
  (\bibinfo{collaboration}{UKQCD}), \bibinfo{year}{1997},
  \bibinfo{journal}{Phys. Lett.} \textbf{\bibinfo{volume}{B401}},
  \bibinfo{pages}{308}.

\bibitem[{\citenamefont{Lahde}(2003)}]{Lahde:2002wj}
\bibinfo{author}{\bibnamefont{Lahde}, \bibfnamefont{T.~A.}},
  \bibinfo{year}{2003}, \bibinfo{journal}{Nucl. Phys.}
  \textbf{\bibinfo{volume}{A714}}, \bibinfo{pages}{183}.

\bibitem[{\citenamefont{Lepage}(1983)}]{Lepage:1983tk}
\bibinfo{author}{\bibnamefont{Lepage}, \bibfnamefont{G.~P.}},
  \bibinfo{year}{1983}, in \emph{\bibinfo{booktitle}{1983 International
  Symposium on Lepton-Photon Interactions at High Energies}}, edited by
  \bibinfo{editor}{\bibfnamefont{D.}~\bibnamefont{Cassel}} and
  \bibinfo{editor}{\bibfnamefont{D.}~\bibnamefont{Kreinick}}
  (\bibinfo{publisher}{Cornell Univ. Lab. Nucl. Studies, Ithaca, NY, 1983}).

\bibitem[{\citenamefont{Lepage}(2005)}]{Lepage:2005eg}
\bibinfo{author}{\bibnamefont{Lepage}, \bibfnamefont{G.~P.}},
  \bibinfo{year}{2005}, \bibinfo{journal}{Annals Phys.}
  \textbf{\bibinfo{volume}{315}}, \bibinfo{pages}{193}.

\bibitem[{Li \emph{et~al.}(2005)\citenamefont{Li} \emph{et~al.}}]{Li:2005uga}
\bibinfo{author}{\bibnamefont{Li}, \bibfnamefont{Z.}}, \emph{et~al.}
  (\bibinfo{collaboration}{CLEO}), \bibinfo{year}{2005},
  \bibinfo{journal}{Phys. Rev.} \textbf{\bibinfo{volume}{D71}},
  \bibinfo{pages}{111103}.

\bibitem[{\citenamefont{Liao and Manke}(2002)}]{Liao:2002rj}
\bibinfo{author}{\bibnamefont{Liao}, \bibfnamefont{X.}}, and
  \bibinfo{author}{\bibfnamefont{T.}~\bibnamefont{Manke}},
  \bibinfo{year}{2002}, \eprint{hep-lat/0210030}.

\bibitem[{\citenamefont{Llanes-Estrada}(2005)}]{LlanesEstrada:2005hz}
\bibinfo{author}{\bibnamefont{Llanes-Estrada}, \bibfnamefont{F.~J.}},
  \bibinfo{year}{2005}, \bibinfo{journal}{Phys. Rev.}
  \textbf{\bibinfo{volume}{D72}}, \bibinfo{pages}{031503}.

\bibitem[{\citenamefont{Luke and Manohar}(1997)}]{Luke:1996hj}
\bibinfo{author}{\bibnamefont{Luke}, \bibfnamefont{M.~E.}}, and
  \bibinfo{author}{\bibfnamefont{A.~V.} \bibnamefont{Manohar}},
  \bibinfo{year}{1997}, \bibinfo{journal}{Phys. Rev.}
  \textbf{\bibinfo{volume}{D55}}, \bibinfo{pages}{4129}.

\bibitem[{MacKay \emph{et~al.}(1984)\citenamefont{MacKay}
  \emph{et~al.}}]{MacKay:1984kv}
\bibinfo{author}{\bibnamefont{MacKay}, \bibfnamefont{W.~W.}}, \emph{et~al.}
  (\bibinfo{collaboration}{CUSB}), \bibinfo{year}{1984},
  \bibinfo{journal}{Phys. Rev.} \textbf{\bibinfo{volume}{D29}},
  \bibinfo{pages}{2483}.

\bibitem[{\citenamefont{Maiani}
  \emph{et~al.}(2005{\natexlab{a}})\citenamefont{Maiani, Piccinini, Polosa, and
  Riquer}}]{Maiani:2004vq}
\bibinfo{author}{\bibnamefont{Maiani}, \bibfnamefont{L.}},
  \bibinfo{author}{\bibfnamefont{F.}~\bibnamefont{Piccinini}},
  \bibinfo{author}{\bibfnamefont{A.~D.} \bibnamefont{Polosa}}, and
  \bibinfo{author}{\bibfnamefont{V.}~\bibnamefont{Riquer}},
  \bibinfo{year}{2005}{\natexlab{a}}, \bibinfo{journal}{Phys. Rev.}
  \textbf{\bibinfo{volume}{D71}}, \bibinfo{pages}{014028}.

\bibitem[{\citenamefont{Maiani}
  \emph{et~al.}(2005{\natexlab{b}})\citenamefont{Maiani, Riquer, Piccinini, and
  Polosa}}]{Maiani:2005pe}
\bibinfo{author}{\bibnamefont{Maiani}, \bibfnamefont{L.}},
  \bibinfo{author}{\bibfnamefont{V.}~\bibnamefont{Riquer}},
  \bibinfo{author}{\bibfnamefont{F.}~\bibnamefont{Piccinini}}, and
  \bibinfo{author}{\bibfnamefont{A.~D.} \bibnamefont{Polosa}},
  \bibinfo{year}{2005}{\natexlab{b}}, \bibinfo{journal}{Phys. Rev.}
  \textbf{\bibinfo{volume}{D72}}, \bibinfo{pages}{031502}.

\bibitem[{\citenamefont{Maltman}(1991)}]{Maltman:1990mp}
\bibinfo{author}{\bibnamefont{Maltman}, \bibfnamefont{K.}},
  \bibinfo{year}{1991}, \bibinfo{journal}{Phys. Rev.}
  \textbf{\bibinfo{volume}{D44}}, \bibinfo{pages}{751}.

\bibitem[{\citenamefont{Maltoni}(2000)}]{Maltoni:2000km}
\bibinfo{author}{\bibnamefont{Maltoni}, \bibfnamefont{F.}},
  \bibinfo{year}{2000}, in \emph{\bibinfo{booktitle}{Proceedings of the 5th
  Workshop on Quantum Chromodynamics (QCD 2000)}}, edited by
  \bibinfo{editor}{\bibfnamefont{H.~M.} \bibnamefont{Fried}},
  \bibinfo{editor}{\bibfnamefont{B.}~\bibnamefont{Muller}}, and
  \bibinfo{editor}{\bibfnamefont{Y.}~\bibnamefont{Gabellini}}
  (\bibinfo{publisher}{World Scientific, Singapore, 2000}),
  \eprint{hep-ph/0007003}.

\bibitem[{Manke \emph{et~al.}(2000)\citenamefont{Manke}
  \emph{et~al.}}]{Manke:2000dg}
\bibinfo{author}{\bibnamefont{Manke}, \bibfnamefont{T.}}, \emph{et~al.}
  (\bibinfo{collaboration}{CP-PACS}), \bibinfo{year}{2000},
  \bibinfo{journal}{Phys. Rev.} \textbf{\bibinfo{volume}{D62}},
  \bibinfo{pages}{114508}.

\bibitem[{\citenamefont{Marsiske}(2006)}]{Marsiske:2006mh}
\bibinfo{author}{\bibnamefont{Marsiske}, \bibfnamefont{H.}},
  \bibinfo{year}{2006}, in \emph{\bibinfo{booktitle}{Proceedings of the Fourth
  International Conference on Flavor Physics and CP Violation (FPCP 2006)}}
  (\bibinfo{address}{Vancouver, B.C., Canada, April 9-12, 2006}),
  \bibinfo{note}{eConf {\bf C060409}, 211 (2006)}, \eprint{hep-ex/0605117}.

\bibitem[{\citenamefont{Martin and Richard}(1982)}]{Martin:1982nw}
\bibinfo{author}{\bibnamefont{Martin}, \bibfnamefont{A.}}, and
  \bibinfo{author}{\bibfnamefont{J.~M.} \bibnamefont{Richard}},
  \bibinfo{year}{1982}, \bibinfo{journal}{Phys. Lett.}
  \textbf{\bibinfo{volume}{B115}}, \bibinfo{pages}{323}.

\bibitem[{\citenamefont{McClary and Byers}(1983)}]{McClary:1983xw}
\bibinfo{author}{\bibnamefont{McClary}, \bibfnamefont{R.}}, and
  \bibinfo{author}{\bibfnamefont{N.}~\bibnamefont{Byers}},
  \bibinfo{year}{1983}, \bibinfo{journal}{Phys. Rev.}
  \textbf{\bibinfo{volume}{D28}}, \bibinfo{pages}{1692}.

\bibitem[{\citenamefont{McNeile} \emph{et~al.}(2002)\citenamefont{McNeile,
  Michael, and Pennanen}}]{McNeile:2002az}
\bibinfo{author}{\bibnamefont{McNeile}, \bibfnamefont{C.}},
  \bibinfo{author}{\bibfnamefont{C.}~\bibnamefont{Michael}}, and
  \bibinfo{author}{\bibfnamefont{P.}~\bibnamefont{Pennanen}}
  (\bibinfo{collaboration}{UKQCD}), \bibinfo{year}{2002},
  \bibinfo{journal}{Phys. Rev.} \textbf{\bibinfo{volume}{D65}},
  \bibinfo{pages}{094505}.

\bibitem[{\citenamefont{Messiah}(1999)}]{Messiah:Vol2}
\bibinfo{author}{\bibnamefont{Messiah}, \bibfnamefont{A.}},
  \bibinfo{year}{1999}, \emph{\bibinfo{title}{Quantum Mechanics}},
  volume~\bibinfo{volume}{2} (\bibinfo{publisher}{Dover Publications, Inc.},
  \bibinfo{address}{Mineola, NY, USA}), \bibinfo{note}{p.554}.

\bibitem[{\citenamefont{Metreveli}(2007)}]{Metreveli:2007sj}
\bibinfo{author}{\bibnamefont{Metreveli}, \bibfnamefont{Z.}},
  \bibinfo{year}{2007}, in \emph{\bibinfo{booktitle}{Proceedings of the Charm
  2007 Workshop}} (\bibinfo{address}{Cornell University, August 5-8, 2007}),
  \bibinfo{note}{eConf {\bf C070805}}, \eprint{arXiv:0710.1884 [hep-ex]}.

\bibitem[{\citenamefont{Miller} \emph{et~al.}(1990)\citenamefont{Miller,
  Nefkens, and Slaus}}]{Miller:1990iz}
\bibinfo{author}{\bibnamefont{Miller}, \bibfnamefont{G.~A.}},
  \bibinfo{author}{\bibfnamefont{B.~M.~K.} \bibnamefont{Nefkens}}, and
  \bibinfo{author}{\bibfnamefont{I.}~\bibnamefont{Slaus}},
  \bibinfo{year}{1990}, \bibinfo{journal}{Phys. Rept.}
  \textbf{\bibinfo{volume}{194}}, \bibinfo{pages}{1}.

\bibitem[{\citenamefont{Moxhay and Rosner}(1983)}]{Moxhay:1983vu}
\bibinfo{author}{\bibnamefont{Moxhay}, \bibfnamefont{P.}}, and
  \bibinfo{author}{\bibfnamefont{J.~L.} \bibnamefont{Rosner}},
  \bibinfo{year}{1983}, \bibinfo{journal}{Phys. Rev.}
  \textbf{\bibinfo{volume}{D28}}, \bibinfo{pages}{1132}.

\bibitem[{\citenamefont{Novikov} \emph{et~al.}(1978)\citenamefont{Novikov,
  Okun, Shifman, Vainshtein, Voloshin, and Zakharov}}]{Novikov:1977dq}
\bibinfo{author}{\bibnamefont{Novikov}, \bibfnamefont{V.~A.}},
  \bibinfo{author}{\bibfnamefont{L.~B.} \bibnamefont{Okun}},
  \bibinfo{author}{\bibfnamefont{M.~A.} \bibnamefont{Shifman}},
  \bibinfo{author}{\bibfnamefont{A.~I.} \bibnamefont{Vainshtein}},
  \bibinfo{author}{\bibfnamefont{M.~B.} \bibnamefont{Voloshin}}, and
  \bibinfo{author}{\bibfnamefont{V.~I.} \bibnamefont{Zakharov}},
  \bibinfo{year}{1978}, \bibinfo{journal}{Phys. Rept.}
  \textbf{\bibinfo{volume}{41}}, \bibinfo{pages}{1}.

\bibitem[{Okamoto \emph{et~al.}(2002)\citenamefont{Okamoto}
  \emph{et~al.}}]{Okamoto:2001jb}
\bibinfo{author}{\bibnamefont{Okamoto}, \bibfnamefont{M.}}, \emph{et~al.}
  (\bibinfo{collaboration}{CP-PACS}), \bibinfo{year}{2002},
  \bibinfo{journal}{Phys. Rev.} \textbf{\bibinfo{volume}{D65}},
  \bibinfo{pages}{094508}.

\bibitem[{Oreglia \emph{et~al.}(1982)\citenamefont{Oreglia}
  \emph{et~al.}}]{Oreglia:1981fx}
\bibinfo{author}{\bibnamefont{Oreglia}, \bibfnamefont{M.}}, \emph{et~al.}
  (\bibinfo{collaboration}{Crystal Ball}), \bibinfo{year}{1982},
  \bibinfo{journal}{Phys. Rev.} \textbf{\bibinfo{volume}{D25}},
  \bibinfo{pages}{2259}.

\bibitem[{Pakhlova \emph{et~al.}(2008)\citenamefont{Pakhlova}
  \emph{et~al.}}]{Pakhlova:2007fq}
\bibinfo{author}{\bibnamefont{Pakhlova}, \bibfnamefont{G.}}, \emph{et~al.}
  (\bibinfo{collaboration}{Belle}), \bibinfo{year}{2008},
  \bibinfo{journal}{Phys. Rev. Lett.} \textbf{\bibinfo{volume}{100}},
  \bibinfo{pages}{062001}.

\bibitem[{Pedlar \emph{et~al.}(2005)\citenamefont{Pedlar}
  \emph{et~al.}}]{Pedlar:2005px}
\bibinfo{author}{\bibnamefont{Pedlar}, \bibfnamefont{T.~K.}}, \emph{et~al.}
  (\bibinfo{collaboration}{CLEO}), \bibinfo{year}{2005},
  \bibinfo{journal}{Phys. Rev.} \textbf{\bibinfo{volume}{D72}},
  \bibinfo{pages}{051108}.

\bibitem[{\citenamefont{Peskin}(1979)}]{Peskin:1979va}
\bibinfo{author}{\bibnamefont{Peskin}, \bibfnamefont{M.~E.}},
  \bibinfo{year}{1979}, \bibinfo{journal}{Nucl. Phys.}
  \textbf{\bibinfo{volume}{B156}}, \bibinfo{pages}{365}.

\bibitem[{\citenamefont{Petrelli} \emph{et~al.}(1998)\citenamefont{Petrelli,
  Cacciari, Greco, Maltoni, and Mangano}}]{Petrelli:1997ge}
\bibinfo{author}{\bibnamefont{Petrelli}, \bibfnamefont{A.}},
  \bibinfo{author}{\bibfnamefont{M.}~\bibnamefont{Cacciari}},
  \bibinfo{author}{\bibfnamefont{M.}~\bibnamefont{Greco}},
  \bibinfo{author}{\bibfnamefont{F.}~\bibnamefont{Maltoni}}, and
  \bibinfo{author}{\bibfnamefont{M.~L.} \bibnamefont{Mangano}},
  \bibinfo{year}{1998}, \bibinfo{journal}{Nucl. Phys.}
  \textbf{\bibinfo{volume}{B514}}, \bibinfo{pages}{245}.

\bibitem[{\citenamefont{Quigg and Rosner}(1977)}]{Quigg:1977dd}
\bibinfo{author}{\bibnamefont{Quigg}, \bibfnamefont{C.}}, and
  \bibinfo{author}{\bibfnamefont{J.~L.} \bibnamefont{Rosner}},
  \bibinfo{year}{1977}, \bibinfo{journal}{Phys. Lett.}
  \textbf{\bibinfo{volume}{B71}}, \bibinfo{pages}{153}.

\bibitem[{\citenamefont{Quigg and Rosner}(1979)}]{Quigg:1979vr}
\bibinfo{author}{\bibnamefont{Quigg}, \bibfnamefont{C.}}, and
  \bibinfo{author}{\bibfnamefont{J.~L.} \bibnamefont{Rosner}},
  \bibinfo{year}{1979}, \bibinfo{journal}{Phys. Rept.}
  \textbf{\bibinfo{volume}{56}}, \bibinfo{pages}{167}.

\bibitem[{\citenamefont{Quigg and Rosner}(1981)}]{Quigg:1981bj}
\bibinfo{author}{\bibnamefont{Quigg}, \bibfnamefont{C.}}, and
  \bibinfo{author}{\bibfnamefont{J.~L.} \bibnamefont{Rosner}},
  \bibinfo{year}{1981}, \bibinfo{journal}{Phys. Rev.}
  \textbf{\bibinfo{volume}{D23}}, \bibinfo{pages}{2625}.

\bibitem[{\citenamefont{Radford and Repko}(2007)}]{Radford:2007vd}
\bibinfo{author}{\bibnamefont{Radford}, \bibfnamefont{S.~F.}}, and
  \bibinfo{author}{\bibfnamefont{W.~W.} \bibnamefont{Repko}},
  \bibinfo{year}{2007}, \bibinfo{journal}{Phys. Rev.}
  \textbf{\bibinfo{volume}{D75}}, \bibinfo{pages}{074031}.

\bibitem[{\citenamefont{Rosner}(2001)}]{Rosner:2001nm}
\bibinfo{author}{\bibnamefont{Rosner}, \bibfnamefont{J.~L.}},
  \bibinfo{year}{2001}, \bibinfo{journal}{Phys. Rev.}
  \textbf{\bibinfo{volume}{D64}}, \bibinfo{pages}{094002}.

\bibitem[{\citenamefont{Rosner}(2004)}]{Rosner:2004ac}
\bibinfo{author}{\bibnamefont{Rosner}, \bibfnamefont{J.~L.}},
  \bibinfo{year}{2004}, \bibinfo{journal}{Phys. Rev.}
  \textbf{\bibinfo{volume}{D70}}, \bibinfo{pages}{094023}.

\bibitem[{\citenamefont{Rosner}(2005)}]{Rosner:2004wy}
\bibinfo{author}{\bibnamefont{Rosner}, \bibfnamefont{J.~L.}},
  \bibinfo{year}{2005}, \bibinfo{journal}{Ann. Phys.}
  \textbf{\bibinfo{volume}{319}}, \bibinfo{pages}{1}.

\bibitem[{\citenamefont{Rosner}(2006{\natexlab{a}})}]{Rosner:2006vc}
\bibinfo{author}{\bibnamefont{Rosner}, \bibfnamefont{J.~L.}},
  \bibinfo{year}{2006}{\natexlab{a}}, \bibinfo{journal}{Phys. Rev.}
  \textbf{\bibinfo{volume}{D74}}, \bibinfo{pages}{076006}.

\bibitem[{\citenamefont{Rosner}(2006{\natexlab{b}})}]{Rosner:2005gf}
\bibinfo{author}{\bibnamefont{Rosner}, \bibfnamefont{J.~L.}},
  \bibinfo{year}{2006}{\natexlab{b}}, \bibinfo{journal}{AIP Conf. Proc.}
  \textbf{\bibinfo{volume}{815}}, \bibinfo{pages}{218}, \bibinfo{note}{eprint
  hep-ph/0508155}.

\bibitem[{Rosner \emph{et~al.}(2005)\citenamefont{Rosner}
  \emph{et~al.}}]{Rosner:2005ry}
\bibinfo{author}{\bibnamefont{Rosner}, \bibfnamefont{J.~L.}}, \emph{et~al.}
  (\bibinfo{collaboration}{CLEO}), \bibinfo{year}{2005},
  \bibinfo{journal}{Phys. Rev. Lett.} \textbf{\bibinfo{volume}{95}},
  \bibinfo{pages}{102003}.

\bibitem[{Rosner \emph{et~al.}(2006)\citenamefont{Rosner}
  \emph{et~al.}}]{Rosner:2005eu}
\bibinfo{author}{\bibnamefont{Rosner}, \bibfnamefont{J.~L.}}, \emph{et~al.}
  (\bibinfo{collaboration}{CLEO}), \bibinfo{year}{2006},
  \bibinfo{journal}{Phys. Rev. Lett.} \textbf{\bibinfo{volume}{96}},
  \bibinfo{pages}{092003}.

\bibitem[{Rubin \emph{et~al.}(2005)\citenamefont{Rubin}
  \emph{et~al.}}]{Rubin:2005px}
\bibinfo{author}{\bibnamefont{Rubin}, \bibfnamefont{P.}}, \emph{et~al.}
  (\bibinfo{collaboration}{CLEO}), \bibinfo{year}{2005},
  \bibinfo{journal}{Phys. Rev.} \textbf{\bibinfo{volume}{D72}},
  \bibinfo{pages}{092004}.

\bibitem[{\citenamefont{Schonfeld} \emph{et~al.}(1980)\citenamefont{Schonfeld,
  Kwong, Rosner, Quigg, and Thacker}}]{Schonfeld:1979cd}
\bibinfo{author}{\bibnamefont{Schonfeld}, \bibfnamefont{J.~F.}},
  \bibinfo{author}{\bibfnamefont{W.}~\bibnamefont{Kwong}},
  \bibinfo{author}{\bibfnamefont{J.~L.} \bibnamefont{Rosner}},
  \bibinfo{author}{\bibfnamefont{C.}~\bibnamefont{Quigg}}, and
  \bibinfo{author}{\bibfnamefont{H.~B.} \bibnamefont{Thacker}},
  \bibinfo{year}{1980}, \bibinfo{journal}{Ann. Phys.}
  \textbf{\bibinfo{volume}{128}}, \bibinfo{pages}{1}.

\bibitem[{\citenamefont{Sebastian} \emph{et~al.}(1992)\citenamefont{Sebastian,
  Grotch, and Ridener}}]{Sebastian:1992xq}
\bibinfo{author}{\bibnamefont{Sebastian}, \bibfnamefont{K.~J.}},
  \bibinfo{author}{\bibfnamefont{H.}~\bibnamefont{Grotch}}, and
  \bibinfo{author}{\bibfnamefont{F.~L.} \bibnamefont{Ridener}},
  \bibinfo{year}{1992}, \bibinfo{journal}{Phys. Rev.}
  \textbf{\bibinfo{volume}{D45}}, \bibinfo{pages}{3163}.

\bibitem[{\citenamefont{Seth}(2004)}]{Seth:2004qc}
\bibinfo{author}{\bibnamefont{Seth}, \bibfnamefont{K.~K.}},
  \bibinfo{year}{2004}, \bibinfo{journal}{Phys. Rev.}
  \textbf{\bibinfo{volume}{D69}}, \bibinfo{pages}{097503}.

\bibitem[{\citenamefont{Seth}(2005)}]{Seth:2005ny}
\bibinfo{author}{\bibnamefont{Seth}, \bibfnamefont{K.~K.}},
  \bibinfo{year}{2005}, \bibinfo{journal}{Phys. Rev.}
  \textbf{\bibinfo{volume}{D72}}, \bibinfo{pages}{017501}.

\bibitem[{Severini \emph{et~al.}(2004)\citenamefont{Severini}
  \emph{et~al.}}]{Severini:2003qw}
\bibinfo{author}{\bibnamefont{Severini}, \bibfnamefont{H.}}, \emph{et~al.}
  (\bibinfo{collaboration}{CLEO}), \bibinfo{year}{2004},
  \bibinfo{journal}{Phys. Rev. Lett.} \textbf{\bibinfo{volume}{92}},
  \bibinfo{pages}{222002}.

\bibitem[{\citenamefont{Siegert}(1937)}]{Siegert:1937yt}
\bibinfo{author}{\bibnamefont{Siegert}, \bibfnamefont{A.~J.~F.}},
  \bibinfo{year}{1937}, \bibinfo{journal}{Phys. Rev.}
  \textbf{\bibinfo{volume}{52}}, \bibinfo{pages}{787}.

\bibitem[{\citenamefont{Skwarnicki}(2005)}]{Skwarnicki:2005pq}
\bibinfo{author}{\bibnamefont{Skwarnicki}, \bibfnamefont{T.}},
  \bibinfo{year}{2005}, in \emph{\bibinfo{booktitle}{40th Rencontres de Moriond
  On QCD and High Energy Hadronic Interactions}} (\bibinfo{address}{La Thuile,
  Aosta Valley, March 12-19, 2005}), \eprint{hep-ex/0505050}.

\bibitem[{Sokolov \emph{et~al.}(2007)\citenamefont{Sokolov}
  \emph{et~al.}}]{Sokolov:2006sd}
\bibinfo{author}{\bibnamefont{Sokolov}, \bibfnamefont{A.}}, \emph{et~al.}
  (\bibinfo{collaboration}{Belle}), \bibinfo{year}{2007},
  \bibinfo{journal}{Phys. Rev.} \textbf{\bibinfo{volume}{D75}},
  \bibinfo{pages}{071103}.

\bibitem[{\citenamefont{Stubbe and Martin}(1991)}]{Stubbe:1991qw}
\bibinfo{author}{\bibnamefont{Stubbe}, \bibfnamefont{J.}}, and
  \bibinfo{author}{\bibfnamefont{A.}~\bibnamefont{Martin}},
  \bibinfo{year}{1991}, \bibinfo{journal}{Phys. Lett.}
  \textbf{\bibinfo{volume}{B271}}, \bibinfo{pages}{208}.

\bibitem[{\citenamefont{Sucher}(1978)}]{Sucher:1978wq}
\bibinfo{author}{\bibnamefont{Sucher}, \bibfnamefont{J.}},
  \bibinfo{year}{1978}, \bibinfo{journal}{Rept. Prog. Phys.}
  \textbf{\bibinfo{volume}{41}}, \bibinfo{pages}{1781}.

\bibitem[{\citenamefont{Swanson}(2004{\natexlab{a}})}]{Swanson:2004pp}
\bibinfo{author}{\bibnamefont{Swanson}, \bibfnamefont{E.~S.}},
  \bibinfo{year}{2004}{\natexlab{a}}, \bibinfo{journal}{Phys. Lett.}
  \textbf{\bibinfo{volume}{B598}}, \bibinfo{pages}{197}.

\bibitem[{\citenamefont{Swanson}(2004{\natexlab{b}})}]{Swanson:2003tb}
\bibinfo{author}{\bibnamefont{Swanson}, \bibfnamefont{E.~S.}},
  \bibinfo{year}{2004}{\natexlab{b}}, \bibinfo{journal}{Phys. Lett.}
  \textbf{\bibinfo{volume}{B588}}, \bibinfo{pages}{189}.

\bibitem[{\citenamefont{Swanson}(2006)}]{Swanson:2006ap}
\bibinfo{author}{\bibnamefont{Swanson}, \bibfnamefont{E.~S.}},
  \bibinfo{year}{2006}, \bibinfo{journal}{AIP Conf. Proc.}
  \textbf{\bibinfo{volume}{870}}, \bibinfo{pages}{349}.

\bibitem[{\citenamefont{Thacker}
  \emph{et~al.}(1978{\natexlab{a}})\citenamefont{Thacker, Quigg, and
  Rosner}}]{Thacker:1977aq}
\bibinfo{author}{\bibnamefont{Thacker}, \bibfnamefont{H.~B.}},
  \bibinfo{author}{\bibfnamefont{C.}~\bibnamefont{Quigg}}, and
  \bibinfo{author}{\bibfnamefont{J.~L.} \bibnamefont{Rosner}},
  \bibinfo{year}{1978}{\natexlab{a}}, \bibinfo{journal}{Phys. Rev.}
  \textbf{\bibinfo{volume}{D18}}, \bibinfo{pages}{274}.

\bibitem[{\citenamefont{Thacker}
  \emph{et~al.}(1978{\natexlab{b}})\citenamefont{Thacker, Quigg, and
  Rosner}}]{Thacker:1977ar}
\bibinfo{author}{\bibnamefont{Thacker}, \bibfnamefont{H.~B.}},
  \bibinfo{author}{\bibfnamefont{C.}~\bibnamefont{Quigg}}, and
  \bibinfo{author}{\bibfnamefont{J.~L.} \bibnamefont{Rosner}},
  \bibinfo{year}{1978}{\natexlab{b}}, \bibinfo{journal}{Phys. Rev.}
  \textbf{\bibinfo{volume}{D18}}, \bibinfo{pages}{287}.

\bibitem[{\citenamefont{Tornqvist}(2003)}]{Tornqvist:2003na}
\bibinfo{author}{\bibnamefont{Tornqvist}, \bibfnamefont{N.~A.}},
  \bibinfo{year}{2003}, \eprint{hep-ph/0308277}.

\bibitem[{Uehara \emph{et~al.}(2006)\citenamefont{Uehara}
  \emph{et~al.}}]{Uehara:2005qd}
\bibinfo{author}{\bibnamefont{Uehara}, \bibfnamefont{S.}}, \emph{et~al.}
  (\bibinfo{collaboration}{Belle}), \bibinfo{year}{2006},
  \bibinfo{journal}{Phys. Rev. Lett.} \textbf{\bibinfo{volume}{96}},
  \bibinfo{pages}{082003}.

\bibitem[{Uehara \emph{et~al.}(2008)\citenamefont{Uehara}
  \emph{et~al.}}]{Uehara:2007vb}
\bibinfo{author}{\bibnamefont{Uehara}, \bibfnamefont{S.}}, \emph{et~al.}
  (\bibinfo{collaboration}{Belle}), \bibinfo{year}{2008},
  \bibinfo{journal}{Eur. Phys. J.} \textbf{\bibinfo{volume}{C53}},
  \bibinfo{pages}{1}.

\bibitem[{\citenamefont{Van~Royen and Weisskopf}(1967)}]{VanRoyen:1967nq}
\bibinfo{author}{\bibnamefont{Van~Royen}, \bibfnamefont{R.}}, and
  \bibinfo{author}{\bibfnamefont{V.~F.} \bibnamefont{Weisskopf}},
  \bibinfo{year}{1967}, \bibinfo{journal}{Nuovo Cim.}
  \textbf{\bibinfo{volume}{A50}}, \bibinfo{pages}{617},
  \bibinfo{note}{[Erratum-ibid.\ A {\bf 51}, 583 (1967)]}.

\bibitem[{\citenamefont{Voloshin}(1975)}]{Voloshin:1975yb}
\bibinfo{author}{\bibnamefont{Voloshin}, \bibfnamefont{M.~B.}},
  \bibinfo{year}{1975}, \bibinfo{journal}{JETP Lett.}
  \textbf{\bibinfo{volume}{21}}, \bibinfo{pages}{347}, \bibinfo{note}{[Pisma
  Zh. Eksp. Teor. Fiz. {\bf 21}, 733 (1975)]}.

\bibitem[{\citenamefont{Voloshin}(1979)}]{Voloshin:1978hc}
\bibinfo{author}{\bibnamefont{Voloshin}, \bibfnamefont{M.~B.}},
  \bibinfo{year}{1979}, \bibinfo{journal}{Nucl. Phys.}
  \textbf{\bibinfo{volume}{B154}}, \bibinfo{pages}{365}.

\bibitem[{\citenamefont{Voloshin}(1986)}]{Voloshin:1985em}
\bibinfo{author}{\bibnamefont{Voloshin}, \bibfnamefont{M.~B.}},
  \bibinfo{year}{1986}, \bibinfo{journal}{Sov. J. Nucl. Phys.}
  \textbf{\bibinfo{volume}{43}}, \bibinfo{pages}{1011}, \bibinfo{note}{[Yad.
  Fiz. {\bf 43}, 1571 (1986)]}.

\bibitem[{\citenamefont{Voloshin}(2003)}]{Voloshin:2003kn}
\bibinfo{author}{\bibnamefont{Voloshin}, \bibfnamefont{M.~B.}},
  \bibinfo{year}{2003}, \bibinfo{journal}{Phys. Lett.}
  \textbf{\bibinfo{volume}{B562}}, \bibinfo{pages}{68}.

\bibitem[{\citenamefont{Voloshin}(2004)}]{Voloshin:2004hs}
\bibinfo{author}{\bibnamefont{Voloshin}, \bibfnamefont{M.~B.}},
  \bibinfo{year}{2004}, \bibinfo{journal}{Mod. Phys. Lett.}
  \textbf{\bibinfo{volume}{A19}}, \bibinfo{pages}{2895}.

\bibitem[{\citenamefont{Voloshin}(2005)}]{Voloshin:2004nu}
\bibinfo{author}{\bibnamefont{Voloshin}, \bibfnamefont{M.~B.}},
  \bibinfo{year}{2005}, \bibinfo{journal}{Phys. Atom. Nucl.}
  \textbf{\bibinfo{volume}{68}}, \bibinfo{pages}{771}, \bibinfo{note}{[Yad.
  Fiz. {\bf 68}, 804 (2005)], eprint hep-ph/0601044}.

\bibitem[{\citenamefont{Voloshin}(2006)}]{Voloshin:2006ce}
\bibinfo{author}{\bibnamefont{Voloshin}, \bibfnamefont{M.~B.}},
  \bibinfo{year}{2006}, \bibinfo{journal}{Phys. Rev.}
  \textbf{\bibinfo{volume}{D74}}, \bibinfo{pages}{054022}.

\bibitem[{Wang \emph{et~al.}(2007)\citenamefont{Wang}
  \emph{et~al.}}]{Wang:2007ea}
\bibinfo{author}{\bibnamefont{Wang}, \bibfnamefont{X.~L.}}, \emph{et~al.}
  (\bibinfo{collaboration}{Belle}), \bibinfo{year}{2007},
  \bibinfo{journal}{Phys. Rev. Lett.} \textbf{\bibinfo{volume}{99}},
  \bibinfo{pages}{142002}.

\bibitem[{\citenamefont{Yan}(1980)}]{Yan:1980uh}
\bibinfo{author}{\bibnamefont{Yan}, \bibfnamefont{T.-M.}},
  \bibinfo{year}{1980}, \bibinfo{journal}{Phys. Rev.}
  \textbf{\bibinfo{volume}{D22}}, \bibinfo{pages}{1652}.

\bibitem[{Yao \emph{et~al.}(2006)\citenamefont{Yao} \emph{et~al.}}]{Yao:2006px}
\bibinfo{author}{\bibnamefont{Yao}, \bibfnamefont{W.~M.}}, \emph{et~al.}
  (\bibinfo{collaboration}{Particle Data Group}), \bibinfo{year}{2006},
  \bibinfo{journal}{J. Phys.} \textbf{\bibinfo{volume}{G33}},
  \bibinfo{pages}{1}.

\bibitem[{\citenamefont{Ye}(2006)}]{Ye:2006QWG}
\bibinfo{author}{\bibnamefont{Ye}, \bibfnamefont{S.}}
  (\bibinfo{collaboration}{Belle}), \bibinfo{year}{2006}, in
  \emph{\bibinfo{booktitle}{Quarkonium Working Group Meeting}}
  (\bibinfo{address}{Brookhaven Natl.\ Lab., June 27-30, 2006}),
  \bibinfo{note}{[unpublished]}.

\bibitem[{Yuan \emph{et~al.}(2007)\citenamefont{Yuan}
  \emph{et~al.}}]{Yuan:2007sj}
\bibinfo{author}{\bibnamefont{Yuan}, \bibfnamefont{C.~Z.}}, \emph{et~al.}
  (\bibinfo{collaboration}{Belle}), \bibinfo{year}{2007},
  \bibinfo{journal}{Phys. Rev. Lett.} \textbf{\bibinfo{volume}{99}},
  \bibinfo{pages}{182004}.

\bibitem[{\citenamefont{Zambetakis and Byers}(1983)}]{Zambetakis:1983te}
\bibinfo{author}{\bibnamefont{Zambetakis}, \bibfnamefont{V.}}, and
  \bibinfo{author}{\bibfnamefont{N.}~\bibnamefont{Byers}},
  \bibinfo{year}{1983}, \bibinfo{journal}{Phys. Rev.}
  \textbf{\bibinfo{volume}{D28}}, \bibinfo{pages}{2908}.

\bibitem[{\citenamefont{Zeng} \emph{et~al.}(1995)\citenamefont{Zeng, Van~Orden,
  and Roberts}}]{Zeng:1994vj}
\bibinfo{author}{\bibnamefont{Zeng}, \bibfnamefont{J.}},
  \bibinfo{author}{\bibfnamefont{J.~W.} \bibnamefont{Van~Orden}}, and
  \bibinfo{author}{\bibfnamefont{W.}~\bibnamefont{Roberts}},
  \bibinfo{year}{1995}, \bibinfo{journal}{Phys. Rev.}
  \textbf{\bibinfo{volume}{D52}}, \bibinfo{pages}{5229}.

\bibitem[{\citenamefont{Zhang} \emph{et~al.}(1991)\citenamefont{Zhang,
  Sebastian, and Grotch}}]{Zhang:1991et}
\bibinfo{author}{\bibnamefont{Zhang}, \bibfnamefont{X.}},
  \bibinfo{author}{\bibfnamefont{K.~J.} \bibnamefont{Sebastian}}, and
  \bibinfo{author}{\bibfnamefont{H.}~\bibnamefont{Grotch}},
  \bibinfo{year}{1991}, \bibinfo{journal}{Phys. Rev.}
  \textbf{\bibinfo{volume}{D44}}, \bibinfo{pages}{1606}.

\bibitem[{\citenamefont{Zhu}(2005)}]{Zhu:2005hp}
\bibinfo{author}{\bibnamefont{Zhu}, \bibfnamefont{S.-L.}},
  \bibinfo{year}{2005}, \bibinfo{journal}{Phys. Lett.}
  \textbf{\bibinfo{volume}{B625}}, \bibinfo{pages}{212}.

\bibitem[{\citenamefont{Zhu}(2006)}]{Zhu:2006uc}
\bibinfo{author}{\bibnamefont{Zhu}, \bibfnamefont{Y.-S.}}
  (\bibinfo{collaboration}{BES}), \bibinfo{year}{2006}, \bibinfo{journal}{AIP
  Conf. Proc.} \textbf{\bibinfo{volume}{814}}, \bibinfo{pages}{580}.

\end{thebibliography}

\end{document}